\documentclass[aps,a4paper,showpacs,superscriptaddress,twocolumn]{revtex4}

\usepackage{hyperref}
\usepackage{graphicx}
\usepackage[sort&compress]{natbib} 
\usepackage[english]{babel}
\usepackage{mathrsfs}
\usepackage{cancel}
\usepackage{amsmath}
\usepackage{amssymb}
\usepackage{bbm}
\usepackage{subfigure}

\begin{document}
\title{The analytic properties of the quark propagator from an effective infrared interaction model}
\date{\today}
\author{Andreas Windisch}
\email[Electronic address: ]{windisch@physics.wustl.edu}
\affiliation{Department of Physics, Washington University in St. Louis, MO 63130, USA}

\begin{abstract}
In this paper I investigate the analytic properties of the quark propagator Dyson-Schwinger equation (DSE) in the Landau gauge. In the quark self-energy, the combined gluon propagator and quark-gluon vertex is modeled by an effective interaction (the so-called Maris-Tandy interaction), where the ultraviolet term is neglected. This renders the loop integrand of the quark self-energy analytic on the cut-plane $-\pi<\arg(x)<\pi$ of the square of the external momentum. Exploiting the simplicity of the truncation, I study solutions of the quark propagator in the domain $x\in[-5.1,0]\mbox{GeV}^2\times i[0,10.2]\mbox{GeV}^2$. Because of a complex conjugation symmetry, this region fully covers the parabolic integration domain for Bethe-Salpeter equations (BSEs) for bound state masses of up to 4.5 GeV. 
Employing a novel numerical technique that is based on highly parallel computation on graphics processing units (GPUs), I extract more than 6500 poles in this region, which arise as the bare quark mass is varied over a wide range of closely spaced values. The poles are grouped in 23 individual 'trajectories', that capture the movement of the poles in the complex region as the bare mass is varied. The raw data of the pole locations and -residues is provided as supplemental material, which can be used to parametrize solutions of the complex quark propagator for a wide range of bare mass values and for large bound state masses. 
This study is a first step towards an extension of previous work on the analytic continuation of perturbative one-loop integrals, with the long-term goal of establishing a framework that allows for the numerical extraction of the analytic properties of the quark propagator with a truncation that extends beyond the rainbow by making adequate adjustments in the contour of the radial integration of the quark self-energy.
\end{abstract}
\maketitle

\section{\label{sec:intro}Introduction}

A successful approach to describe mesons as quark-antiquark bound states is the framework of
Dyson-Schwinger equations (DSEs) and Bethe-Salpeter equations (BSEs) \cite{Alkofer:2000wg,Fischer:2006ub,Roberts:2007jh,Sanchis-Alepuz:2015tha}, which complements lattice based studies, see e.g.\ \cite{Liu:2012ze,Thomas:2014dpa,Lang:2015sba}. Also baryons are described in the functional approach, see \cite{Eichmann:2016yit} for a recent review.
In order to solve the BSE, the quark propagator has to be computed in a (parabolically bounded) region in the complex plane of the square of the external momentum. 
Even though a thorough treatment of the quark propagator in principle involves simultaneous computation of the dressed quark-gluon vertex, together with the propagators and vertices of the Yang-Mills sector (see \cite{Williams:2015cvx}), it is still interesting to employ a truncation of the quark DSE where the (tree-level approximated) quark-gluon vertex and the gluon propagator are modeled by an effective interaction \cite{Maris:1999nt,Alkofer:2002bp,Qin:2011dd}. 
The analytic properties of the quark DSE
that arises when only the infrared (IR) part of the Maris-Tandy interaction \cite{Maris:1999nt} is taken into account has been the subject of a thorough study \cite{Dorkin:2013rsa,Dorkin:2014lxa}. Neglecting the ultra violet (UV) part of the interaction can be justified for light meson masses. 
In \cite{Dorkin:2013rsa} it has been found, that for meson masses of less than $M_{q\bar{q}}\sim 1$ GeV, the quark propagator is analytic in the complex domain relevant for the BSE, and it is fairly simple to evaluate the quark DSE at the required complex momenta. 
For larger meson masses, however, the IR Maris-Tandy modeled quark propagator features
(complex conjugate and real) poles within the parabolic region in the complex plane where the BSE has to be evaluated. For a consistent treatment, knowledge of the location and
residues of those poles is important \cite{Blank:2010pa,Dorkin:2013rsa,Dorkin:2014lxa}. Some of the pole locations, together with their residues, have been extracted in \cite{Dorkin:2013rsa,Dorkin:2014lxa}, where the parabolic region under consideration corresponded to bound state masses of up to $M_{q\bar{q}}\sim 3.5$ GeV. 

Employing the infrared part of the Maris-Tandy interaction \cite{Maris:1999nt}, the results presented in this paper are:
\begin{itemize}
\item \textit{All} poles within the parabolic region for bound state masses of up to $M_{q\bar{q}}= 4.5$ GeV have been identified, see Section \ref{sec:results}. Their positions and residues have been extracted by employing a novel, Graphics Processing Unit (GPU) based numerical framework that allows for quick, reliable and automatic extraction of the poles in the scalar and vector part of the propagator, see Section \ref{sec:numerical_procedures}.
\item The bare quark mass $m_0$ is varied from 5 MeV to 2500 MeV in increments of 5 MeV, and the movement of the poles in the complex plane is captured in 'trajectories'. For these 500 different mass values, a total of more than 6500 poles has been extracted, see Section \ref{sec:results}.
\item The bare mass dependence of the position of a given pole, as well as the corresponding mass dependence of the residues in the scalar and vector part of the quark propagator are presented as plots for quick reference in Sections \ref{sec:pole_location_discussion} and \ref{sec:residues}.
\item  The behavior of the poles on the real axis is studied in detail. It is found, that for bare mass values above $m_0\sim 455$ MeV, the segment of the real axis enclosed by the parabola for bound state masses of $M_{q\bar{q}}= 4.5$ GeV is free of real poles, see Section \ref{sec:poles_on_real_axis}.
\item The positions and residues of all poles found in this study are available as supplemental material published alongside with this paper, see Section \ref{sec:Supplemental_Material}. In this section, I also provide a step-by-step guide that shows how the data can be used.
\end{itemize}

The numerical framework developed throughout this study is a direct continuation of previous works, where a technique to solve perturbative one loop integrals in the complex domain has been established, see \cite{Windisch:2012zd,Windisch:2012sz} and \cite{Windisch:2013mg}. The paper at hand is considered to be the first step towards the long term goal of developing a non-perturbative framework that allows for the extraction of the analytic properties of the quark propagator for more complicate truncation schemes, such that issues of positivity violation and confinement, as recently studied for the Yang-Mills sector \cite{Strauss:2012dg}, can be addressed. The main goal is to investigate the impact of certain tensor structures in the quark-gluon vertex on the property of positivity violation, as suggested in \cite{Alkofer:2003jj}.

\section{\label{sec:rainbow_quark}The rainbow truncated Landau gauge quark propagator Dyson-Schwinger equation}

\begin{figure}[hbt]
\label{fig1}
\centering
\includegraphics[width=9cm]{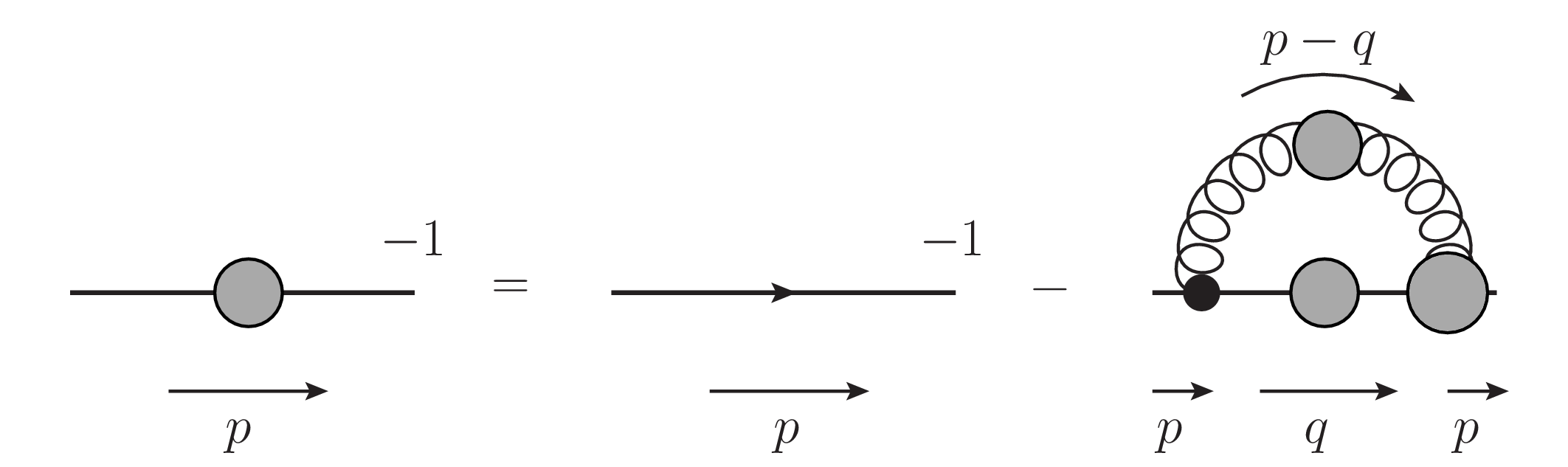}
\caption{A diagrammatic representation of the Landau gauge quark propagator Dyson-Schwinger equation. The momentum routing is shown explicitly.}
\end{figure}

A diagrammatic representation of the quark propagator Dyson-Schwinger equation is shown in Figure \ref{fig1}.

The dressed (inverse) quark propagator on the left is composed of all possible ways of propagation: the 'bare' propagation, without the occurrence of any intermediate interaction, and the quark self-energy constituted by the quark emitting a gluon via the tree-level vertex, after which both, the quark and the gluon propagate with all quantum corrections as 'dressed' quantities, to finally rejoin in all possible ways through a dressed vertex. Every dressed quantity is depicted by a big blob, while the bare propagator is just a line without a blob, and the tree-level vertex is denoted by a small black dot. The algebraic expression corresponding to this diagrammatic representation requires the following ingredients.
The (inverse) quark propagator is a Lorentz scalar with Dirac structure and one associated momentum, such that it can be spanned by two basis elements,

\begin{eqnarray}
S^{-1}\left(p\right) & = & \delta^{\alpha\beta}\left(i\cancel{p}\ A\left(p^{2}\right)+B\left(p^{2}\right)\mathbbm{1}_D\right)\label{eq:3},
\end{eqnarray}

where $A\left(p^2\right)$ and $B\left(p^2\right)$ are the dressing functions. Note that this study is carried out in Euclidean space. Here I consider one flavor only, and the propagator is also diagonal in color space, as indicated by the $\delta^{\alpha\beta}$.
This is readily inverted, to give
 
\begin{eqnarray}
S\left(p\right) & = & \delta^{\alpha\beta}\frac{-i\cancel{p}\ A\left(p^{2}\right)+B\left(p^{2}\right)\mathbbm{1}_D}{p^{2}A^{2}\left(p^{2}\right)+B^{2}\left(p^{2}\right)}.\label{eq:4}
\end{eqnarray}

Consequently, the bare inverse quark propagator reads
\begin{equation}
S_{0}^{-1}\left(p^{2}\right)=\delta^{\alpha\beta}\left(Z_{2}\left(i\cancel{p}+Z_{m}m_{0}\right)\right)\label{eq:5},
\end{equation}

where I furthermore introduced the wave function- and mass renormalization constants $Z_2$ and $Z_m$.
The free Landau gauge gluon propagator reads

\begin{eqnarray}
D_{free}^{\mu\nu}\left(p\right) & = & \delta^{ab}\frac{1}{p^{2}}\left(\delta^{\mu\nu}-\frac{p^{\mu}p^{\nu}}{p^{2}}\right),\label{eq:6}
\end{eqnarray}

where the indices $a$ and $b$ are associated with color in the adjoint representation.
Finally, the bare quark-gluon vertex is given by

\begin{equation}
\Gamma_0^\mu = Z_{1F}i g \left(t_a\right)_{\alpha\beta}\gamma^\mu,
\label{bare_vertex}
\end{equation}

with $t_a$ being the $a^{th}$ generator of $SU(3)_c$, $\alpha$ and $\beta$ are color indices with respect to the fundamental representation, and $Z_{1F}$ is the renormalization constant associated with the coupling.
The rainbow truncated quark propagator DSE uses a bare vertex instead of the dressed one, that is,

\begin{equation}
i g \left(t_a\right)_{\alpha\beta} \Gamma^\mu \rightarrow i g \left(t_a\right)_{\alpha\beta}\gamma^\mu.
\label{dressed_vertex}
\end{equation}

This simplifies the complexity of the quark equation considerably, since, in Landau gauge and in the vacuum, the dressed vertex still requires eight (transverse) tensor structures as basis elements. Because a bare vertex is used here, an Ansatz for the combined gluon- and quark-gluon interaction term is employed. Keeping this general for now, the combined term is written as

\begin{equation}
\label{eq:model_interaction}
Z_{1F}g^2D^{\mu\nu}(k)\Gamma^\nu(q,p)=:k^2\mathcal{G}\left(k^2\right)D_{free}^{\mu\nu}(k)\gamma^\nu,
\end{equation}
where $k=p-q$, and $\mathcal{G}\left(k^2\right)$ is an effective interaction term that I specify later.

Having collected all the ingredients, the algebraic expression for the rainbow truncated quark-propagator Dyson-Schwinger equation becomes 

\begin{eqnarray}
S^{-1}\left(p\right) & = & S_{0}^{-1}\left(p\right)+\int\frac{d^{4}q}{\left(2\pi\right)^{4}}\left[\mathcal{G}\left(\left(p-q\right)^{2}\right)\right.\nonumber \\
 &  & \times\left.\left(p-q\right)^{2}D_{free}^{\mu\nu}\left(p-q\right)\gamma^{\mu}S\left(q\right)\gamma^{\nu}\right].\label{eq:7}
\end{eqnarray}

where the change of sign in the self-energy term comes from the two imaginary units going with the two vertices.
The color structure of the equation evaluates to the quadratic Casimir operator and contributes a prefactor of $\frac{4}{3}$ in front of the quark self energy, see Appendix \ref{app:A}.
The coupling renormalization constant has been absorbed into the model of the effective interaction,
thus, there are only two renormalization constants to be fixed: the remaining
wave function renormalization constant $Z_{2}$ and the mass renormalization
constant $Z_{m}.$ 
In order to solve the equation numerically it is convenient
to isolate the quark propagator dressing functions $A$ and
$B$, which is achieved by multiplying the equation on both sides from the left with an appropriate term, and performing a Dirac trace afterwards,

\begin{eqnarray}
A(p) & = & Z_2 + \frac{1}{4p^{2}}\mbox{Tr}_{D}\left\{-i\cancel{p}\Sigma(p) \right\} ,\label{eq:8}\\
B(p) & = & Z_2 Z_m m_0 + \frac{1}{4}\mbox{Tr}_{D}\left\{ \Sigma(p)\right\} ,\label{eq:9}
\end{eqnarray}

where $\Sigma(p)$ is the quark self energy after dealing with color space,

\begin{eqnarray}
\label{sigma}
\Sigma(p) &=& \frac{4}{3}\int\frac{d^{4}q}{\left(2\pi\right)^{4}}\left[\mathcal{G}\left(\left(p-q\right)^{2}\right)\right.\nonumber \\
 &  & \times\left.\left(p-q\right)^{2}D_{free}^{\mu\nu}\left(p-q\right)\gamma^{\mu}S\left(q\right)\gamma^{\nu}\right].
\end{eqnarray}

The traces of the self-energy in Dirac space are presented in Appendix \ref{app:A}, see equations (\ref{sigma_A_1}) and (\ref{sigma_B_1}). The two coupled integral equations for the dressing functions $A$ and $B$ are then given by

\begin{eqnarray}
A\left(p^2\right) & = & Z_2 + \frac{4}{3p^2}\int\frac{d^4q}{(2\pi)^4}\frac{A\left(q^2\right)\mathcal{G}\left((p-q)^2\right)}{q^2A^2\left(q^2\right)+B^2\left(q^2\right)}\nonumber\\
&&\times\left(2(q.p)+\frac{(p^2+q^2)(q.p)-2p^2q^2}{(p-q)^2}\right)\nonumber,\\
\end{eqnarray}

\begin{eqnarray}
B\left(p^2\right) & = & Z_2 Z_m m_0 + \frac{4}{3}\int\frac{d^4q}{(2\pi)^4}\frac{3B\left(q^2\right)\mathcal{G}\left((p-q)^2\right)}{q^2A^2\left(q^2\right)+B^2\left(q^2\right)}.\nonumber\\ 
\end{eqnarray}

Switching to hyperspherical coordinates and integrating the two trivial
angles (see Appendix \ref{app:A} for details), the equations become

\begin{eqnarray}
A\left(x\right) & = & Z_{2}+ \frac{1}{6\pi^3}\int_\varepsilon^\Lambda dy y\frac{A\left(y\right)}{yA^2\left(y\right)+B^2\left(y\right)}\\
&&\times\int_{-1}^{+1}dz\sqrt{1-z^2}\mathcal{G}\left(x+y-2\sqrt{x}\sqrt{y}z\right)\nonumber\\
&&\times\left(2\frac{\sqrt{y}}{\sqrt{x}}z+\frac{(1+\frac{y}{x})\sqrt{x}\sqrt{y}z-2y}{x+y-2\sqrt{x}\sqrt{y}z}\right)\nonumber,
\end{eqnarray}

\begin{eqnarray}
B\left(x\right) & = & Z_{2}Z_{m}m_{0} + \frac{1}{6\pi^3}\int_\varepsilon^\Lambda dy y\frac{3B\left(y\right)}{yA^2\left(y\right)+B^2\left(y\right)}\nonumber\\
&&\times\int_{-1}^{+1}dz\sqrt{1-z^2}\mathcal{G}\left(x+y-2\sqrt{x}\sqrt{y}z\right)\nonumber.\\
\end{eqnarray}

In order to fix the renormalization constants $Z_{2}$ and $Z_{m}$, one can
demand that the dressing functions $A$ and $B$ become one and $m_{0}$ at
the renormalization point respectively, by employing a momentum subtraction (MOM) renormalization
scheme, see e.g. \cite{Fischer:2003zc}. However, in this study I only consider the infrared part of the Maris-Tandy model. 
I thus do not apply such a renormalization scheme, but simply put the renormalization
constants to one. An explicit evaluation of the self-energy at a renormalization scale of $\zeta=19$ GeV revealed that the integrals to be subtracted evaluate to values that are indeed negligibly small.


\section{\label{sec:MT_IR}The infrared part of the Maris-Tandy interaction}

In this section I discuss the structure of the quark propagator DSE as arising from the IR part of the Maris-Tandy (MT) interaction model, \cite{Maris:1999nt}. The full interaction model is given by 
\begin{eqnarray}
\label{eq:MT_model}
Z_{1F}\ g^{2}\frac{\mathscr{G}}{k^{2}} & = & \frac{4\pi^{2}}{\omega^{6}}Dk^{2}e^{-\frac{k^{2}}{\omega^{2}}}\label{eq:1}\\
 &  & +4\pi^{2}\frac{\frac{12}{33-2N_{f}}}{\frac{1}{2}\ln\left[e^{2}-1+\left(1+\frac{k^{2}}{\Lambda_{QCD}^{2}}\right)^{2}\right]}\mathscr{F}\left(k^{2}\right),\nonumber 
\end{eqnarray}
with
\begin{eqnarray}
\mathscr{F}\left(k^{2}\right) & = & \frac{1}{k^{2}}\left(1-e^{-\frac{k^{2}}{4m_{t}^{2}}}\right),\label{eq:2}
\end{eqnarray}
and the parameters are $N_{f}=4$, $\Lambda_{QCD}^{Nf=4}=0.234$
GeV, $\omega=0.3$ GeV, $D=1.25$ $\mbox{GeV}^{2}$ and $m_{t}=0.5$
GeV.
As discussed in \cite{Windisch:2013dxa}, the angular integral could induce branch cuts in the complex plane of the radial integration variable once the external momentum square becomes a complex number. If this is the case, the radial integration contour must be deformed in order to avoid the branch cuts. 
Here, however, similar to the studies \cite{Alkofer:2002bp,Dorkin:2013rsa,Dorkin:2014lxa}, I only consider the IR part of the model, that is, the first term in equation (\ref{eq:MT_model}). This simplifies the computation of the quark propagator DSE in the complex plane significantly, since the integrand of the loop in the quark self-energy is an analytic function in the cut-plane $-\pi< \arg\left(p^2\right)<\pi$. This in turn implies that the contour of the self-energy loop integral can be kept solely real, and it is sufficient to have knowledge of the positive real-axis solution of the dressing functions $A$ and $B$. This can be easily verified by following the procedure presented in \cite{Windisch:2013dxa}.
At this point, a further remark that concerns the more recently introduced Qin-Chang (QC) interaction \cite{Qin:2011dd} is in order. It would, of course, be very interesting to perform a similar analysis as the one presented in the sections below in a scenario where the Qin-Chang interaction is employed. However, for the momentum routing used in this study, even if one restricts oneself to the IR part of the QC model, only parts of the quark self-energy integral can be treated as easily as in the case of the MT interaction. In particular, parts of the quark self-energy integrand feature branch cuts that require appropriate contour adjustments away from the real axis. While this can be done in principle, it is beyond the scope of this initial study on the subject, and will be considered in a future publication.
In order to allow for easy comparison with the results presented in \cite{Dorkin:2013rsa}, I will use the parametrization presented by Alkofer, Watson and Weigel (AWW) \cite{Alkofer:2002bp}, which is dimensionally different from the original MT parametrization. In order to emphasize that, I will henceforth add the subscript 'AWW' to the parameters $D$ and $\omega$.
The IR part of the MT interaction in AWW parametrization is given by
\begin{eqnarray}
\mathcal{G}_{\mbox{\tiny{AWW}}}(x)=\frac{4\pi^2D_{\mbox{\tiny{AWW}}}}{\omega_{\mbox{\tiny{AWW}}}^2}x\exp\left\{\frac{-x}{\omega_{\mbox{\tiny{AWW}}}^2}\right\},
\end{eqnarray}
where I use $D_{\mbox{\tiny{AWW}}}=16\ \mbox{GeV}^2$ and $\omega_{\mbox{\tiny{AWW}}}=0.5\ \mbox{GeV}$.
Since the actual structure of the integrals plays an important role once complex arguments are considered, I present the coupled equations for this interaction model explicitly,

\begin{eqnarray}
&&A\left(x\right) =  \\
&&1 + \frac{D_{\mbox{\tiny{AWW}}}}{\omega_{\mbox{\tiny{AWW}}}^2}\int_\varepsilon^\Lambda \frac{dyyA\left(y\right)}{yA^2\left(y\right)+B\left(y\right)}\nonumber\\
&&\times\int_{-1}^{+1}dz\sqrt{1-z^2}\exp\left\{-\frac{(x+y-2\sqrt{x}\sqrt{y}z)}{\omega_{\mbox{\tiny{AWW}}}^2}\right\}\nonumber\\
&&\times\frac{2}{\pi}\left[\sqrt{x}\sqrt{y}z\left(1+\frac{y}{x}\right)-\frac{2}{3}y-\frac{4}{3}yz^2\right],\nonumber
\end{eqnarray}

\begin{eqnarray}
&& B\left(x\right) = \\ 
&&m_{0} + \frac{D_{\mbox{\tiny{AWW}}}}{\omega_{\mbox{\tiny{AWW}}}^2}\int_\varepsilon^\Lambda  \frac{dyyB\left(y\right)}{yA^2\left(y\right)+B^2\left(y\right)}\nonumber\\
&&\times\int_{-1}^{+1}dz\sqrt{1-z^2}\exp\left\{\frac{-(x+y-2\sqrt{x}\sqrt{y}z)}{\omega_{\mbox{\tiny{AWW}}}^2}\right\}\nonumber\\
&&\times\frac{2}{\pi}\left[x+y-2\sqrt{x}\sqrt{y}z\right].\nonumber
\end{eqnarray}

In this particular case, the angular integral can be solved everywhere in the cut-plane $-\pi< \arg\left(p^2\right)<\pi$. The solution can be expressed in terms of \textit{modified Bessel Functions of the First Kind} \cite{Alkofer:2002bp}. After the angular integration (see Appendix \ref{app:ang_int} for details), the coupled equations become  

\begin{eqnarray}
\label{eq:A_final}
&&A\left(x\right) =  \\
&&1 + D_{\mbox{\tiny{AWW}}}\int_\varepsilon^\Lambda \frac{dyyA\left(y\right)}{yA^2\left(y\right)+B^2\left(y\right)}\nonumber\\
&&\times \exp\left\{-\frac{x+y}{\omega_{\mbox{\tiny{AWW}}}^2} \right\}\bigg[\left(1+\frac{y}{x}+\frac{2\omega_{\mbox{\tiny{AWW}}}^2}{x}\right)I_2\left(\frac{2\sqrt{x}\sqrt{y}}{\omega_{\mbox{\tiny{AWW}}}^2}\right)\nonumber\\
&&-2\frac{\sqrt{y}}{\sqrt{x}}I_1\left(\frac{2\sqrt{x}\sqrt{y}}{\omega_{\mbox{\tiny{AWW}}}^2}\right)\bigg],\nonumber
\end{eqnarray}

\begin{eqnarray}
\label{eq:B_final}
&&B\left(x\right) =  \\
&&m_0 + D_{\mbox{\tiny{AWW}}}\int_\varepsilon^\Lambda \frac{dyyB\left(y\right)}{yA^2\left(y\right)+B^2\left(y\right)}\nonumber\\
&&\times \exp\left\{-\frac{x+y}{\omega_{\mbox{\tiny{AWW}}}^2} \right\}\bigg[\left(\frac{\sqrt{x}}{\sqrt{y}}+\frac{\sqrt{y}}{\sqrt{x}}\right)I_1\left(\frac{2\sqrt{x}\sqrt{y}}{\omega_{\mbox{\tiny{AWW}}}^2}\right)\nonumber\\
&&-2I_2\left(\frac{2\sqrt{x}\sqrt{y}}{\omega_{\mbox{\tiny{AWW}}}^2}\right)\bigg]\nonumber,
\end{eqnarray}
where $I_n(z),\ z\in\mathbb{C}\setminus\mathbb{R}^-$ are the modified Bessel Functions of the First Kind.
Equations (\ref{eq:A_final}) and (\ref{eq:B_final}) are the central object of this study. Once real and complex solutions have been obtained, the scalar and vector part of the propagator can be computed,
\begin{eqnarray} 
\sigma_S(p)=\frac{B\left(p\right)}{p^2A^2\left(p\right)+B^2\left(p\right)},
\end{eqnarray} 
\begin{eqnarray} 
\sigma_V(p)=\frac{A\left(p\right)}{p^2A^2\left(p\right)+B^2\left(p\right)}.
\end{eqnarray} 
The real and complex solutions of these quantities are discussed in detail in the following sections.
\section{\label{sec:numerical_procedures}Numerical implementation}
\subsection{Positive real axis}
\begin{figure*}[t]
\centering
\subfigure[][]{
 \label{fig:dsesol_a}
\includegraphics[width=0.45\hsize]{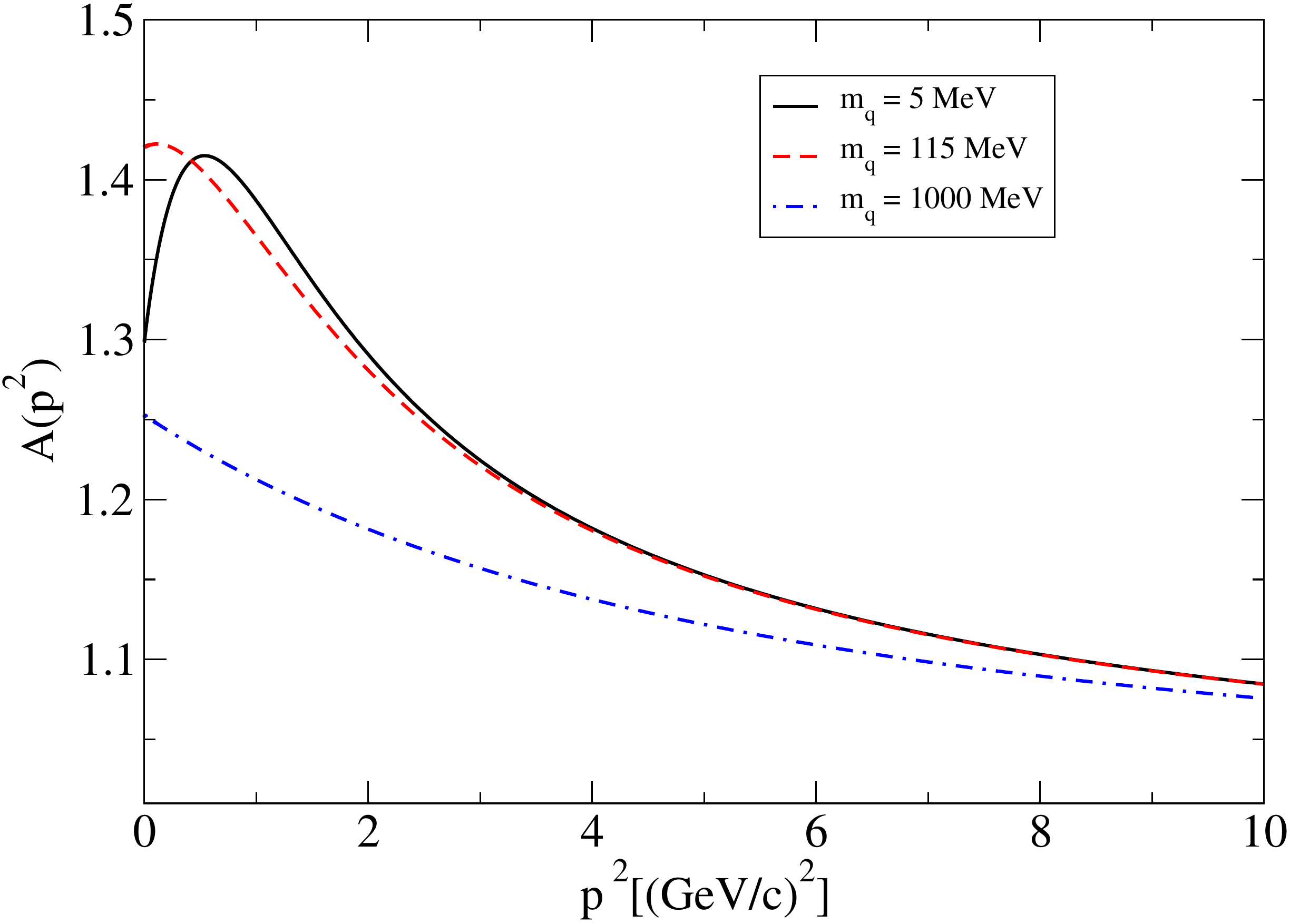}
}\hspace{8pt}
\subfigure[][]{
 \label{fig:dsesol_b}
\includegraphics[width=0.45\hsize]{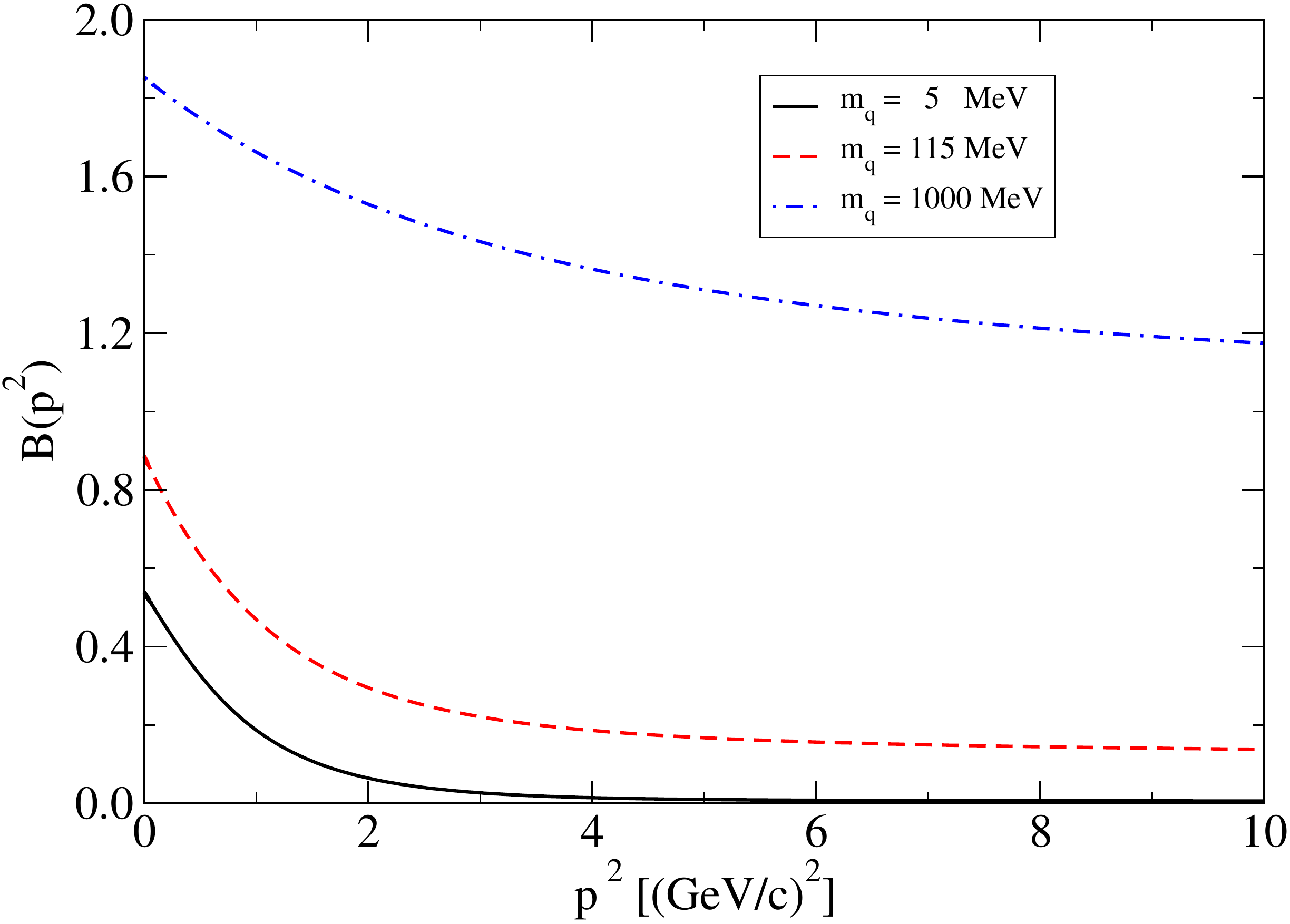}
}\\
\subfigure[][]{
 \label{fig:dsesol_c}
\includegraphics[width=0.45\hsize]{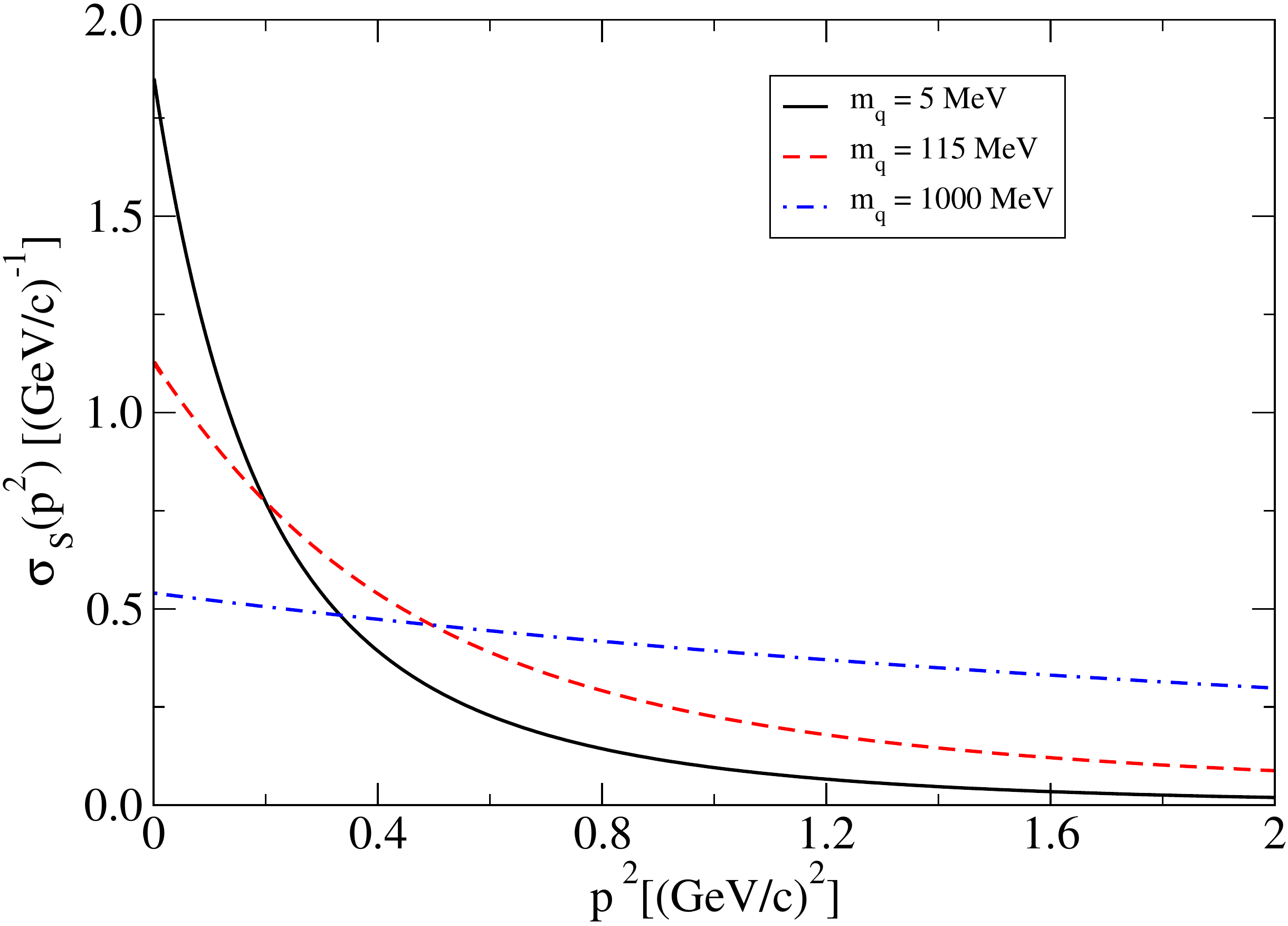}
}
\hspace{8pt}
\subfigure[][]{
 \label{fig:dsesol_d}
\includegraphics[width=0.45\hsize]{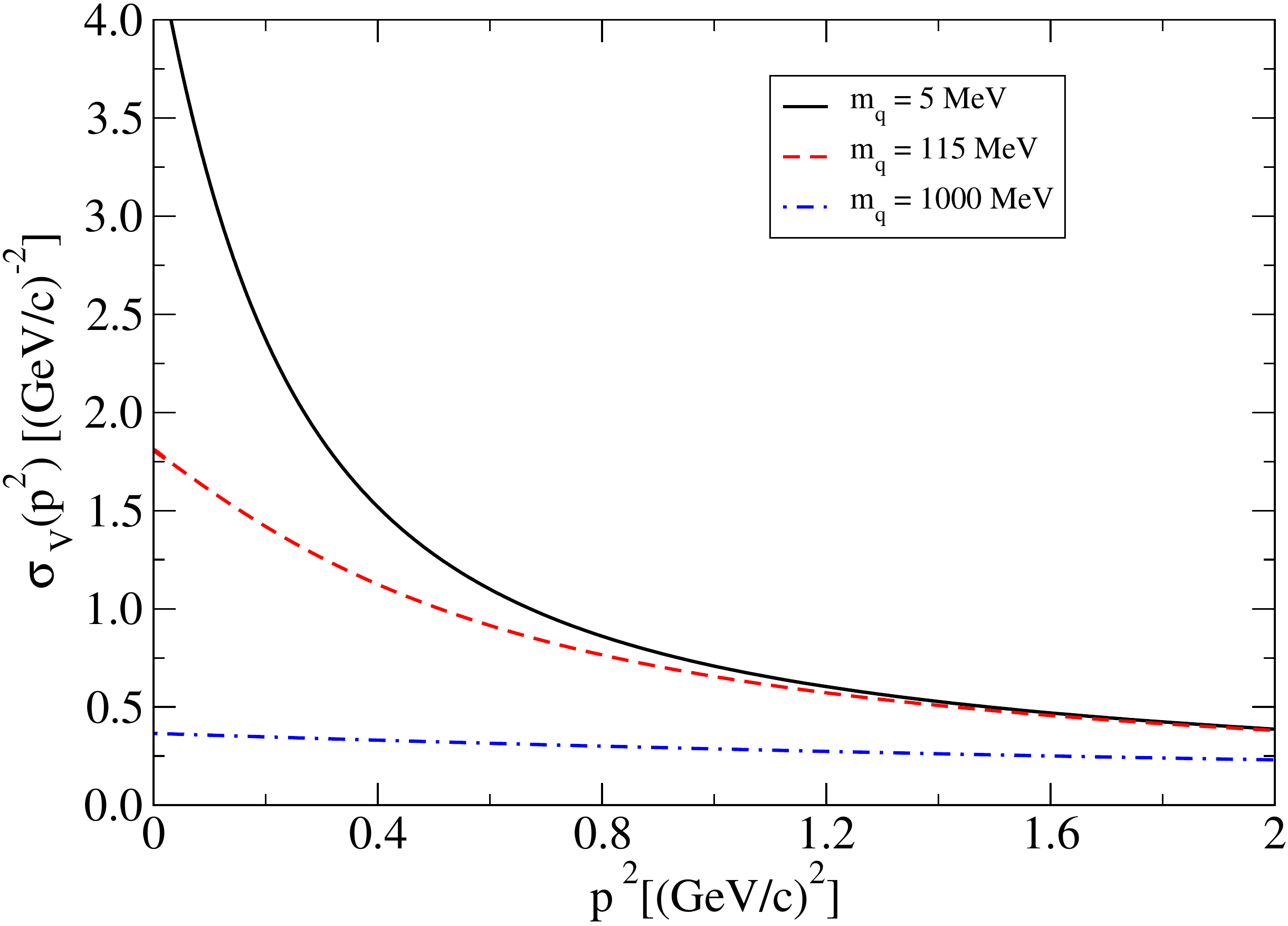}
}
\caption[]{Plots of the quark propagator dressing functions \subref{fig:dsesol_a} $A\left(p^2\right)$ and \subref{fig:dsesol_b} $B\left(p^2\right)$, as well as the alternative functions \subref{fig:dsesol_c} $\sigma_S\left(p^2\right)$ and \subref{fig:dsesol_d} $\sigma_V\left(p^2\right)$. The plots have been generated to provide a comparison with the solutions presented in \cite{Dorkin:2013rsa}, thereby validating the graphics-processing unit (GPU) parallelized code used in this study.}
\label{fig:dsesol}
\end{figure*}

In this study, a novel numerical technique for solving the quark propagator DSE has been implemented, which exploits the parallel computing capabilities of Graphics Processing Units (GPUs). Since this approach is technically more involved, this also introduces a source of errors that is not present if one employs a sequentially executed CPU code.
Thus, as a first step, the real-axis solution is produced and compared with known results. Similar to \cite{Dorkin:2013rsa}, bare masses of $m_0^{u,d}=0.005\ \mbox{GeV}$, $m_0^s=0.115\ \mbox{GeV}$ and $m_0^c=1\ \mbox{GeV}$ have been employed. The positive real axis results are summarized in Figure \ref{fig:dsesol}, which has been arranged in a similar fashion as Figure 2 in \cite{Dorkin:2013rsa} for easy comparison. The results are in very good agreement.
\subfiglabelskip=0pt
\subsection{Complex domain}
\begin{figure}[hbt]
\centering
\includegraphics[width=0.5\textwidth]{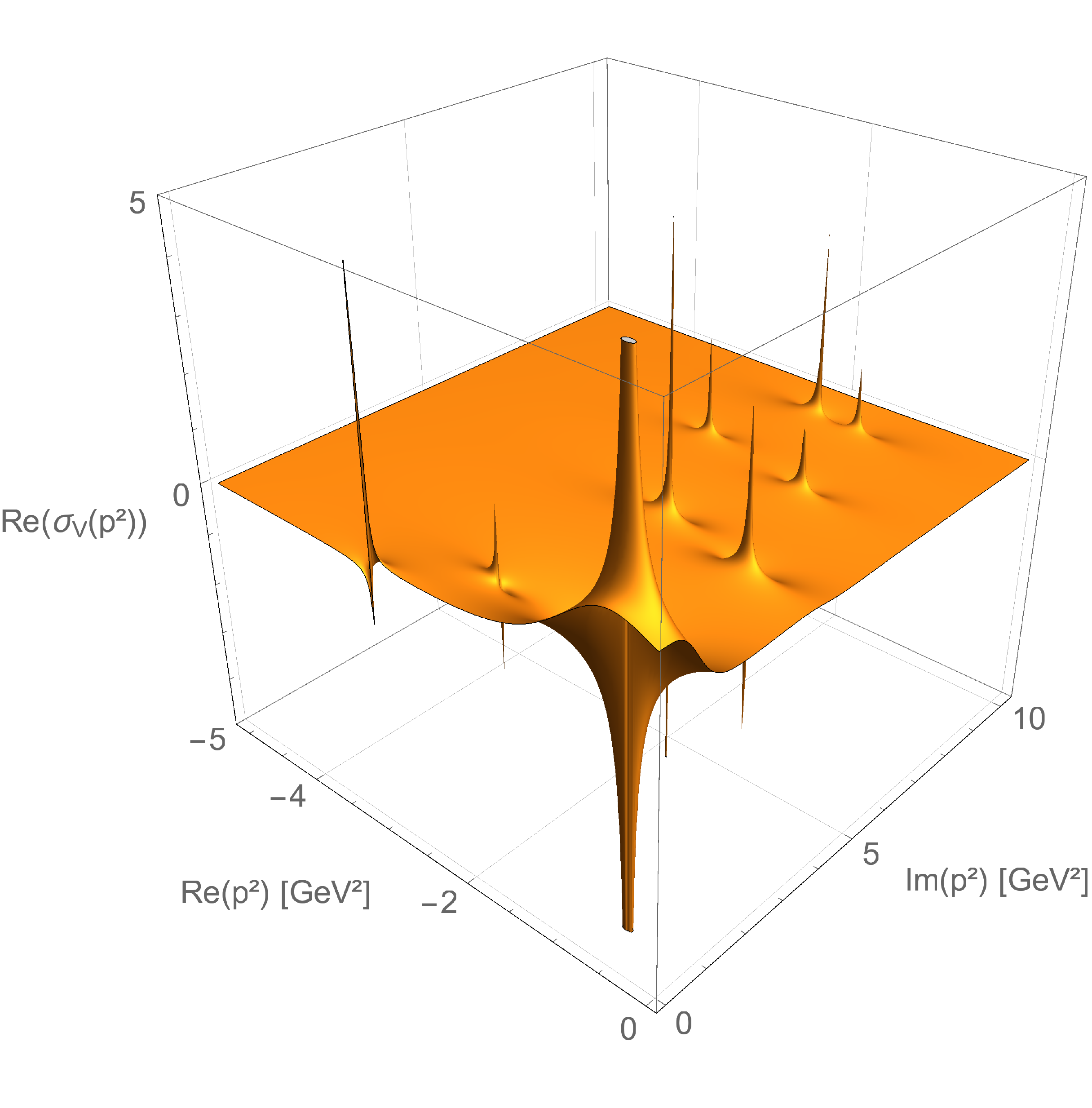}
\caption{This figure shows the real part of the solution for $\sigma_V\left(p^2\right)$ for $m_0=225$ MeV, obtained in the region $p^2\in [-5.1,0]\times i[0,10.2]$ of the square of the external momentum.}
\label{fig:solution_225_MeV}
\end{figure}

The second step and main goal of this study is to compute the solution of the quark propagator in the complex domain.
 To this end, first, the real axis solution is obtained. For the angular integral, both, the exact solution in terms of Bessel functions, as well as explicit numerical integration based on Gauss-Chebyshev quadrature has been used. Even though the iterative procedure that yields the dressing functions $A$ and $B$ has been performed on the GPU, the Bessel functions were pre-computed on the CPU using \cite{Amos:1986app}, because of the lack of an adequate GPU library. Both approaches, exact angular integration and numerical treatment of the angular integral, produce the same result and are consistent within numerical precision. Once the real axis solution has been obtained for a given bare mass value $m_0$, the propagator is evaluated in the rectangular region $[-5.1,0]\mbox{ GeV}^2\times i[0,10.2]\mbox{ GeV}^2$ of the square of the external momentum $x=p^2$, which completely contains half of the parabolic region that extends into the Euclidean timelike domain for meson masses of $M_{q\bar{q}}=4.5$ GeV. The other half follows from a complex conjugation symmetry, which has been verified explicitly by an additional calculation below the real axis, see Section \ref{sec:results}. The rectangular region has been discretized on a $850\times 850$ lattice, which provides a resolution of $6\times10^{-3}\mbox{ GeV}^2$ for the real part, and $12\times10^{-3}\mbox{ GeV}^2$ for the imaginary part. Since only a radial integration has to be performed, even for more than thousand Gauss-Legendre quadrature nodes for the radial integration, the GPU execution for all $850^2$ points takes only a few seconds, thereby producing the complex dressing functions $A$ and $B$. The pre-computation of the Bessel functions takes several minutes. However, once the Bessel functions are known, a solution for a different bare mass value can be obtained within seconds, since the Bessel functions only have to be evaluated for values within the cube $[-5.1,0]\mbox{ GeV}^2\times i[0,10.2]\mbox{ GeV}^2\times [\varepsilon,\Lambda]\mbox{ GeV}^2 $ spanned by the external and internal momenta, but they do not depend on the bare mass.
This allowed for a numerically very efficient scan of a large range of bare mass values, starting at $m_0=5$ MeV, and ranging up to $m_0=2500$ MeV in increments of $5$ MeV. For each of these 500 complex solutions for $A$ and $B$, a pole search- and analysis procedure has been employed.
As an example, Figure \ref{fig:solution_225_MeV} shows the solution for $\Re\sigma_V$ for the arbitrarily chosen bare mass of $m_0=225$ MeV.

\subsection{\label{sec:algorithm}Pole search and -analysis strategy}
\begin{figure}[hbt]
\centering
\includegraphics[width=0.5\textwidth]{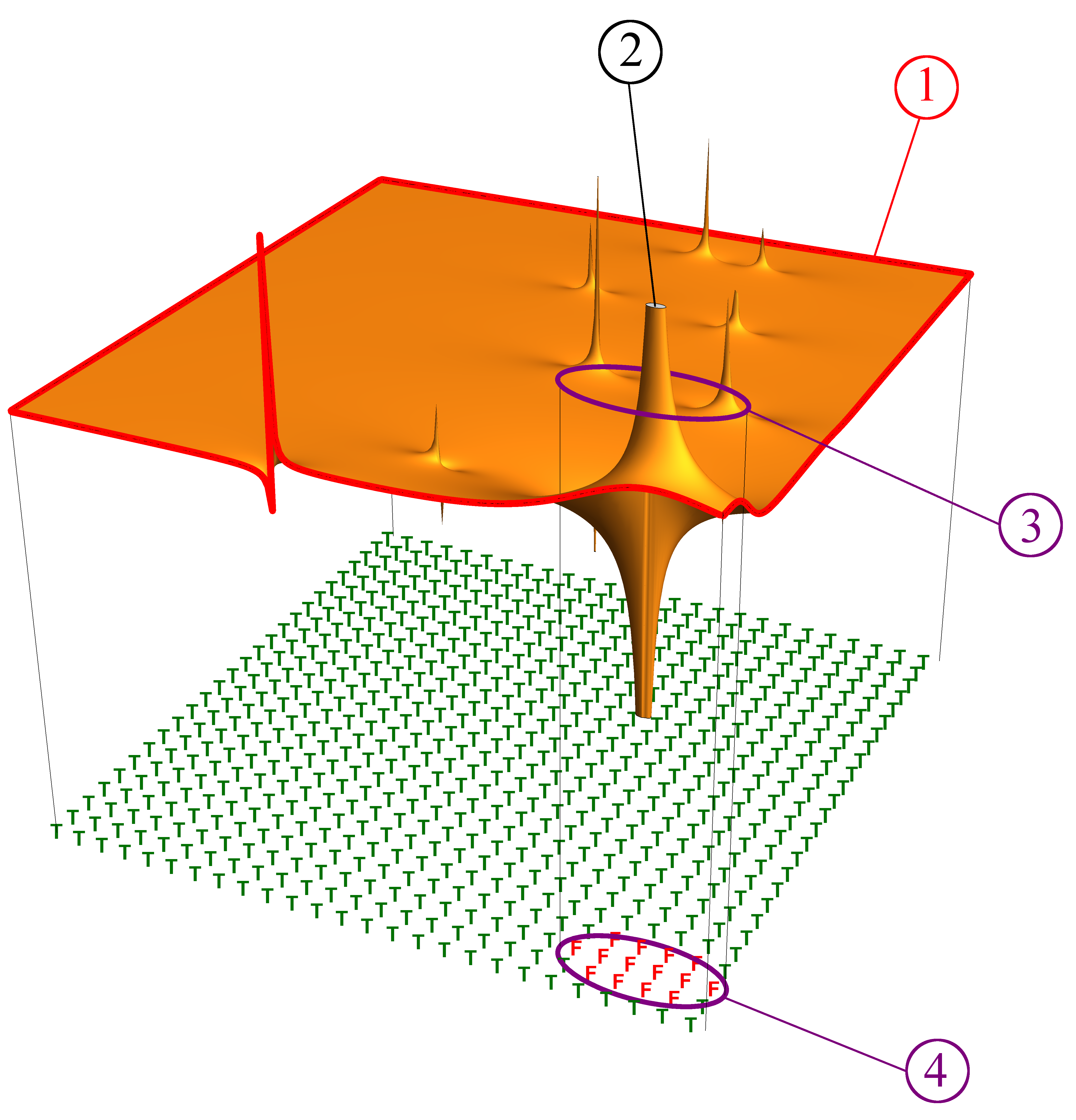}
\caption{The real part of the solution for $\sigma_V\left(p^2\right)$ for $m_0=225$ MeV features one pole on the real axis, and 8 further poles at locations with non-trivial imaginary part. Various steps that are used for the automated identification of the poles are highlighted and are referred to in the discussion of the pole search procedure in the main text.}
\label{fig:pole_search_example}
\end{figure}

Once a complex solution $A$ and $B$ has been obtained for a given bare mass, a pole search and -analysis has been conducted. Instead of using the dressing functions $A$ and $B$, the scalar and vector part of the propagator, $\sigma_S$ and $\sigma_V$, has been computed. Since the real axis solutions for $A$ and $B$ are free of poles, and since the self-energy integrand is analytic everywhere except along the negative real axis, poles arising in the functions $\sigma_S$ and $\sigma_V$ in the complex domain must coincide with the zeros of the denominator $p^2A^2(p)+B^2(p)$. Thus, similar to the strategy used in \cite{Dorkin:2013rsa}, I exploit the Cauchy argument principle to learn how many zeros are to be expected in the complex region in which the solution has been obtained. 
Once the number of zeros is known, a simple, yet effective algorithm searches local maxima and performs a circular contour integral around each local maximum until the number of zeros matches the number of identified poles. Let me discuss the procedure in detail by considering an example. In every step of the procedure, I will refer to Figure \ref{fig:pole_search_example} where each step is shown explicitly for the arbitrarily chosen example of $m_0=225$ MeV.
\begin{itemize}
\item \textbf{STEP 1: Determine number of zeros}\\
After obtaining the complex and real solutions for $A$ and $B$ for a given mass, the first step is to obtain the number of zeros ($N_z$) of the denominator of $\sigma_S$ and $\sigma_V$. This is done by exploiting Cauchy's argument principle,
\begin{eqnarray}
\label{eq:Nz}
N_z &=& \frac{1}{2\pi i}\oint_\gamma d\xi^2\frac{[\xi^2A^2(\xi)+B^2(\xi)]'_{\xi^2}}{\xi^2A^2(\xi)+B^2(\xi)}\\
&=& \frac{1}{2\pi i}\oint_\gamma d\xi^2\frac{A^2(\xi)+2\xi^2A(\xi)A(\xi)'+2B(\xi)B(\xi)'}{\xi^2A^2(\xi)+B^2(\xi)}.\nonumber
\end{eqnarray}
The contour is chosen to be the boundary of the complex region of evaluation, that is, 
\begin{eqnarray}
\gamma:&& (-5.1,0)\rightarrow (0,0)\rightarrow (0,10.2)\rightarrow \\
&&(-5.1,10.2)\rightarrow (-5.1,0),\nonumber
\end{eqnarray}
where the first entry in the tuple corresponds to the real part of $\xi^2$ and the second entry to the imaginary part of $\xi^2$. In the example of $m_0=225$ MeV shown in Figure \ref{fig:pole_search_example}, the contour $\gamma$ is shown in red and  is labeled as item \textcircled{\scriptsize{1}} (note that the figure shows the real part of $\sigma_V$, so, in fact, while the contour is the same, the integrand that enters (\ref{eq:Nz}) is complex and consists only of the denominator).
In the example of Figure \ref{fig:pole_search_example}, this integral evaluates to $N_z=8.503$.
The fact that this number is half integer indicates the existence of a pole sitting exactly on the integration contour, which, in this case, happens to be a segment of the real axis. The eight remaining poles are clearly visible as sharp spikes in the Figure. The expected number of poles to be extracted is thus 9.  
\item \textbf{STEP 2: Find the global maximum of the masked matrix}\\
Since the GPU produces the complex solution on an $850\times 850$ lattice, the local maxima of the real part of $\sigma_V$ can be used to locate the poles. Of course, one can also use the imaginary part, or $\sigma_S$ for that matter, but it suffices to consider $\Re\sigma_V$. Apart from the complex matrices that hold the values for $x,\ A,\ B,\ \sigma_S$, and $ \sigma_V$, a Boolean matrix of the same size is maintained (in Figure \ref{fig:pole_search_example}, the Boolean matrix is shown as an array of green values of 'T' and red values of 'F', located underneath the plot for $\Re\sigma_V$). The initial setup of this matrix is that all of its values are 'True'. Next, the (CUDA-) Fortran \cite{CUDA-Fortran:2016aa} intrinsic function that provides the location of the global maximum is called, where the Boolean matrix serves as a mask in the sense that only those parts of the matrix $\Re\sigma_V$ are considered for the maximum search for which the corresponding Boolean entry is 'True'. Since on initial time of the search all entries are set to 'True', this just yields the global maximum ($z_0$) of the array. In the example shown in Figure \ref{fig:pole_search_example}, the global maximum $z_0$ happens to be the pole labeled as item \textcircled{\scriptsize{2}}. 
\item \textbf{STEP 3: Exploit the residue theorem}\\
The previous step provided a location $z_0$ of a possible pole, thus it has to be checked whether the residue of $\sigma_V$ at this point is zero or non-zero. To be more explicit, the following quantities are computed at point $z_0$,
\begin{eqnarray}
\mbox{Res}(\sigma_S;z_0)&=&\frac{1}{2\pi i}\oint_\gamma\sigma_S(z)dz,\\
\mbox{Res}(\sigma_V;z_0)&=&\frac{1}{2\pi i}\oint_\gamma\sigma_V(z)dz.
\end{eqnarray}
The contour used for this computation is a circle of a fixed radius (I used $r=0.1$ GeV$^2$). In order to increase the precision of the calculation of the residues, the values of the integrands are not interpolated, but recomputed for at least 1024 integration points per contour. The contour is shown in Figure \ref{fig:pole_search_example} as item \textcircled{\scriptsize{3}} (the circular contour appears elliptic in the figure because the extent of the complex region in imaginary direction is twice its real extent, while the lattice is a square). If one of the four values, $\mbox{Res}(\Re\sigma_S),\ \mbox{Res}(\Im\sigma_S),\ \mbox{Res}(\Re\sigma_V)$ or $\mbox{Res}(\Im\sigma_V)$ is greater than $10^{-8}$, $z_0$ is considered to be a singular point, and the location and residues are stored. If all four values are below that number, then $z_0$ is considered as a regular point, and the computed residues, as well as the location are disregarded.
\item \textbf{STEP 4: Update the Boolean mask}\\
Regardless of whether the previous step found a pole or not, the point (area) of consideration has to be 'masked out', such that a repetitive process can be engaged. This is achieved by setting entries of the Boolean matrix that correspond to neighboring points of $z_0$ to 'False', such that a subsequent global maximum search conducted by \textbf{STEP 2} produces the next  pole location with a great chance. It turned out to be practical to use the same radius as for the residue integral to define this neighborhood. Note that this radius also limits the minimal resolution at which two adjacent poles can be identified as being separated. In Figure \ref{fig:pole_search_example}, this step is labeled as item \textcircled{\scriptsize{4}}. Next, one can continue with \textbf{STEP 2}, and run the procedure as long as necessary until all $N_z$ poles have been identified.
\end{itemize}  
 
\section{\label{sec:results}Results}
In this section, the results produced by using the numerical procedures outlined in Section \ref{sec:numerical_procedures} above, are presented. Overall, more than 6500 pole locations for a total of 500 different bare mass values have been found and their residues have been extracted. The full table of all pole locations and residues is provided as supplemental material, see Section \ref{sec:Supplemental_Material}. However, for quick consultations, all locations and residues as functions of the bare mass are included as plots in the main text in this section. The area in the complex plane for which solutions have been sought has been chosen such that the full parabolic region for meson bound state masses of up to $M_{q\bar{q}}=4.5\ \mbox{GeV}$ is covered.

\subsection{Complex conjugate nature of the solution}
\begin{figure}[hbt]
\centering
\includegraphics[width=0.45\textwidth]{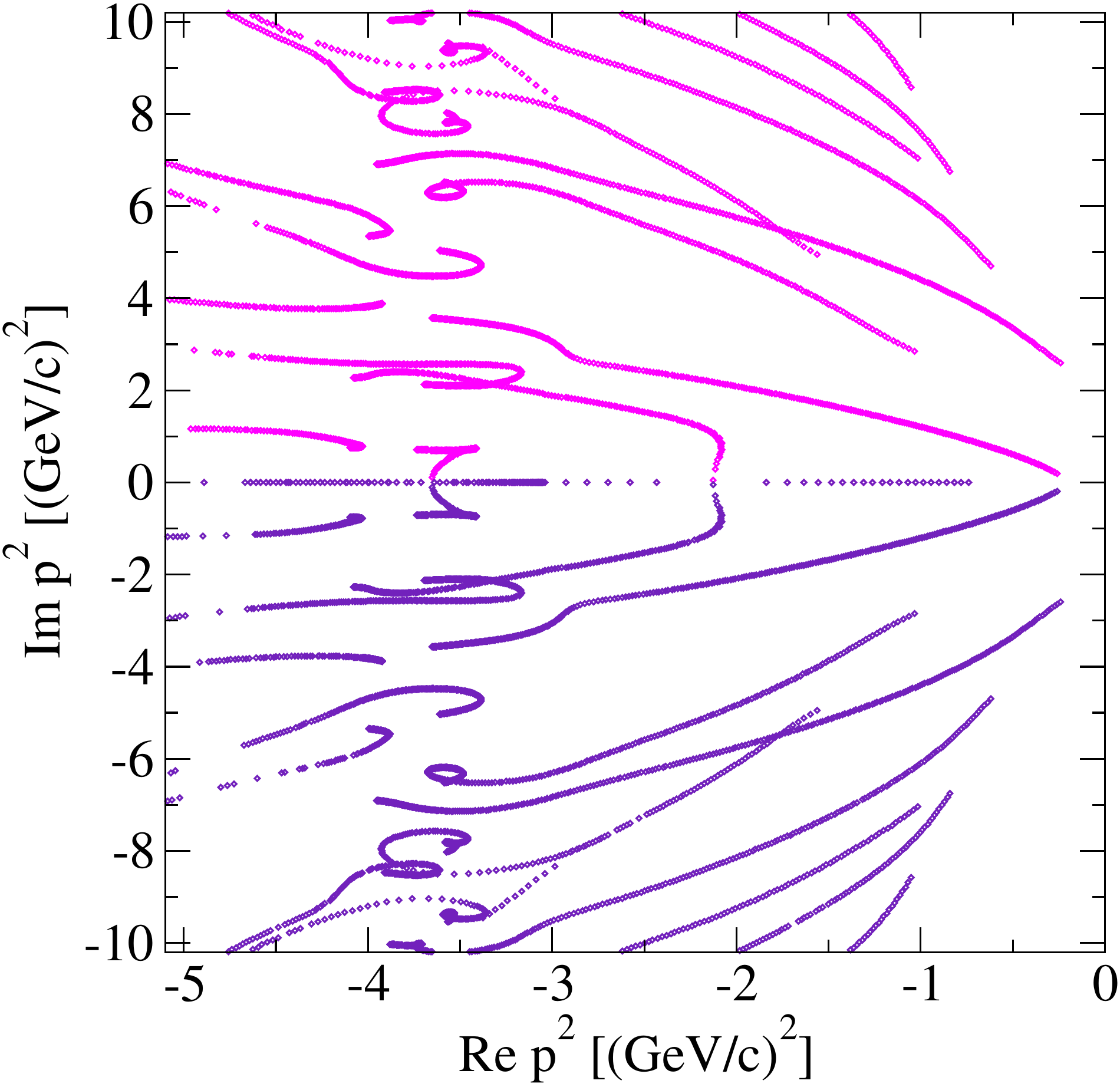}
\caption{All poles that have been detected as the bare mass was varied from 5 MeV to 2500 MeV in increments of 5 MeV. While the following analysis has been restricted to the second quadrant, the third quadrant has also been calculated in order to verify the complex conjugate nature of the poles. Overall, for all mass parameters and in both quadrants together, more than 13000 poles, as well as the corresponding residues in $\sigma_S$ and $\sigma_V$ have been extracted.}
\label{fig:all_poles}
\end{figure}

For the main result of this study, I focused on a rectangular region in the complex plane with negative real parts are positive imaginary parts. The axes have been included and constitute two of the four boundary edges of the lattice. Since the evaluation of the complex solution is performed in a highly parallelized manner, it is computationally rather cheap to explicitly perform a second computation in a rectangular region of the same size as the one for which the results are presented, but for negative real parts and negative imaginary parts. The combined results of both regions are shown in Figure \ref{fig:all_poles}, where the complex conjugate nature of the pole positions is evident. Thus, in the following, the discussion will be based solely on the region above, and including the real axis.

\subsection{\label{sec:pole_location_discussion}Pole locations with varying bare mass \texorpdfstring{$m_0$}{m0}}
\begin{figure*}[t]
\centering
\includegraphics[width=\textwidth]{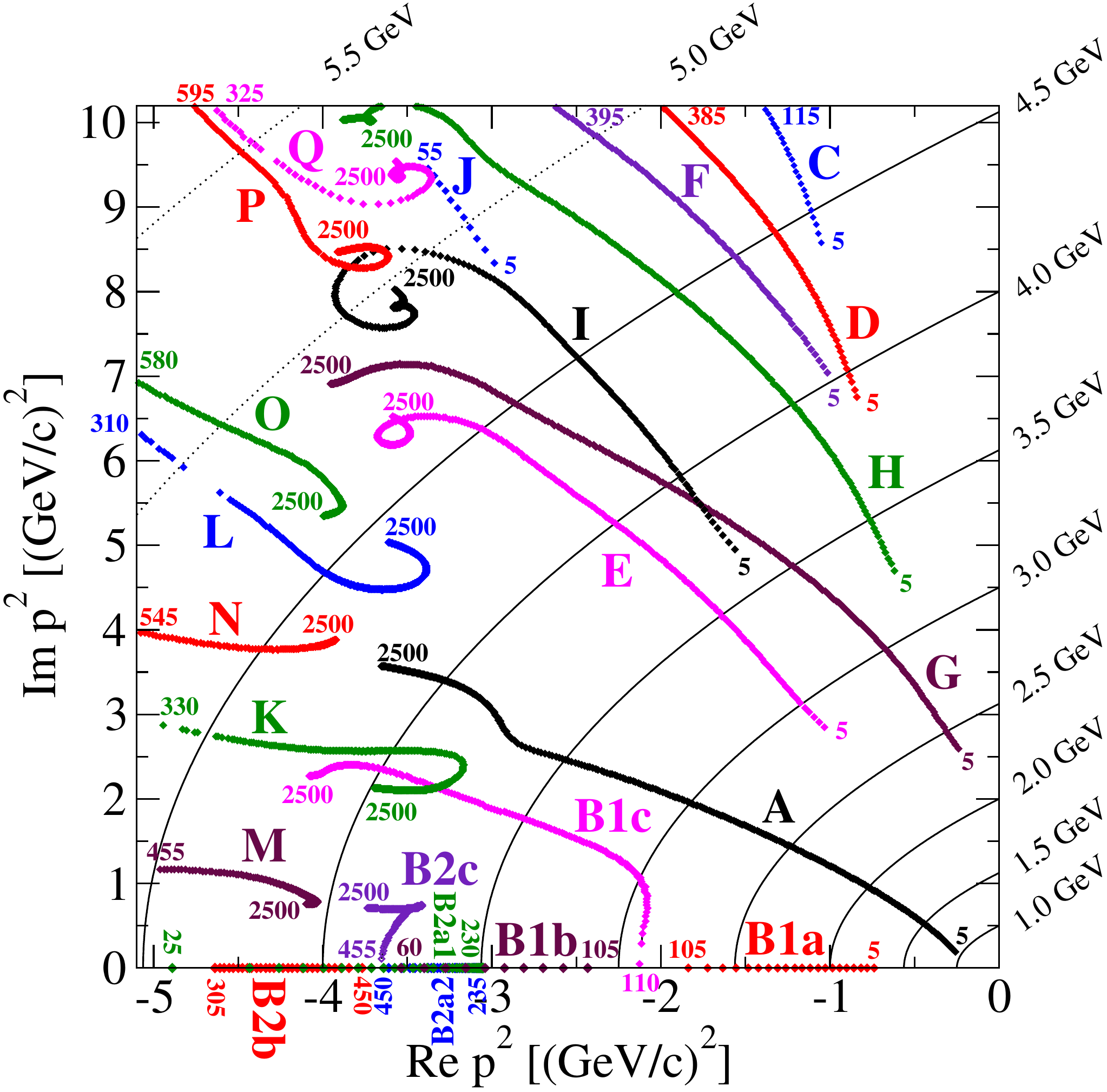}
\caption{This figure shows the main result of this study, obtained in the region $p^2\in [-5.1,0]\times i[0,10.2]$ of the square of the external momentum. For the sake of completeness, the behavior of the poles on the real axis is included. A detailed analysis of the trajectories of the real poles is presented in Section \ref{sec:poles_on_real_axis}. The plot is to be read as follows. Trajectories which do not posses any real valued poles are self-explanatory. The bare mass value of the occurrence of the first pole is explicitly provided (in MeV), and also the last mass value for which a particular trajectory has been tracked. The highest mass value in the computation is 2500 MeV, and, apart from trajectory J, all trajectories which have an endpoint within the domain specified above, end at this value. The scale that emerges around $p^2=-4$ GeV$^2$ is discussed in the main text.}
\label{fig:all_pole_trajectories}
\end{figure*}

Figure \ref{fig:all_pole_trajectories} summarizes the main result of this study. It shows the movement (trajectories) of the poles in the region $[-5.1,0]\ \mbox{GeV}^2\times i[0,10.2]\ \mbox{GeV}^2$, as the bare mass $m_0$ is varied from 5 MeV to 2500 MeV in increments of 5 MeV. Overall, 23 trajectories have been identified and have been assigned a letter (from A to Q) for identification. The letter B (together with appropriate sub-labels) has been used multiple times, since those trajectories are related to one another, yet they have different features which makes an individual identification fruitful. Figure \ref{fig:all_pole_trajectories} contains a lot of information, such that a thorough description is in order. Firstly, as also most prominently evident in Figure \ref{fig:all_poles}, a scale emerges for external momenta with a real part of just above -4 GeV$^2$. This scale could not be attributed to any of the scales involved in the calculation (IR and UV cutoff scale, mass scale), and could possibly indicate a breakdown of the numerical procedure. However, varying the cut-offs or number of integration nodes did not affect the emergence of this behavior. Since this scale emerges in a region that is only relevant for relatively high bound state masses, where the approximation of the interaction that lacks an UV term becomes less justified anyway, and because poles enter this region predominantly for very high bare masses, I did not pursue a more profound analysis of this phenomenon.\\
Having addressed this, let me continue by discussing Figure \ref{fig:all_pole_trajectories}. In addition to the trajectories, the parabolic integration domains as relevant for Bethe-Salpeter equations for bound states of different masses have been included as a guidance, where the respective bound state mass is denoted outside of the plot region. For instance, the region bound by the 3.5 GeV parabola requires, depending on the bare mass, inclusion of the poles on the trajectories A, B1a, B1b, B1c, E, G and H. Each trajectory that does not have a vanishing imaginary part is in principle self explanatory. For example, the trajectory labeled 'A' arises for a bare mass of 5 MeV around the point $x=-0.25$ GeV$^2 +\ i\ 0.19$ GeV$^2$, and ends around the point $x=-3.65$ GeV$^2 +\ i\ 3.57$ GeV$^2$ for a bare mass value of 2500 MeV. Whenever a trajectory enters or leaves the region in which the quark DSE has been solved, the respective mass value for which the region is left/entered is provided near that point. The behavior of the poles on the real axis is more involved and is discussed separately in Section \ref{sec:poles_on_real_axis}.

\begin{figure*}
\centering
\subfigure[][]{
 \label{fig:pos_a}
\includegraphics[width=0.45\hsize]{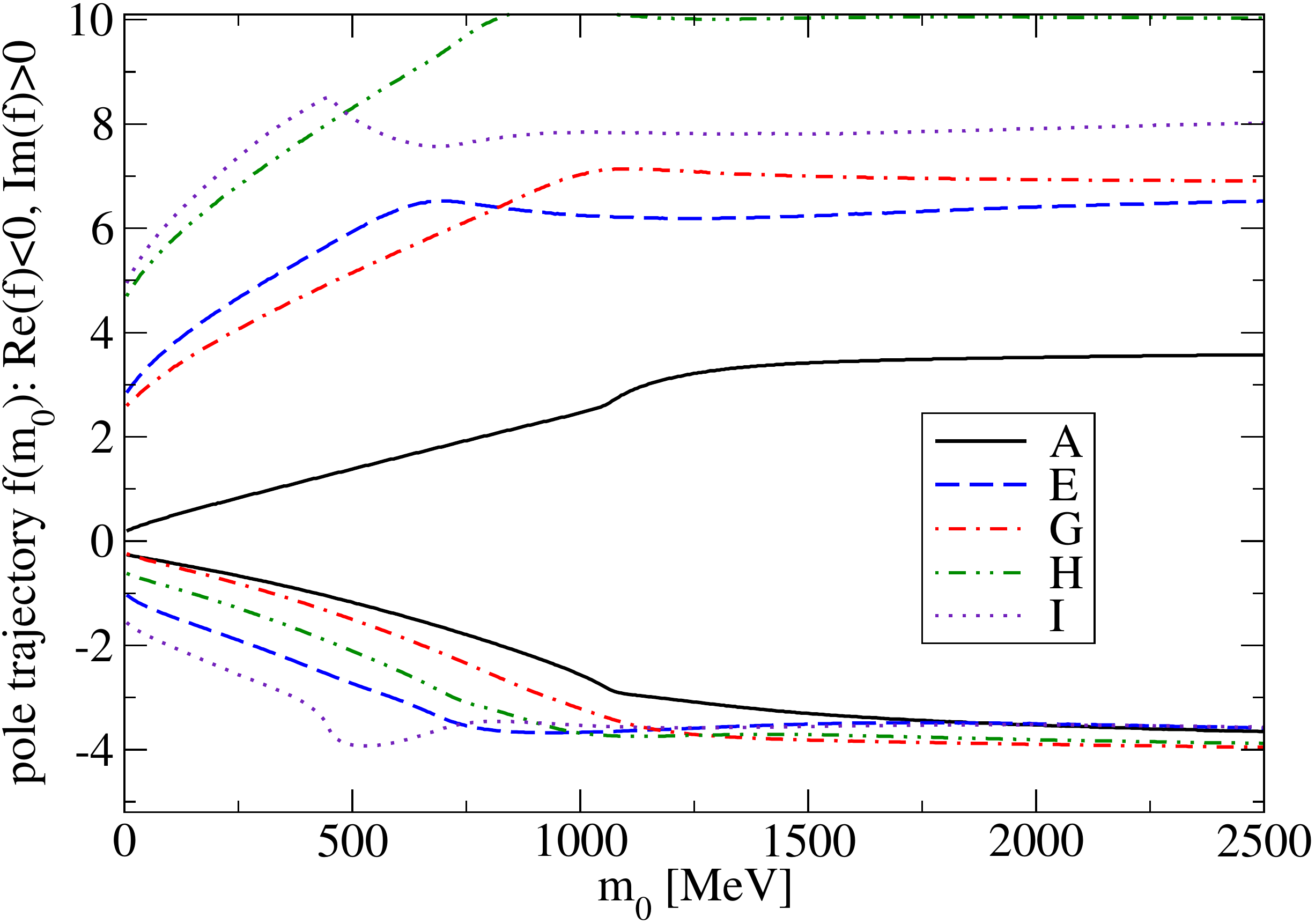}
}\hspace{8pt}
\subfigure[][]{
 \label{fig:pos_b}
\includegraphics[width=0.45\hsize]{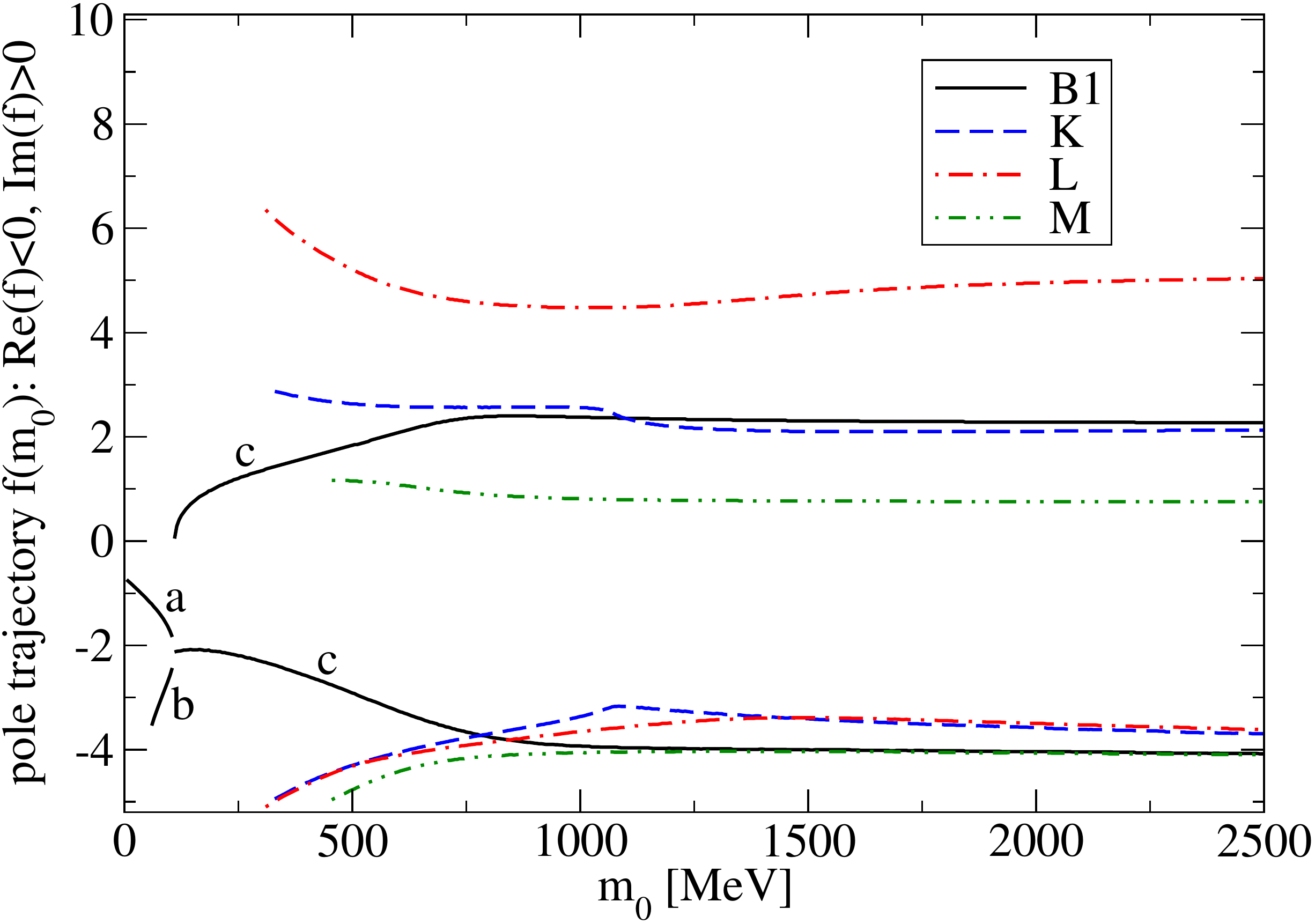}
}\\
\subfigure[][]{
 \label{fig:pos_c}
\includegraphics[width=0.45\hsize]{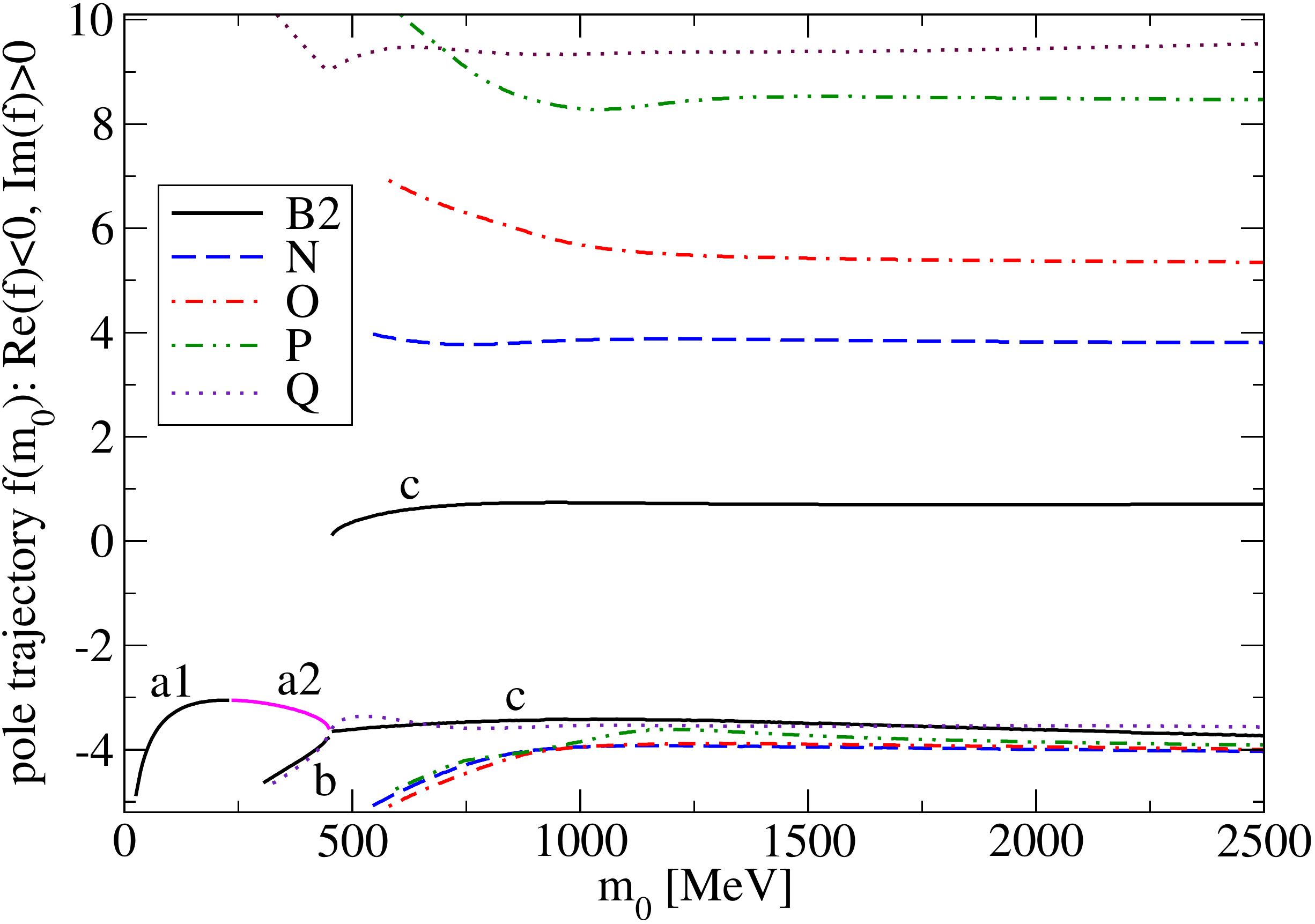}
}
\hspace{8pt}
\subfigure[][]{
 \label{fig:pos_d}
\includegraphics[width=0.45\hsize]{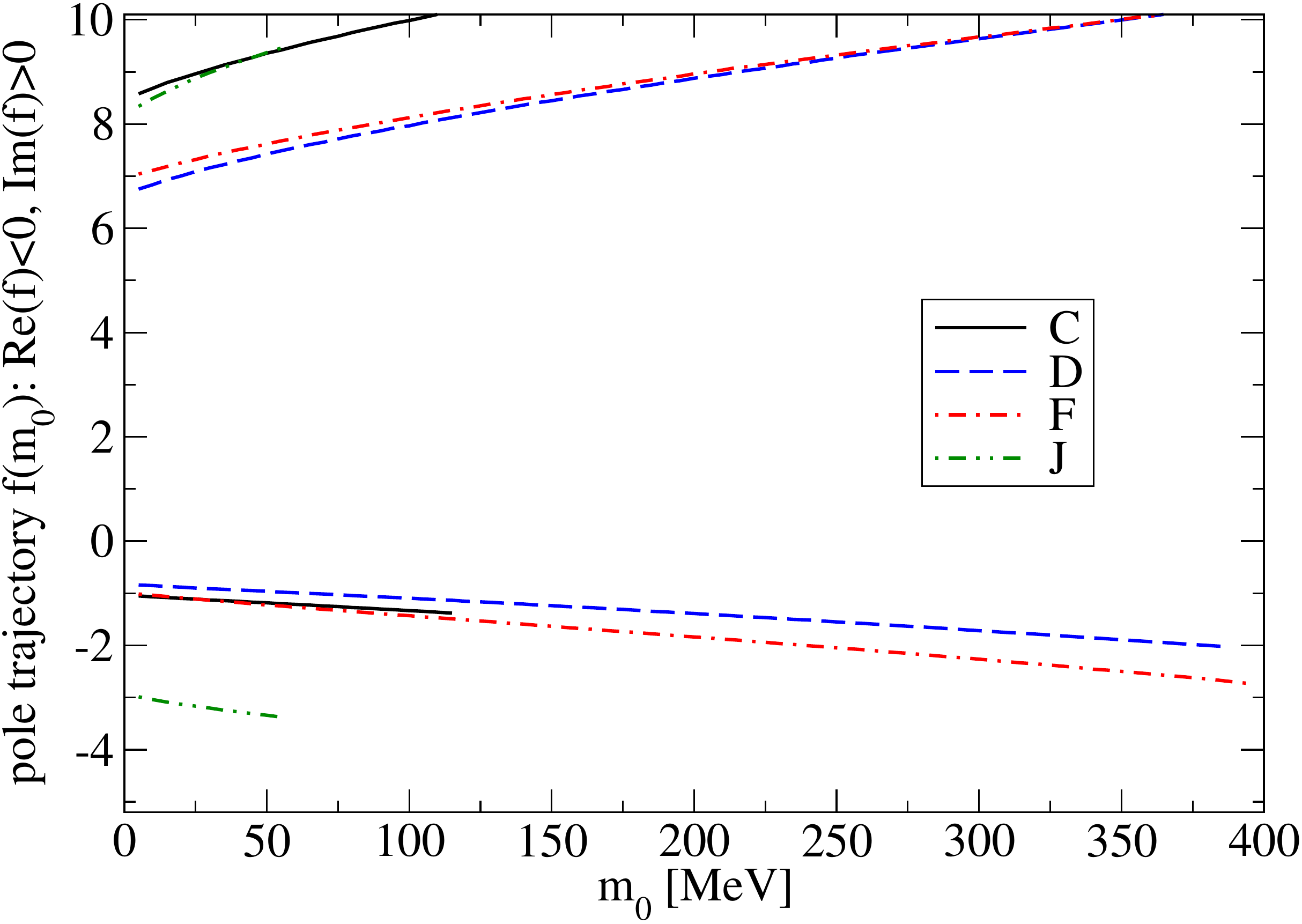}
}
\caption[]{These plots show the real and imaginary part of the trajectories traced out by the poles as the bare mass is varied from 5 MeV to 2500 MeV in increments of 5 MeV. The various trajectories have been grouped together in four sets. Each sub-figure is dedicated to such a set. Since this computation has been performed in the second quadrant, all real parts are negative, and all imaginary parts are positive. Consequently, all graphs below zero correspond to the real part of the respective trajectory, while graphs greater than zero represent the imaginary part of the trajectory. In \subref{fig:pos_a}, the imaginary part of trajectory H left the region of computation, but re-entered after a short while, which is why this graph appears to be cut off. Panel \subref{fig:pos_b} and \subref{fig:pos_c} contain the trajectories B1 and B2, which originate on the negative real axis. As the poles meet, they leave the real axis, as indicated by them developing a non-zero imaginary part, see Section \ref{sec:poles_on_real_axis} for a detailed analysis.}
\label{fig:pos}
\end{figure*}

Finally, as far as the pole locations are concerned, Figure \ref{fig:pos} shows the various pole positions in terms of the real and imaginary part as a function of the bare mass for all trajectories.

\subfiglabelskip=0pt
   
\subsection{\label{sec:poles_on_real_axis} The behavior of the poles on the real axis}
\begin{figure*}
\centering
\subfigure[][]{
 \label{fig:B_a}
\includegraphics[width=0.45\hsize]{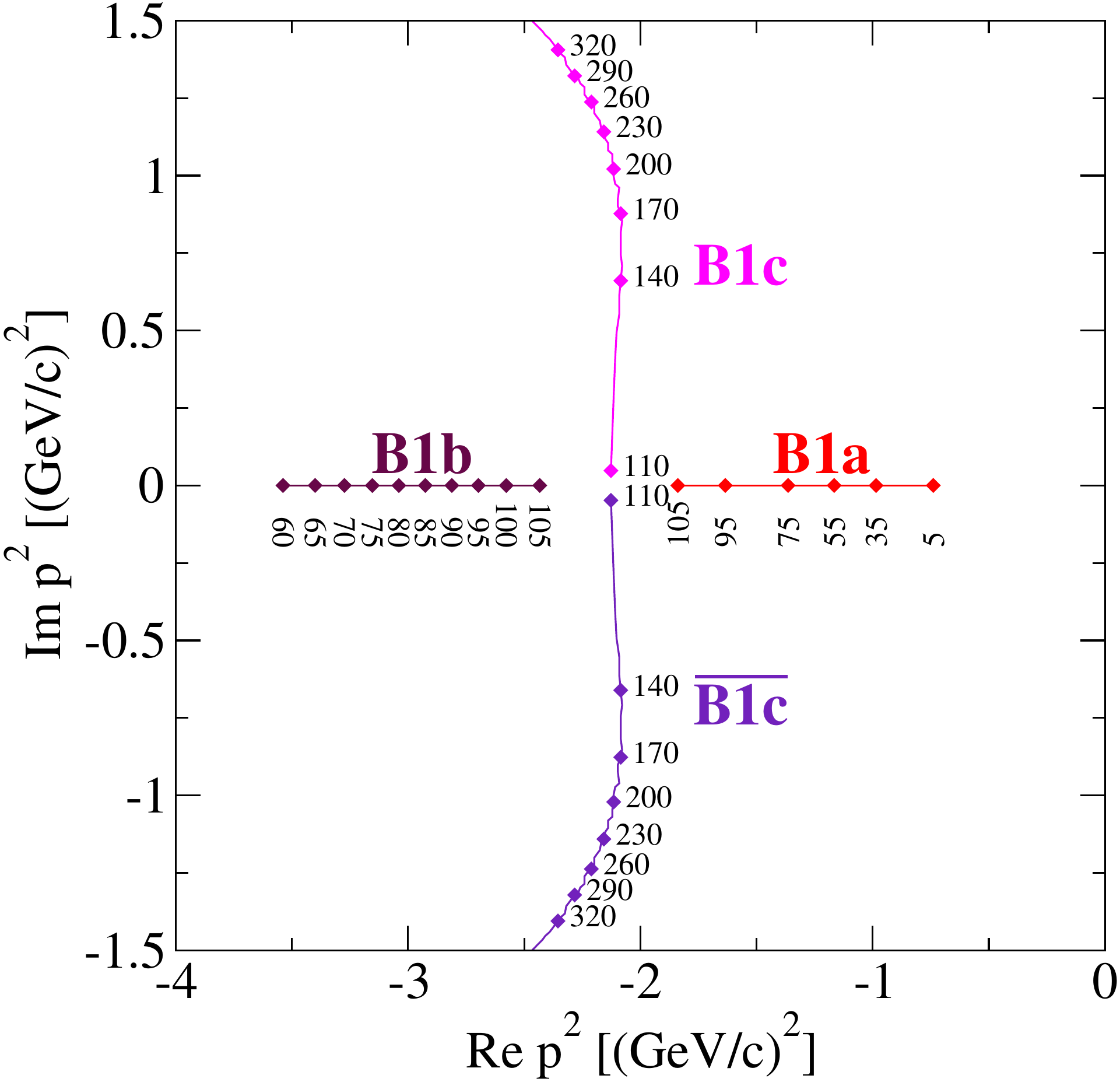}
}\hspace{8pt}
\subfigure[][]{
 \label{fig:B_b}
\includegraphics[width=0.45\hsize]{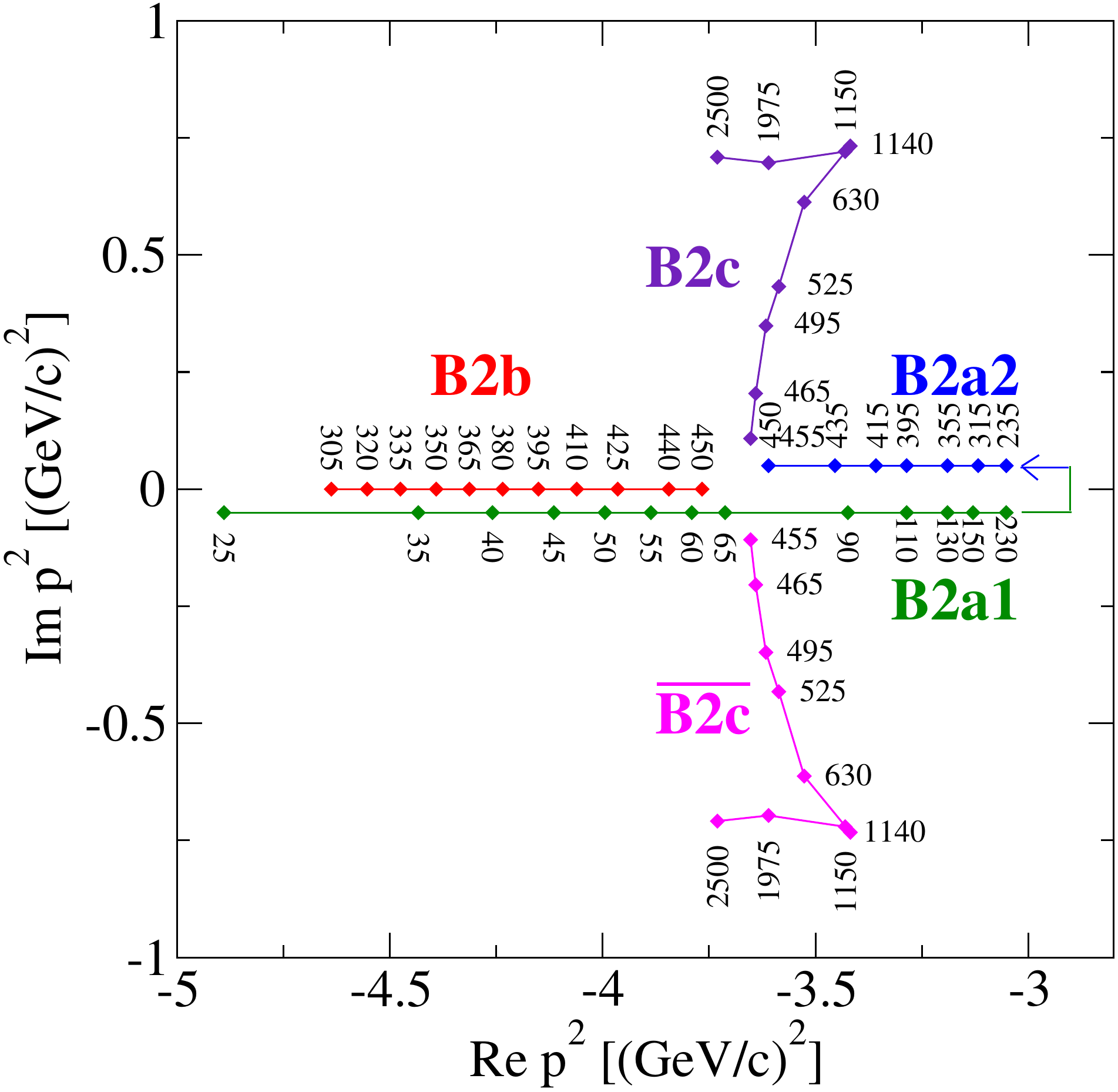}
}
\caption[]{ This figure shows the behavior of the poles in the interval $p^2\in[-5.1,0]$, which 'collide', and 'scatter' off into imaginary direction. In \subref{fig:B_a}, two trajectories, B1a (approaching from the left) and B1b (approaching from the right) come together and  form a new (complex conjugate) trajectory B1c. The collision occurs between a bare mass value of 105 and 110 MeV. In \subref{fig:B_b}, again two trajectories collide. At low mass values, trajectory B2a1 approaches from the left and reverses its direction of movement between a mass value of 230 MeV and 235 MeV, after which it is labeled as B2a2. Because the trajectories B2a1 and B2a2 would overlap in the figure, they have been slightly shifted away from the real axis, even though their imaginary parts are zero. Finally, a pole moving to the right along trajectory B2a1, collides with a pole moving to the left along trajectory B2b, creating a new complex (conjugate) trajectory B2c between mass values of 450 MeV and 455 MeV. Above 455 MeV, there are no real poles in the interval $[-5.1,0]$ for all bare mass values of up to 2500 MeV.}
\label{fig:B}
\end{figure*}

As pointed out in Section \ref{sec:pole_location_discussion}, the behavior of the poles on the real axis is more involved and requires a separate discussion. There are two main 'processes' involving the real axis, where process means that two real poles meet at a certain real value of the square of the external momentum, and 'scatter' off perpendicularly. Thereby, they are tracing out two new trajectories, along which the two poles are heading into positive and negative imaginary direction. These processes are labeled B1 and B2. B1 involves the trajectories B1a, B1b and B1c, see Figure \ref{fig:B_a} for details. In process B1, two trajectories are approaching each other head-on on the real axis. At a bare mass value between 105 MeV $<m_0<$ 110 MeV, the two trajectories 'collide' and create a new, complex (conjugate) trajectory B1c, which leads away from the real axis in imaginary direction. Process B2 is even more complicated, see \ref{fig:B_b}. Similar to the process B1, two trajectories collide, but, prior to the collision, one of the trajectories changes its direction along the real axis at some point. Trajectories involved in this process are B2a1, B2a2, B2b and B2c. Let me start by discussing B2a1. It starts at a bare mass value of 25 far out on the real axis and moves towards the origin. At a value of around $x\approx -3$ GeV$^2$, the trajectory changes direction, and starts moving along the real axis, away from the origin. The trajectory after the change of direction is labeled B2a2. For the sake of clarity, trajectories B2a1 and B2a2 are slightly shifted off the axis in Figure \ref{fig:B_b}, even though they do not have an imaginary part.  At the same time, trajectory B2b approaches from lesser real values and eventually collides with B2a2 between a bare mass value of 450 MeV $<m_0<$ 455 MeV. The new complex (conjugate) trajectory, B1c, again leads away from the real axis in imaginary direction. Thus, the whole region in which the quark DSE has been solved is completely free of poles on the real axis for bare mass values $m_0\gtrsim$ 455 MeV.
The collision of the poles can also be seen in Figure \ref{fig:pos_b} and \ref{fig:pos_c}, where the (real) positions of the two trajectories come together, and a non-zero imaginary component develops.
\subfiglabelskip=0pt

\subsection{\label{sec:residues}The residues of \texorpdfstring{$\sigma_S$}{Sigma S} and \texorpdfstring{$\sigma_V$}{Sigma V} for varying masses \texorpdfstring{$m_0$}{m0}}
\begin{figure*}
\centering
\subfigure[][]{
 \label{fig:Res_A_a}
\includegraphics[width=0.45\hsize]{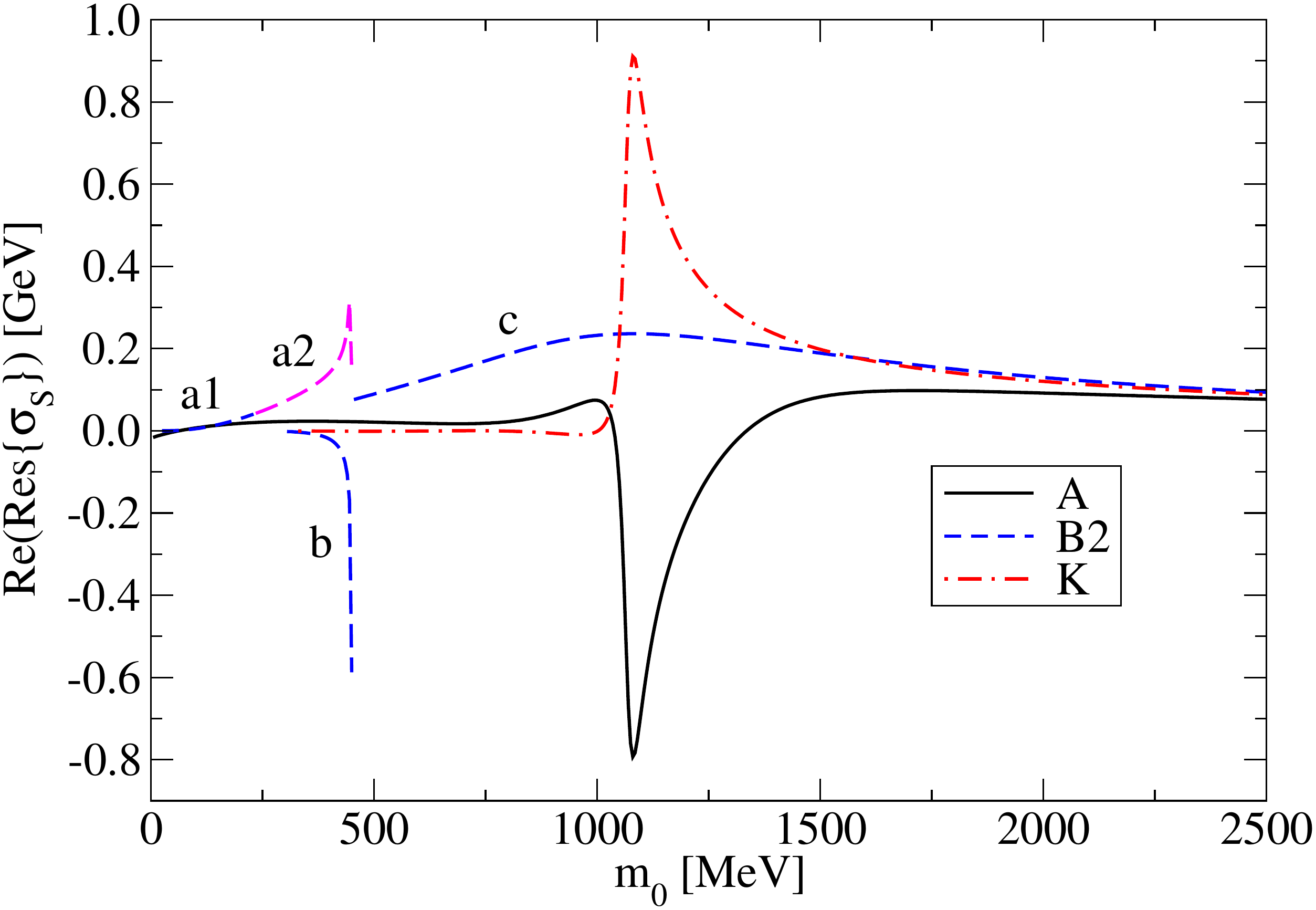}
}\hspace{8pt}
\subfigure[][]{
 \label{fig:Res_A_b}
\includegraphics[width=0.45\hsize]{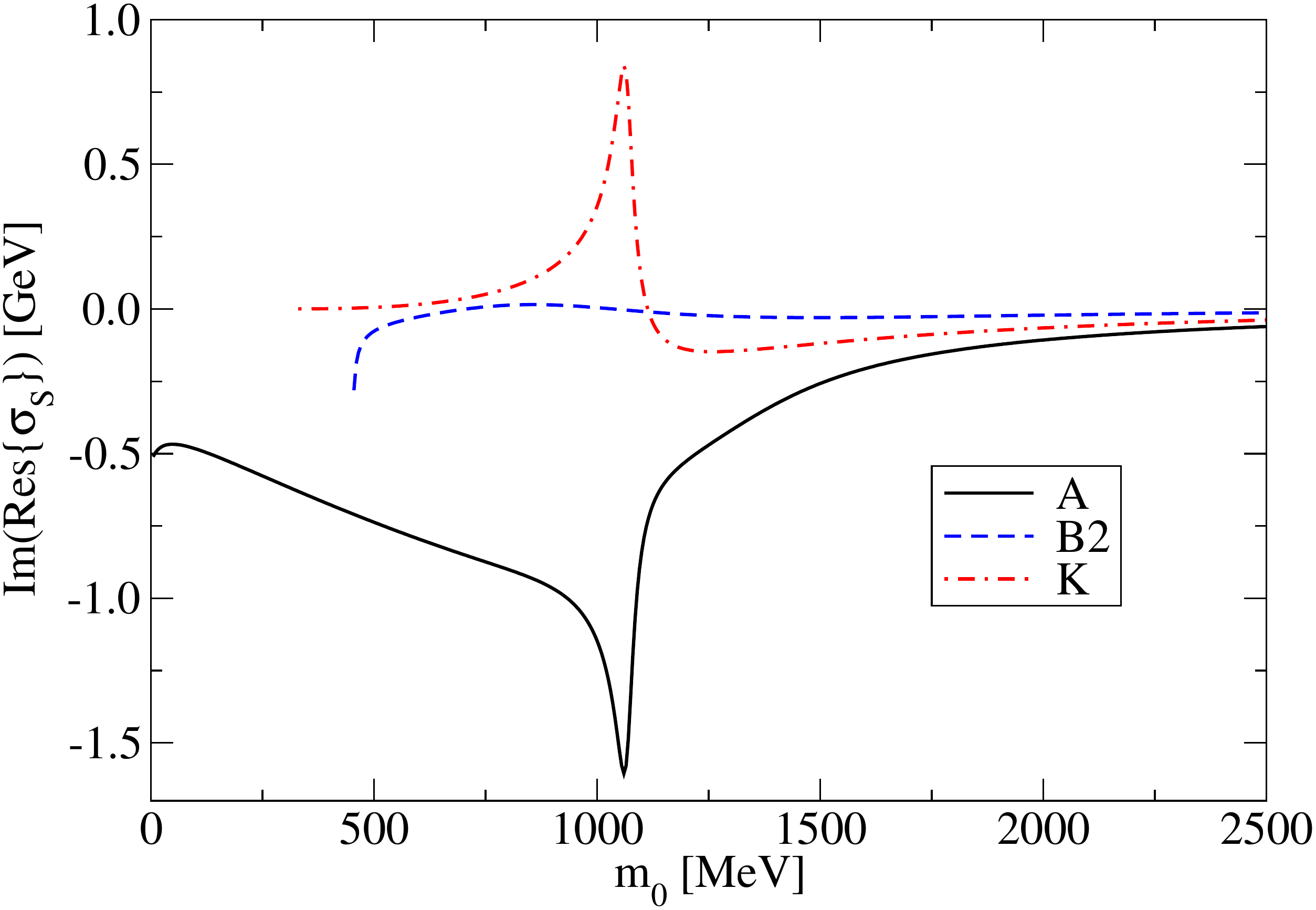}
}\\
\subfigure[][]{
 \label{fig:Res_A_c}
\includegraphics[width=0.45\hsize]{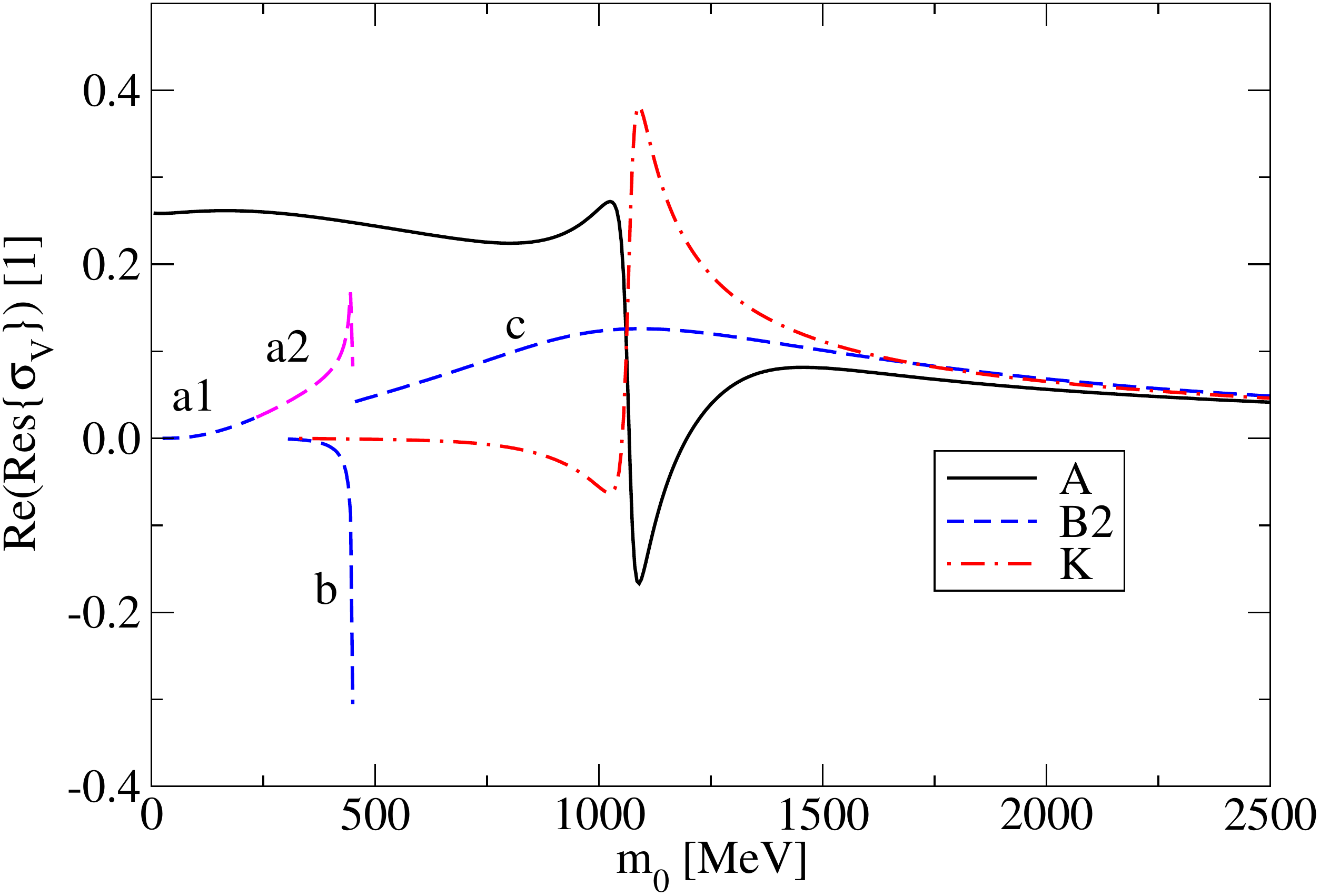}
}
\hspace{8pt}
\subfigure[][]{
 \label{fig:Res_A_d}
\includegraphics[width=0.45\hsize]{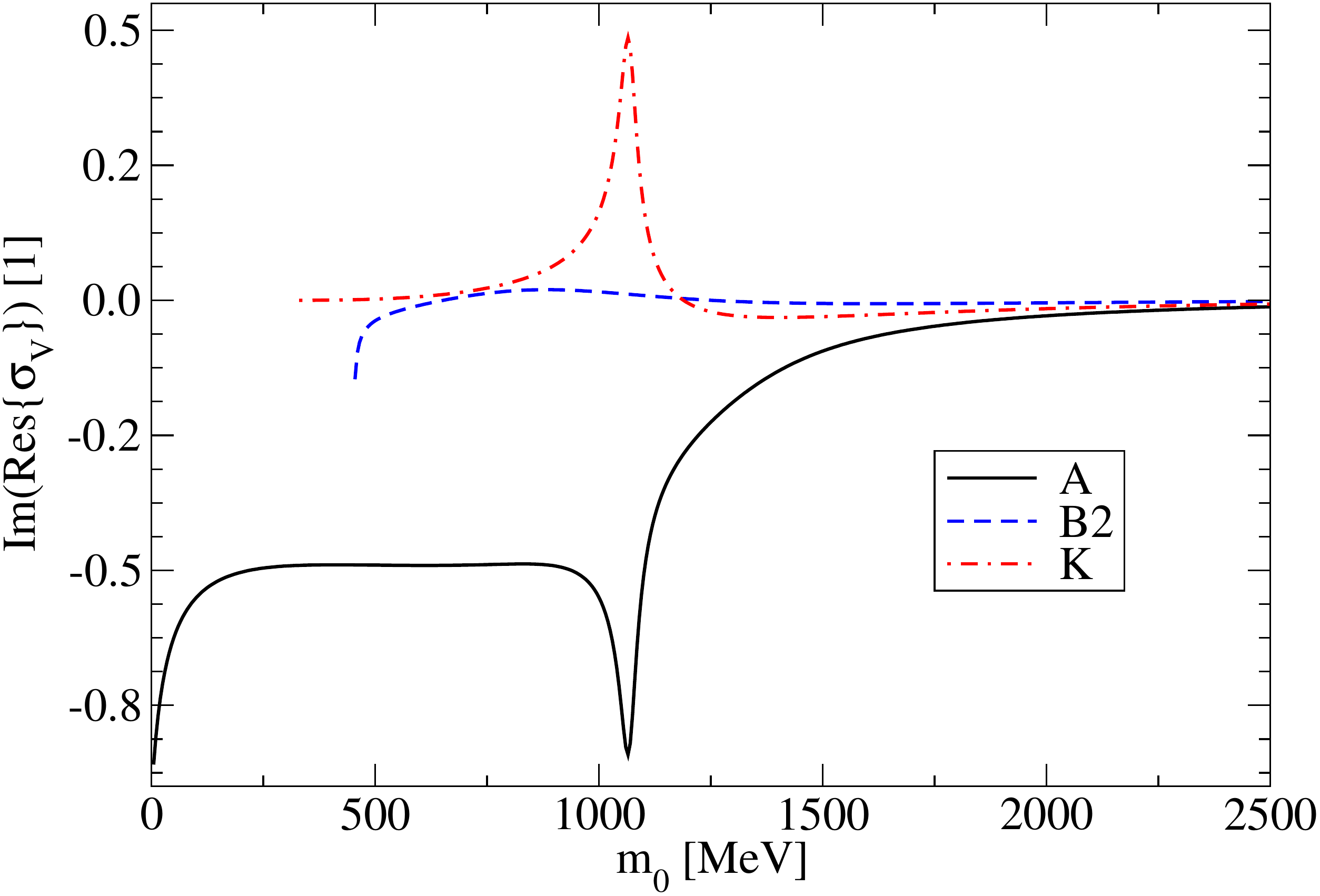}
}\\
\hspace{8pt}
\centering
\subfigure[][]{
 \label{fig:Res_B1_a}
\includegraphics[width=0.45\hsize]{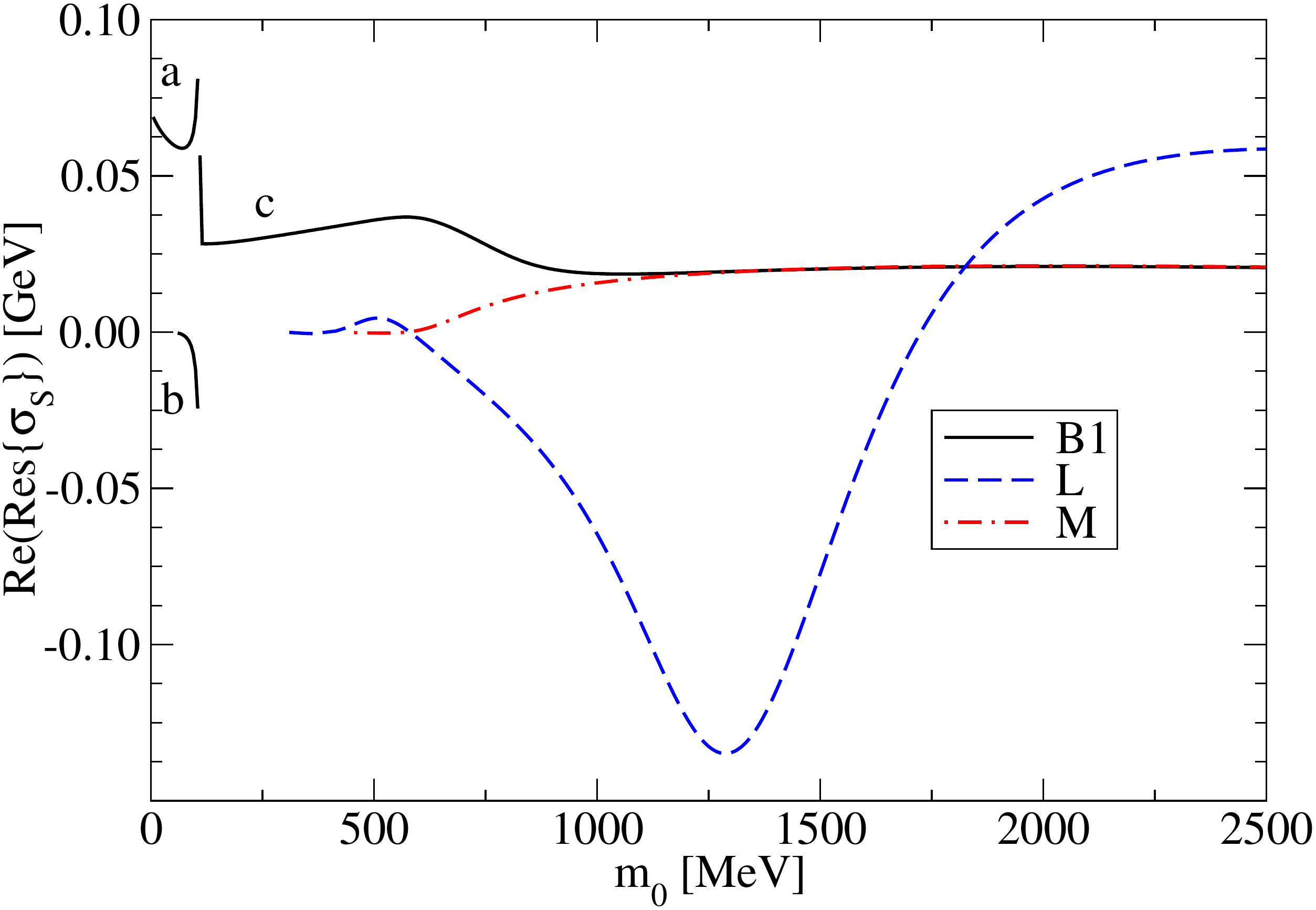}
}\hspace{8pt}
\subfigure[][]{
 \label{fig:Res_B1_b}
\includegraphics[width=0.45\hsize]{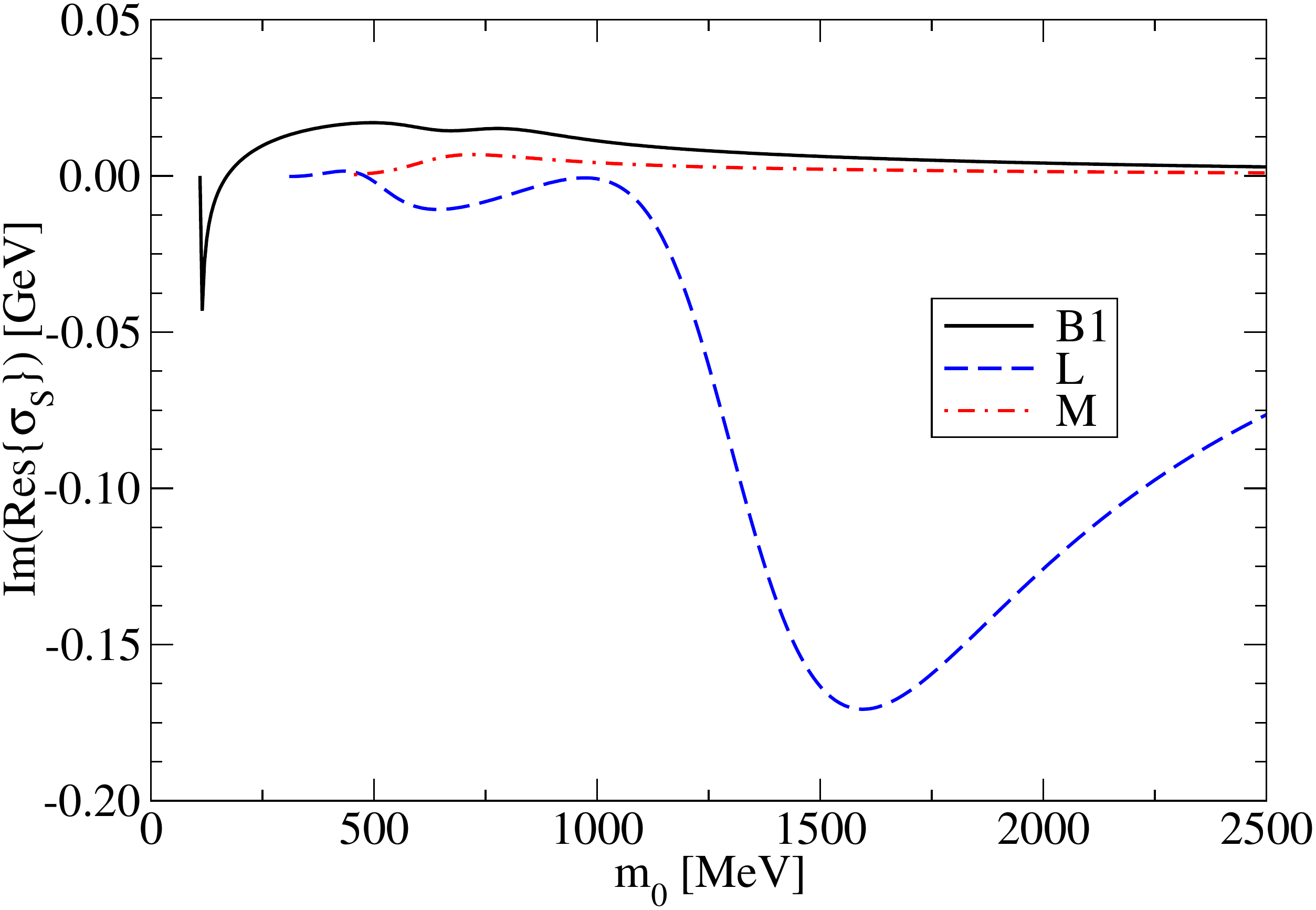}
}\\
\subfigure[][]{
 \label{fig:Res_B1_c}
\includegraphics[width=0.45\hsize]{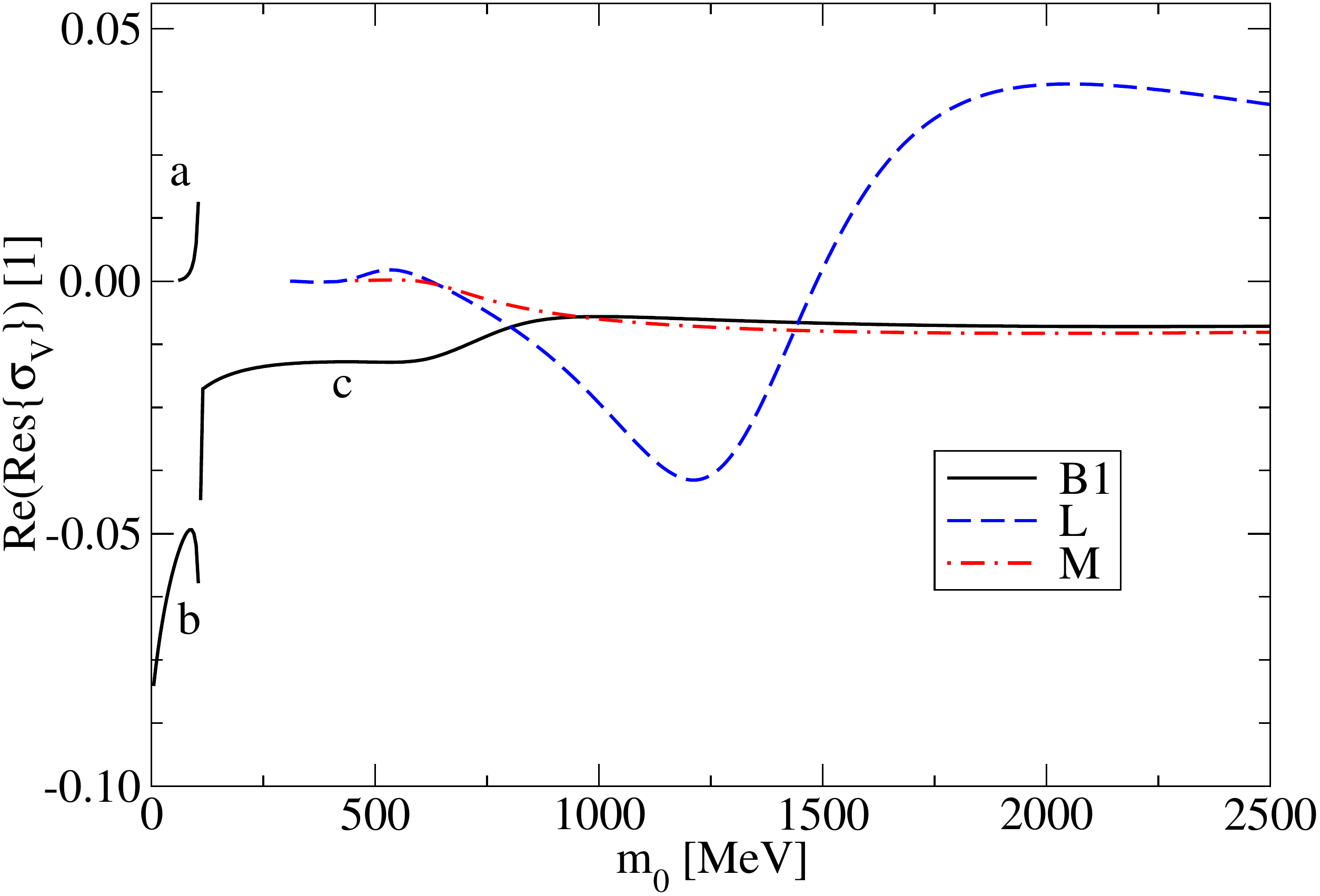}
}
\hspace{8pt}
\subfigure[][]{
 \label{fig:Res_B1_d}
\includegraphics[width=0.45\hsize]{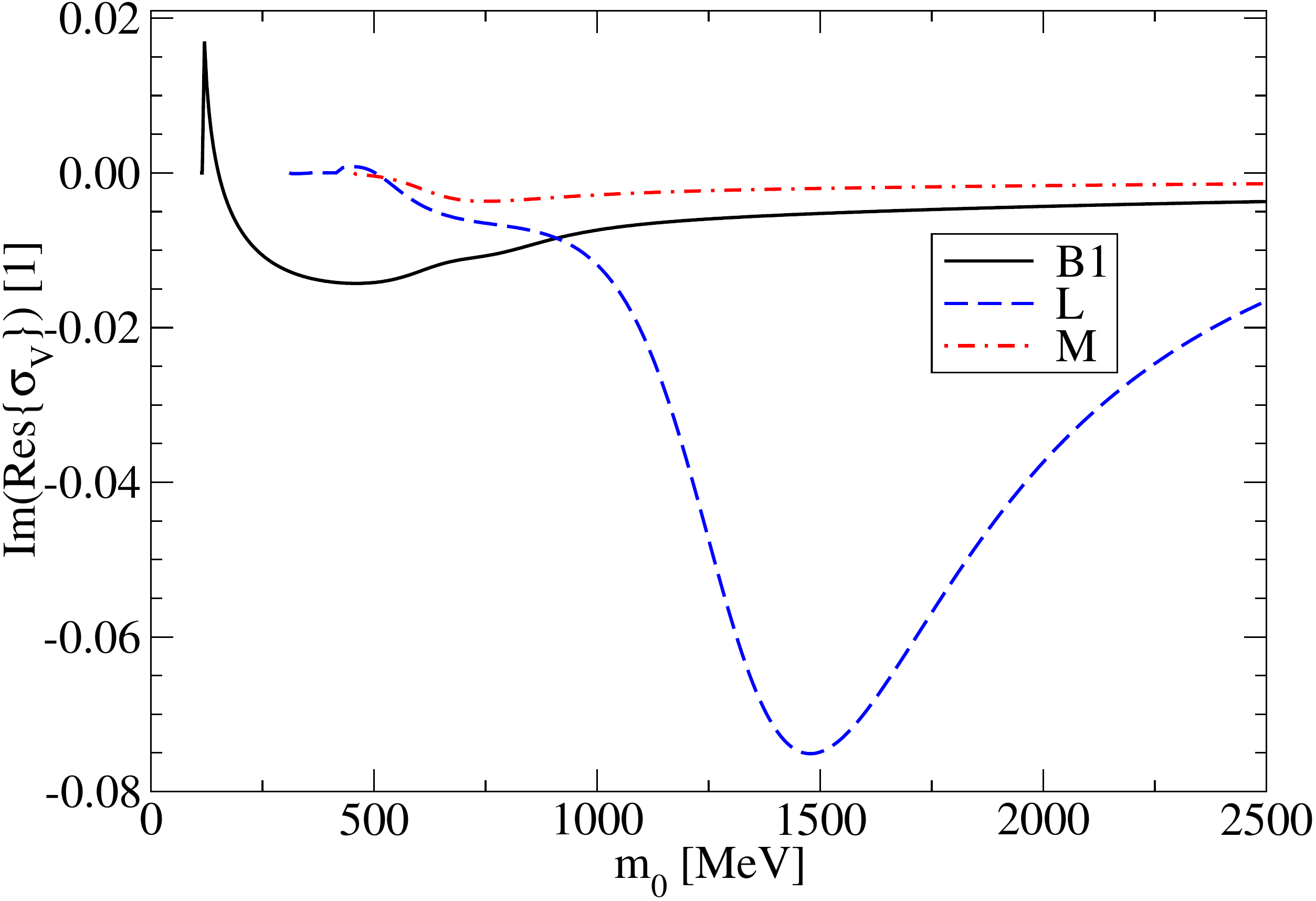}
}
\caption[]{ Residues of $\sigma_S$ and $\sigma_V$ as a function of $m_0$ for trajectories A, B2, K (\subref{fig:Res_A_a}-\subref{fig:Res_A_d}), and B1, L and M (\subref{fig:Res_B1_a}-\subref{fig:Res_B1_d}).}
\label{fig:Res_AB1}
\end{figure*}

\begin{figure*}
\centering
\subfigure[][]{
 \label{fig:Res_E_a}
\includegraphics[width=0.45\hsize]{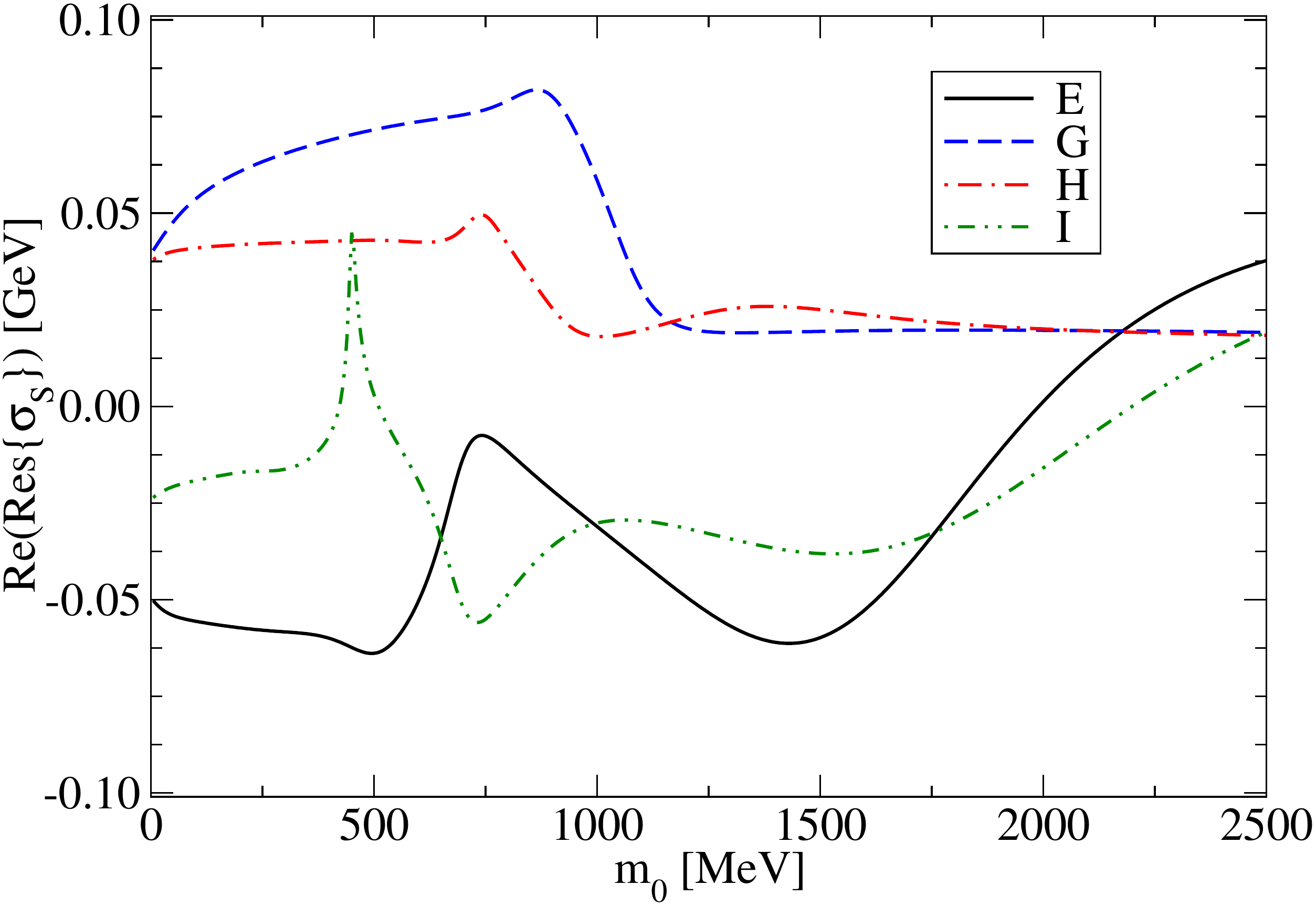}
}\hspace{8pt}
\subfigure[][]{
 \label{fig:Res_E_b}
\includegraphics[width=0.45\hsize]{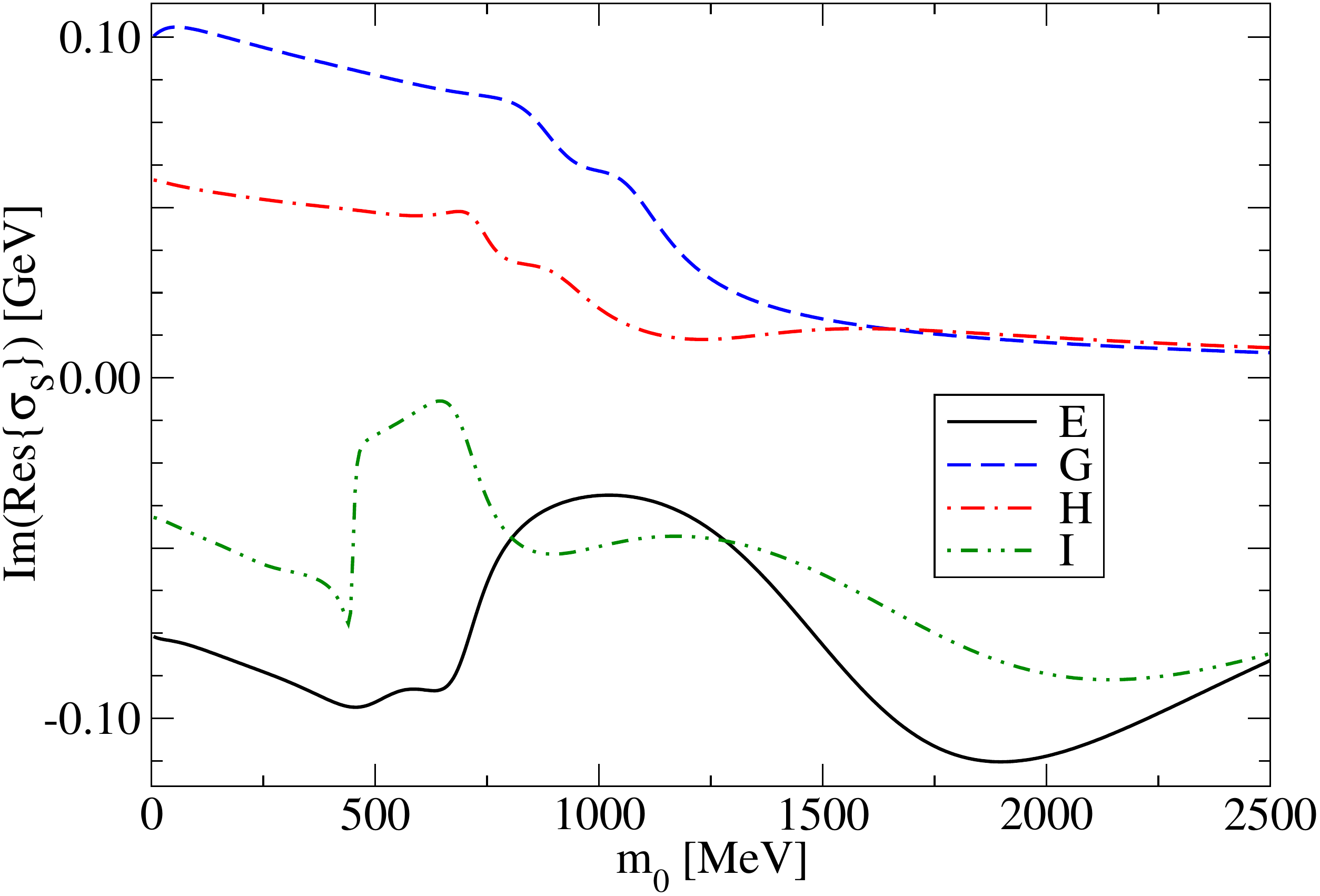}
}\\
\subfigure[][]{
 \label{fig:Res_E_c}
\includegraphics[width=0.45\hsize]{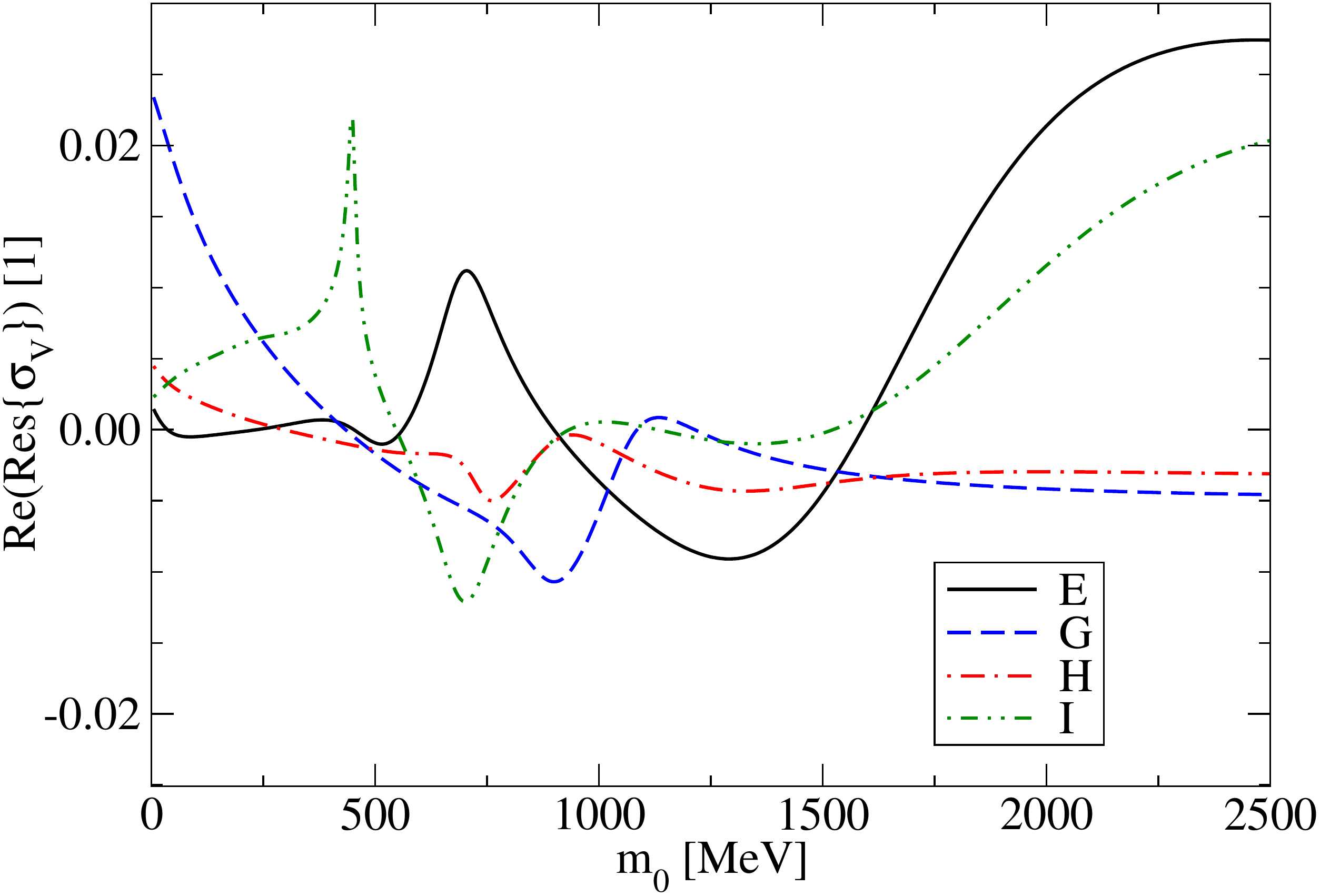}
}
\hspace{8pt}
\subfigure[][]{
 \label{fig:Res_E_d}
\includegraphics[width=0.45\hsize]{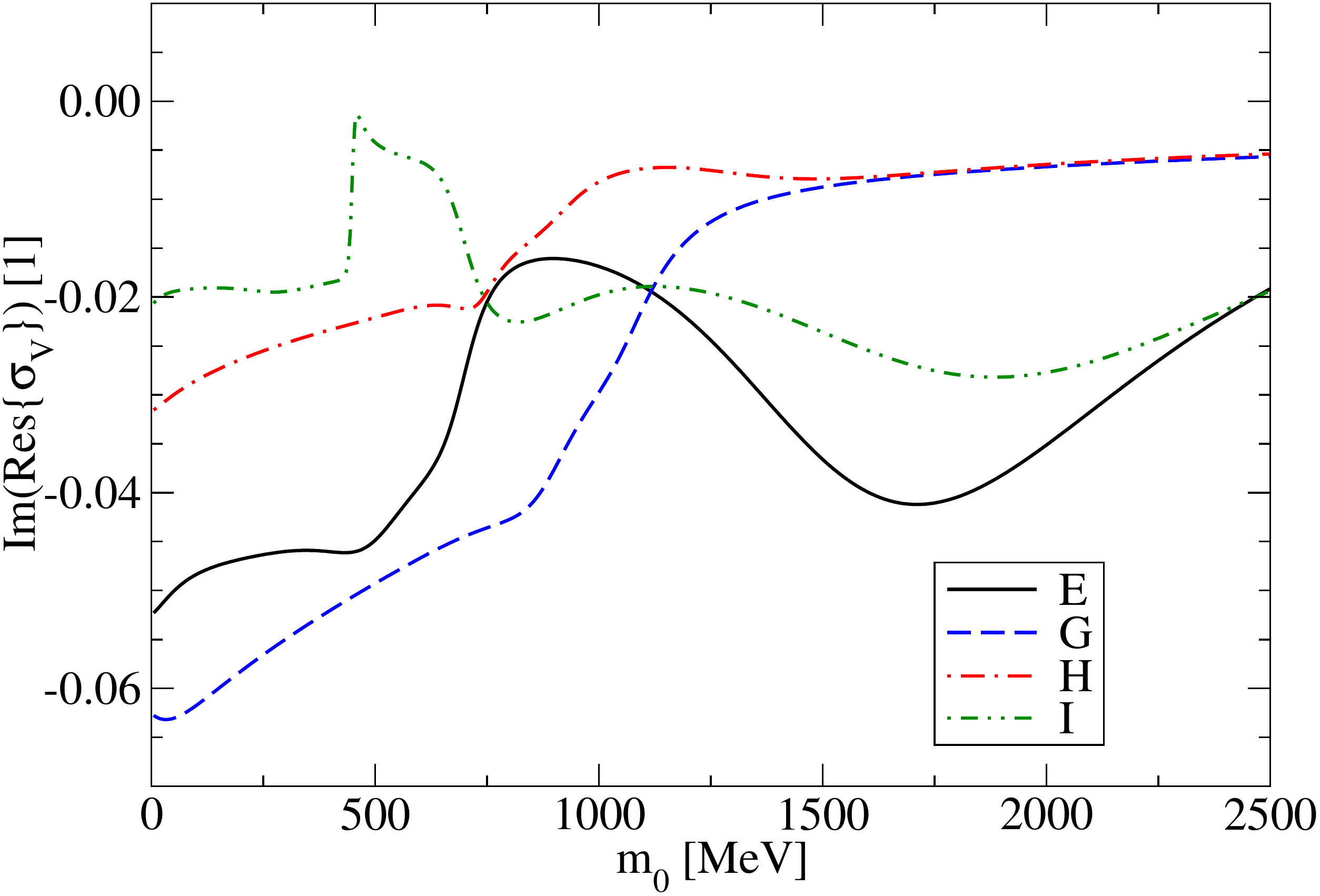}
}\\
\hspace{8pt}
\subfigure[][]{
 \label{fig:Res_N_a}
\includegraphics[width=0.45\hsize]{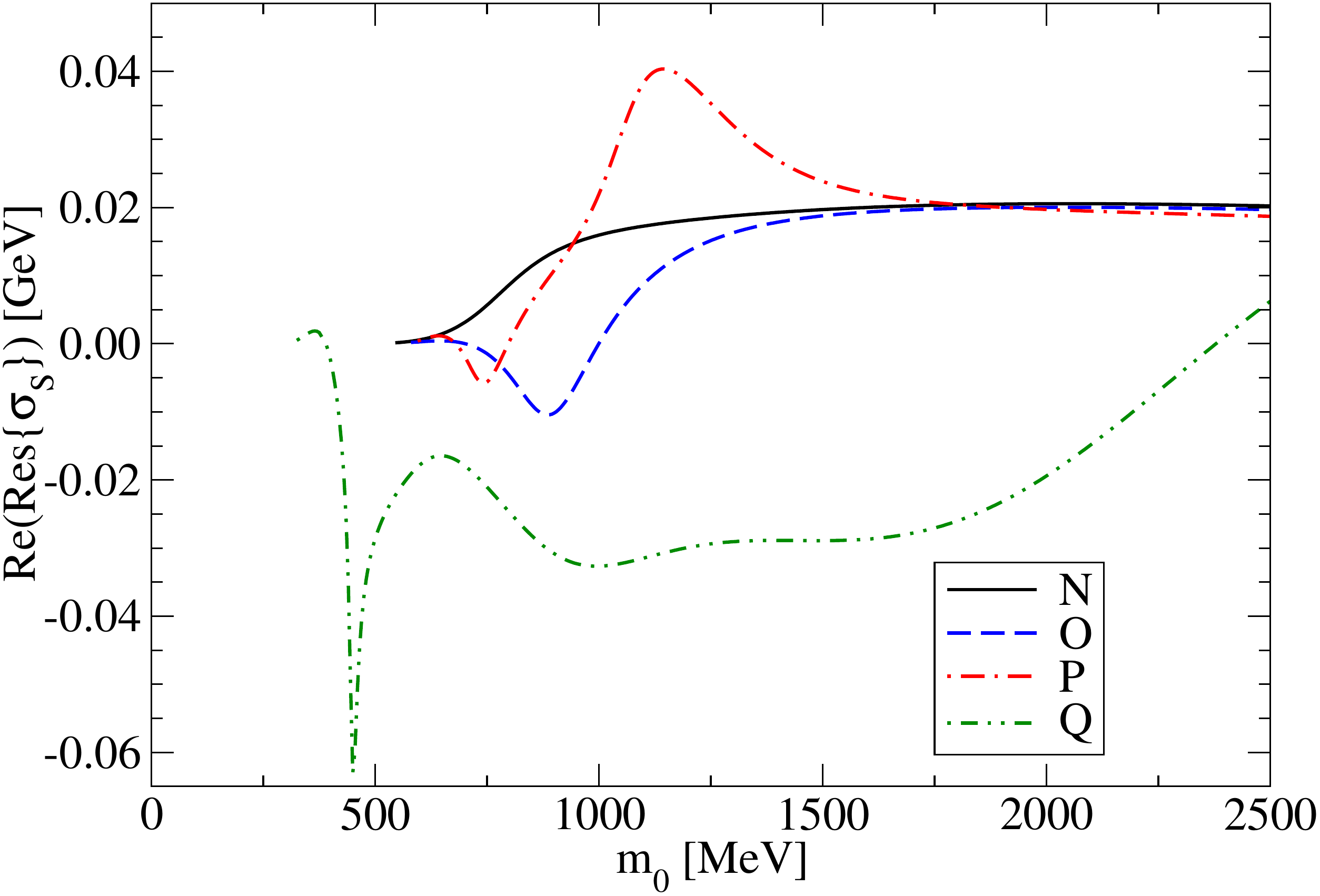}
}\hspace{8pt}
\subfigure[][]{
 \label{fig:Res_N_b}
\includegraphics[width=0.45\hsize]{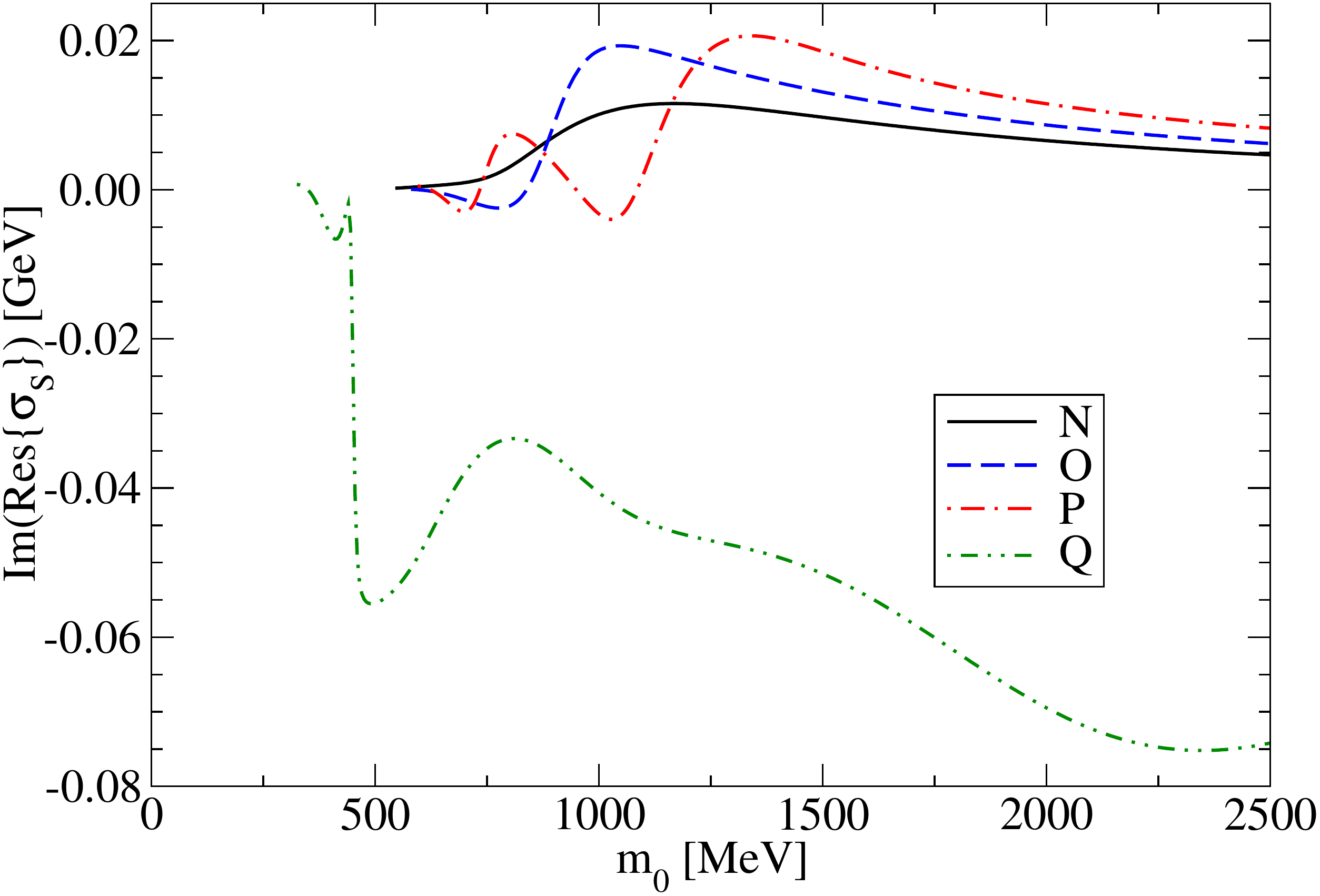}
}\\
\subfigure[][]{
 \label{fig:Res_N_c}
\includegraphics[width=0.45\hsize]{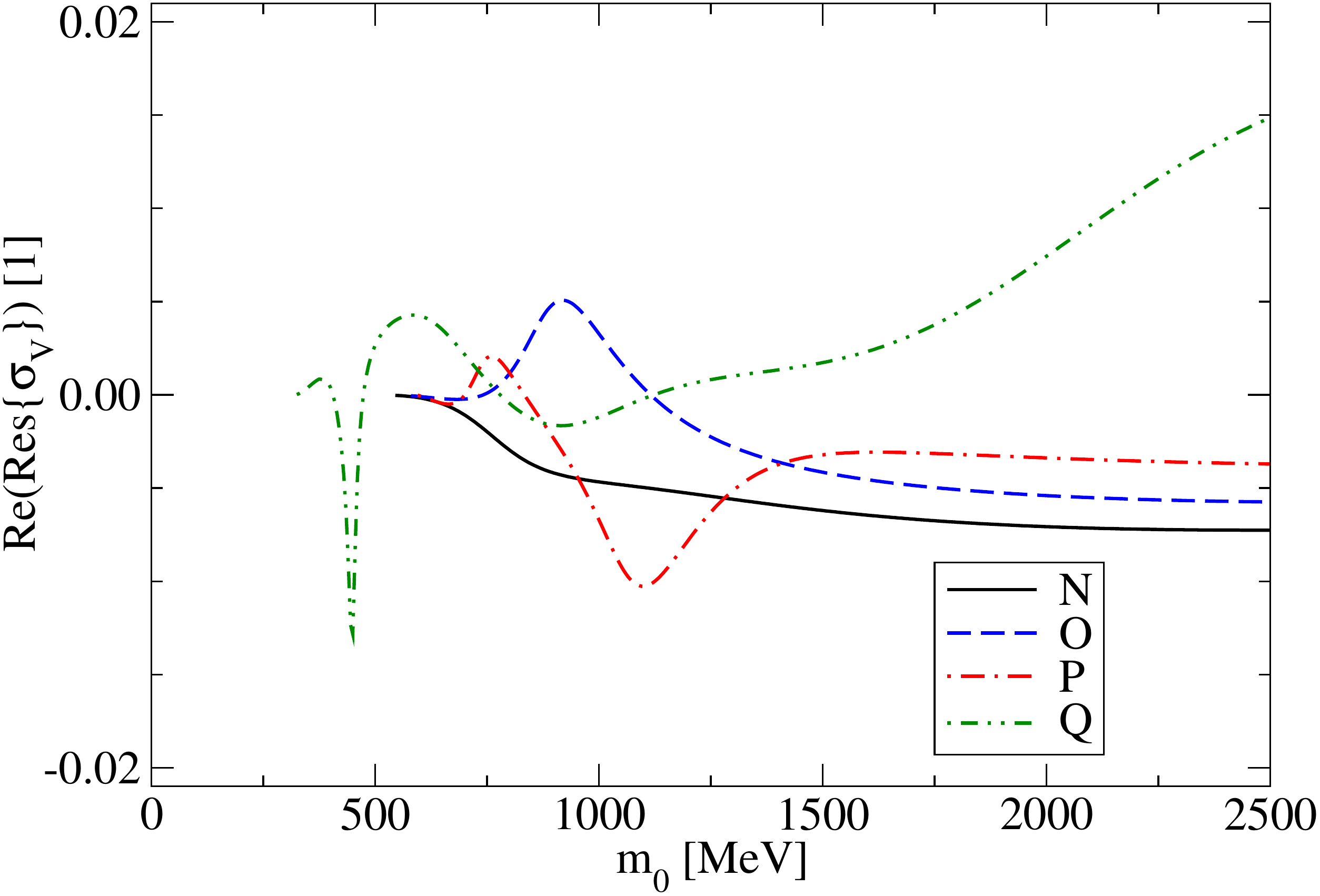}
}
\hspace{8pt}
\subfigure[][]{
 \label{fig:Res_N_d}
\includegraphics[width=0.45\hsize]{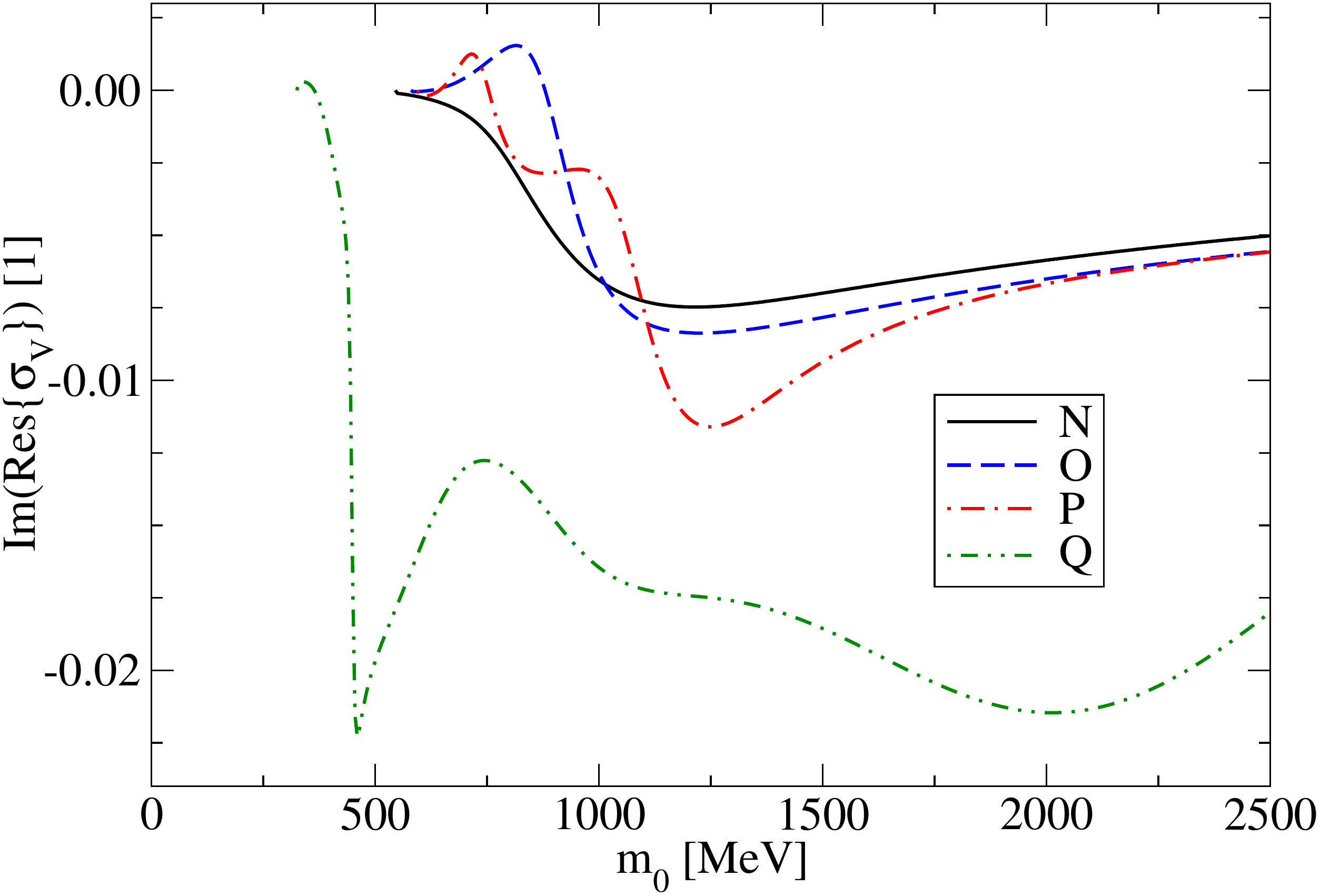}
}
\caption[]{ Residues of $\sigma_S$ and $\sigma_V$ as a function of $m_0$ for trajectories E, G, H, I (\subref{fig:Res_A_a}-\subref{fig:Res_A_d}), and N, O, P, Q (\subref{fig:Res_B1_a}-\subref{fig:Res_B1_d}).}
\label{fig:Res_EN}
\end{figure*}

\begin{figure*}
\centering
\subfigure[][]{
 \label{fig:Res_C_a}
\includegraphics[width=0.45\hsize]{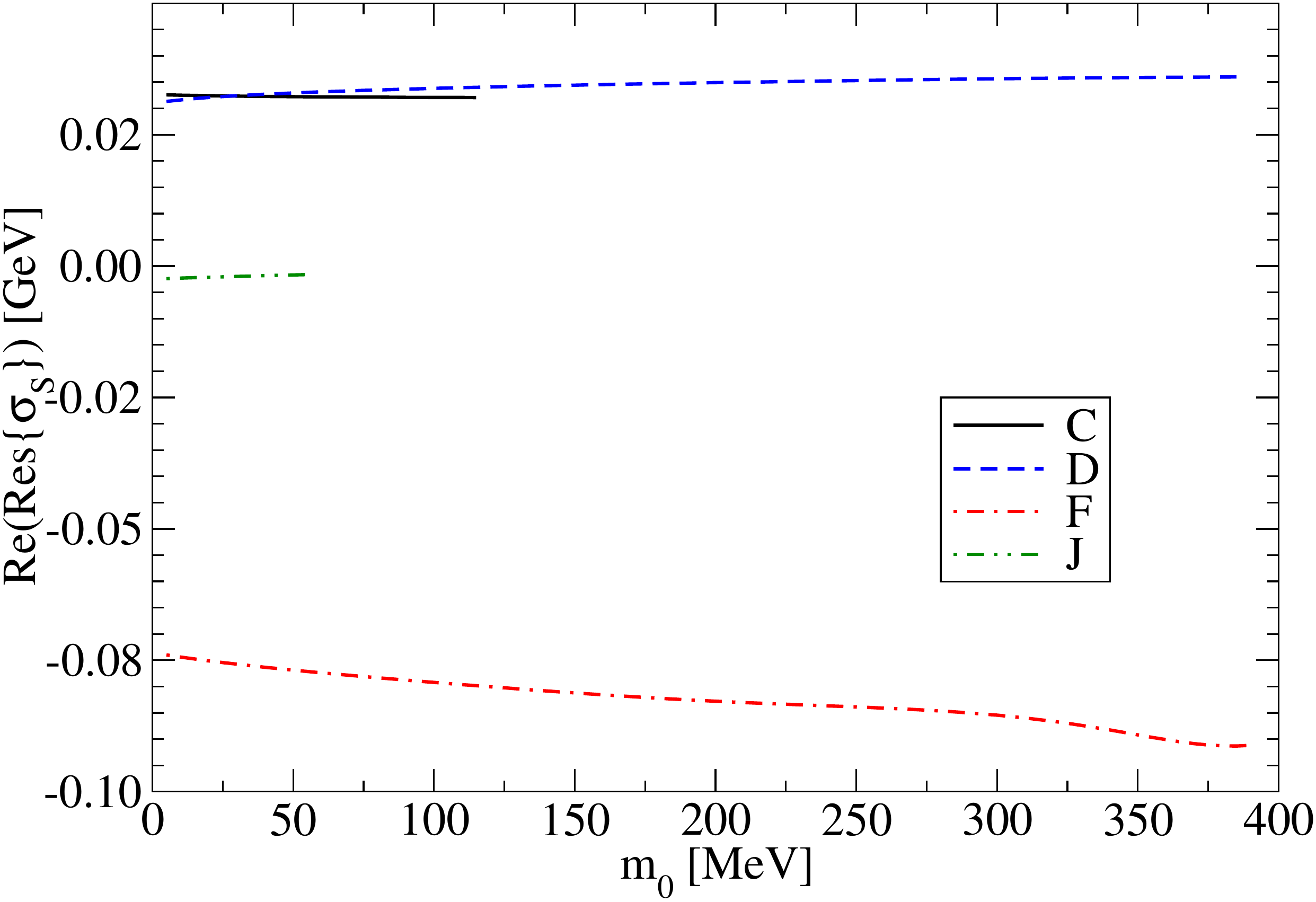}
}\hspace{8pt}
\subfigure[][]{
 \label{fig:Res_C_b}
\includegraphics[width=0.45\hsize]{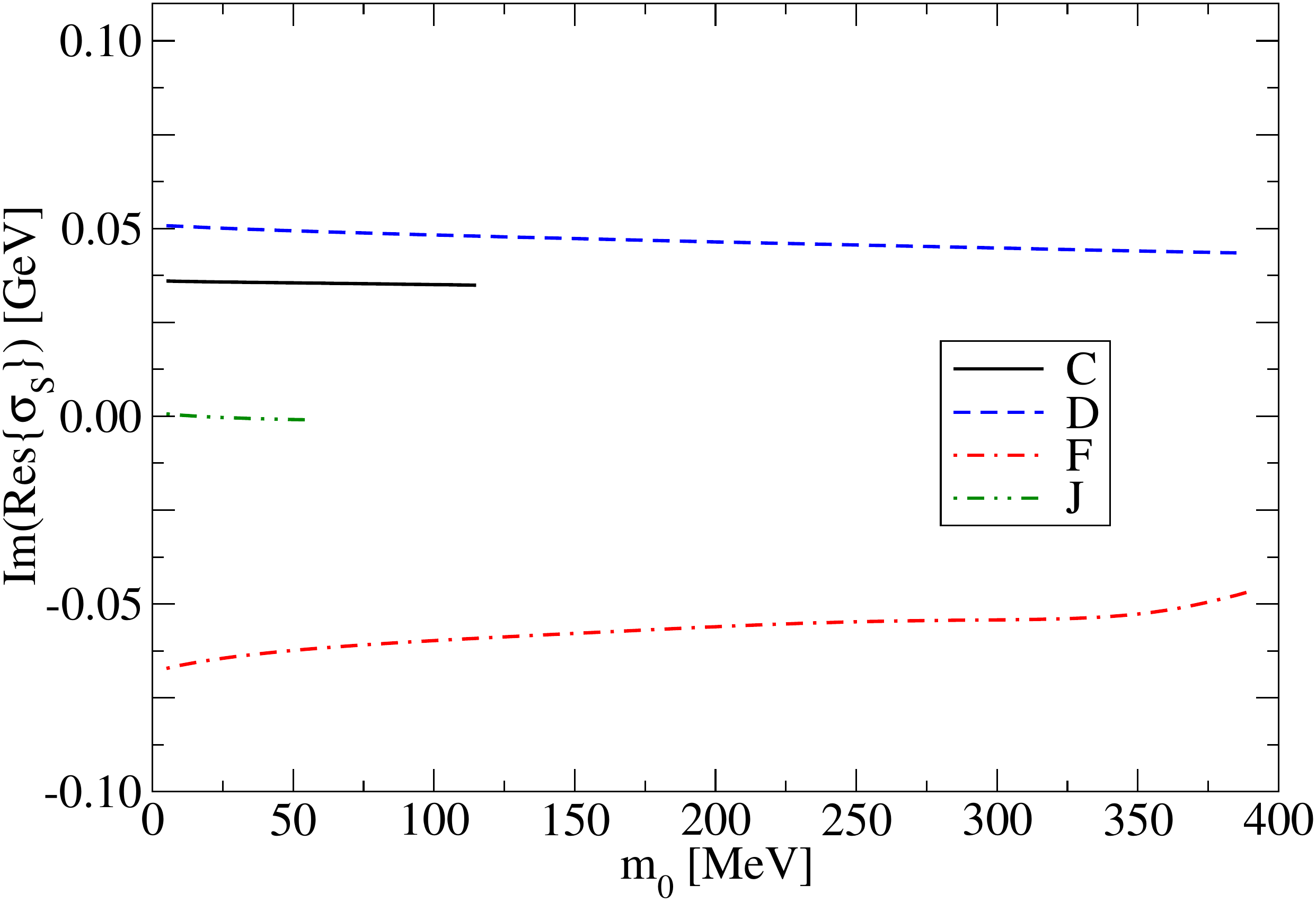}
}\\
\subfigure[][]{
 \label{fig:Res_C_c}
\includegraphics[width=0.45\hsize]{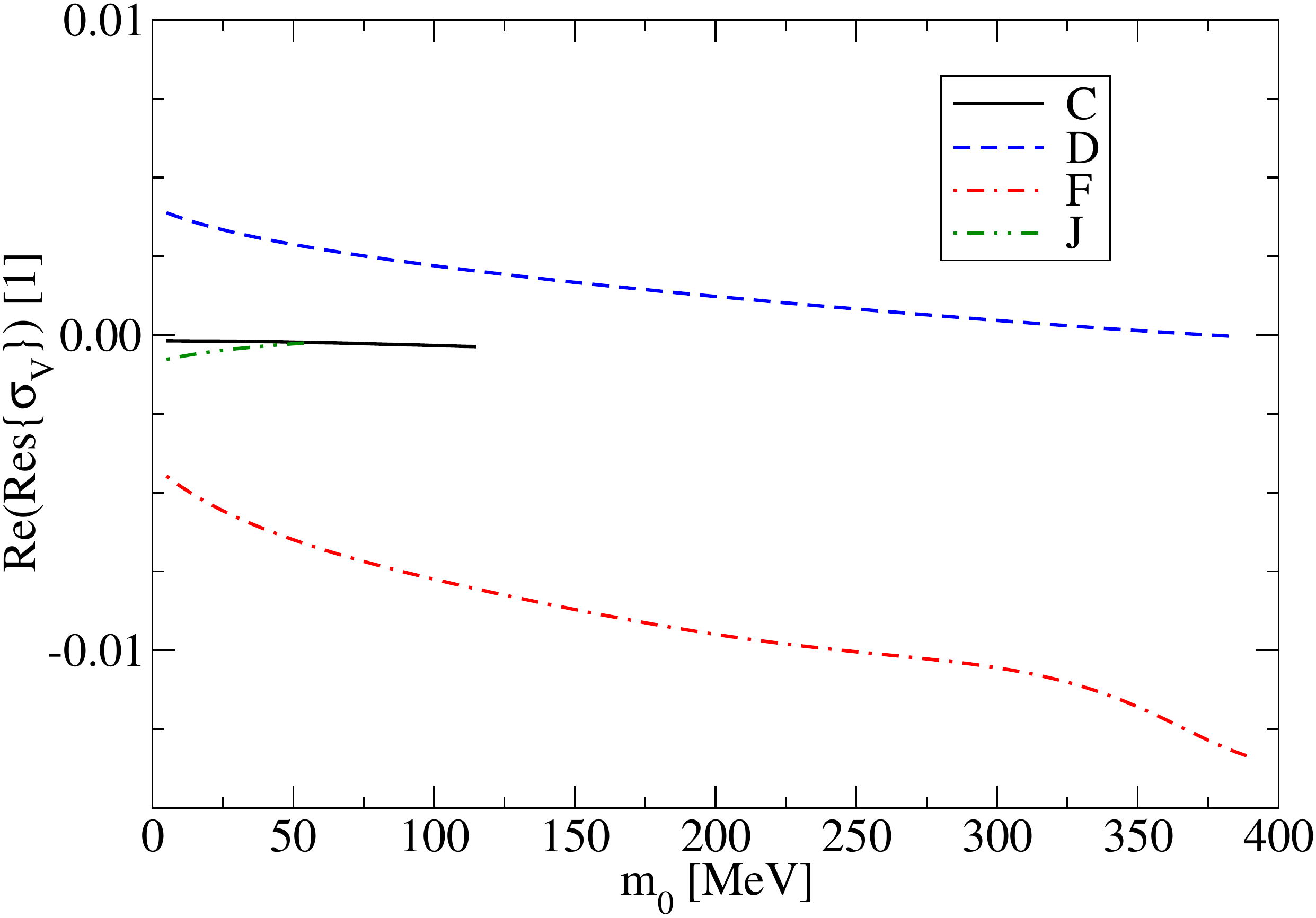}
}
\hspace{8pt}
\subfigure[][]{
 \label{fig:Res_C_d}
\includegraphics[width=0.45\hsize]{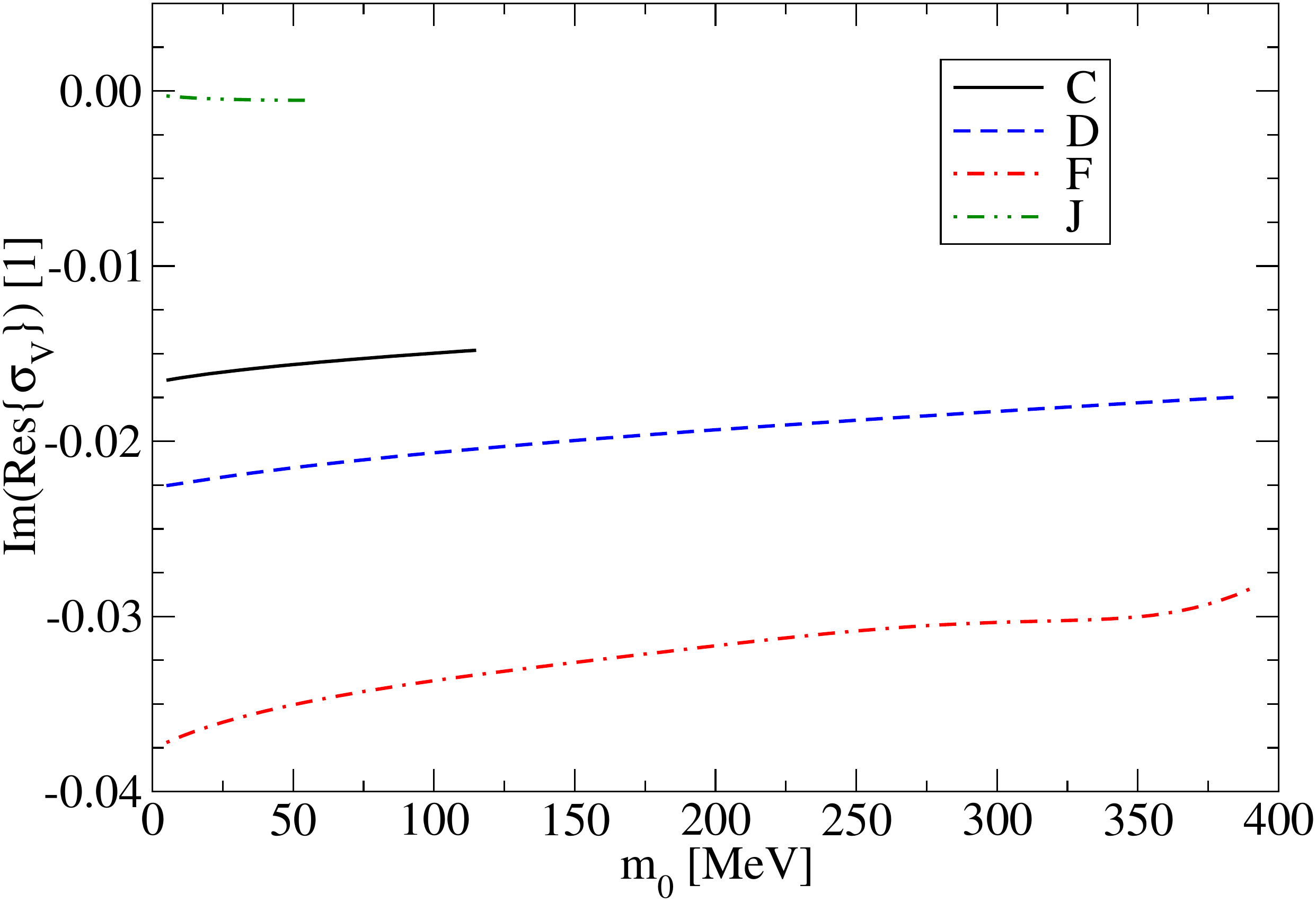}
}
\caption[]{ Residues of $\sigma_S$ and $\sigma_V$ as a function of $m_0$ for trajectories C, D, F, J (\subref{fig:Res_A_a}-\subref{fig:Res_A_d}).}
\label{fig:Res_C}
\end{figure*}

Apart from the pole locations, their residues are also of importance. The residues have been computed by following the procedure outlined in Section \ref{sec:numerical_procedures} above. The complete numerical results are provided as supplemental material, see Section \ref{sec:Supplemental_Material}. In Figures \ref{fig:Res_AB1}, \ref{fig:Res_EN} and \ref{fig:Res_C}, real and imaginary parts of all residues of $\sigma_S$ and $\sigma_V$ are provided as a function of mass. The trajectories are grouped together in such a way that the range in which the respective residue varies is roughly of the same order.  
\subfiglabelskip=0pt

\subfiglabelskip=0pt

\subfiglabelskip=0pt

\subsection{\label{sec:comparison}Comparison with known results and numerical errors}
In order to further validate the findings presented here, a comparison with published results is made. In \cite{Dorkin:2013rsa}, pole locations, as well as residues for $\sigma_S$ and $\sigma_V$ are presented for bare mass values of $m_0=5$ MeV and $m_0=115$ MeV. For the convenience of the reader, I reproduce the relevant table from \cite{Dorkin:2013rsa} here, see Table \ref{tab:Dorkin_table}.
In Table \ref{tab:Andreas_table}, I summarize the poles that have been found by the automated numerical procedure outlined in Section \ref{sec:numerical_procedures}. Because the pole location is derived from the locations of points on the discretized complex region, the pole position has uncertainties of  $\pm6\times 10^{-3}$ GeV$^2$ for the real part, and $\pm12\times 10^{-3}$ GeV$^2$ for the imaginary part. The residues, however, are expected to be of much greater precision, and no estimate for an error can be provided.
\begin{table*}
\centering
\begin{tabular}{c c c c c}
\hline
\hline
$m_0=5$ MeV & 1 & 2 & 3 & 4\\
\hline
pole position $z_0$ & (-0.2588,$\pm$0.19618) & (-0.2418,$\pm$2.597) & (-1.0415,$\pm$2.8535) & (-0.738, 0.0)\\
$\mbox{Res}(\sigma_S,z_0)$ & (-0.016,$\mp$0.511) & (0.04,$\pm$0.10) & (-0.05,$\mp$0.076) & (0.069,0.0)\\
$\mbox{Res}(\sigma_V,z_0)$ & (0.259,$\mp$0.859) & (0.0234,$\mp$0.063) & (0.0014,$\mp$0.052) & (-0.080,0.0)\\
\hline
$m_0=115$ MeV & 1 & 2 & 3 & 4\\
\hline
pole position $z_0$ & (-0.463,$\pm$0.513) & (-0.51,$\pm$3.35) & (-1.45,$\pm$3.82) & (-3.25, 0.0)\\
$\mbox{Res}(\sigma_S,z_0)$ & (0.009,$\mp$0.49) & (0.06,$\pm$0.10) & (-0.056,$\mp$0.08) & (0.007,0.0)\\
$\mbox{Res}(\sigma_V,z_0)$ & (0.26,$\mp$0.54) & (0.013,$\mp$0.06) & (-0.0005,$\mp$0.048) & (0.004,0.0)\\
\hline
\hline
\end{tabular}
\caption{Table of pole positions and -residues as presented in \cite{Dorkin:2013rsa}, denoted as $(\Re x, \Im x)$.}
\label{tab:Dorkin_table}
\bigskip
\begin{tabular}{c c c c c c}
\hline
\hline
$m_0=5$ MeV & 1 & 2 & 3 & 4 & 5\\
\hline
pole position $z_0$ & (-0.258,$\pm$0.192) & (-0.240,$\pm$2.595) & (-1.033,$\pm$2.847) & (-0.739, 0.0) & (-1.562,$\pm$4.950)\\
$\mbox{Res}(\sigma_S,z_0)$ & (-0.016,$\mp$0.511) & (0.040,$\pm$0.100) & (-0.050,$\mp$0.076) & (0.069,0.0) & (-0.024,$\mp$0.041)\\
$\mbox{Res}(\sigma_V,z_0)$ & (0.259,$\mp$0.860) & (0.023,$\mp$0.063) & (0.001,$\mp$0.052) & (-0.080,0.0) & (0.002,$\mp$0.021)\\
\hline
$m_0=5$ MeV & 6 & 7 & 8 & 9 & 10\\
\hline
pole position $z_0$ & (-0.619,$\pm$4.698) & (-1.015,$\pm$7.040) & (-2.986,$\pm$8.338) & (-1.051,$\pm$8.578) & (-0.841,$\pm$6.752) \\
$\mbox{Res}(\sigma_S,z_0)$ & (0.038,$\pm$0.058) & (-0.074,$\mp$0.067) & (-0.002,$\mp$0.001) & (0.033,$\pm$0.036)& (0.031,$\pm$0.051)\\
$\mbox{Res}(\sigma_V,z_0)$ & (0.004,$\mp$0.032) & (-0.004,$\mp$0.037) & (-0.001,$\mp$0.0003) & (-0.0002,$\mp$0.017)& (0.004,$\mp$0.023)\\
\hline
\hline
$m_0=115$ MeV & 1 & 2 & 3 & 4 & 5\\
\hline
pole position $z_0$ & (-0.439,$\pm$0.517) & (-0.505,$\pm$3.364) & (-1.484,$\pm$3.845) & (-3.256, 0.0)& (-1.490,$\pm$8.266)\\
$\mbox{Res}(\sigma_S,z_0)$ & (0.009,$\mp$0.491) & (0.055,$\pm$0.102) & (-0.056,$\mp$0.080) & (0.007,0.0)& (-0.080,$\mp$0.059)\\
$\mbox{Res}(\sigma_V,z_0)$ & (0.261,$\mp$0.539) & (0.013,$\mp$0.061) & (-0.0005,$\mp$0.048) & (0.004,0.0)& (-0.008,$\mp$0.033)\\
\hline
$m_0=115$ MeV & 6 & 7 & 8 & 9 & 10\\
\hline
pole position $z_0$ & (-2.114,$\pm$0.288) & (-0.919,$\pm$5.851) & (-1.135,$\pm$8.122) & (-2.066, 6.283)& (-1.382,$\pm$10.150)\\
$\mbox{Res}(\sigma_S,z_0)$ & (0.028,$\mp$0.043) & (0.041,$\pm$0.055) & (0.034,$\pm$0.048) & (-0.019,$\mp$0.047)& (0.032,$\pm$0.035)\\
$\mbox{Res}(\sigma_V,z_0)$ & (-0.021,$\mp$0.028) & (0.002,$\mp$0.028) & (0.002,$\mp$0.020) & (0.005,$\mp$0.019)& (-0.0004,$\mp$0.015)\\
\hline
\hline
\end{tabular}
\caption{Table of pole positions and -residues as found by the automated numerical procedure described in Section \ref{sec:numerical_procedures} for $m_0=5$ MeV and $m_0=115$ MeV, denoted as $(\Re x, \Im x)$. The poles are numbered in an arbitrary fashion, except for the first four, which have been picked such that they match the poles presented in Table \ref{tab:Dorkin_table} for comparison. The pole numbers chosen for $m_0=5$ MeV and $m_0=115$ MeV are not consistent in the sense that a pole labeled '6' for $m_0=5$ MeV does not necessarily belong to the same trajectory as the pole labeled '6' for $m_0=115$ MeV. For a consistent treatment of the properties of a pole, consult the trajectories discussed above.}
\label{tab:Andreas_table}
\end{table*}
A direct comparison of the pole properties presented in \cite{Dorkin:2013rsa} and the corresponding pole properties that have been extracted by the automated procedure used in this study shows that the results agree remarkably well.
\section{\label{sec:Supplemental_Material}Supplemental Material and usage of the data}
Because the amount of data produced in this study is too overwhelming to be included in the main text in an economic fashion, I provide the main data as supplemental material in form of the ASCII text file '\texttt{pole\_trajectory\_data.txt}'. This file contains all pole locations and residues for all 23 pole trajectories discussed in the text. For a given trajectory, the data is arranged in seven columns, which are summarized in Table \ref{tab:data_organization}.
\begin{table*}
\centering
\begin{tabular}{c|c|c|c|c|c|c|c}
\hline
\hline
column & 1 & 2 & 3 & 4 & 5 & 6 & 7\\
\hline
dimension &  [MeV] &  [GeV$^2$] &  [GeV$^2$] &  [GeV] &  [GeV] &  [1] &  [1]\\
\hline
content & $m_0$  & $\Re(z_0)$  & $\Im(z_0)$  & $\Re(\mbox{Res}(\sigma_S,z_0))$  & $\Im(\mbox{Res}(\sigma_S,z_0))$  & $\Re(\mbox{Res}(\sigma_V,z_0))$  & $\Im(\mbox{Res}(\sigma_V,z_0))$ \\
\hline
\hline
\end{tabular}
\caption{Organization of the data in the supplementary text file '\texttt{pole\_trajectory\_data.txt}'.}
\label{tab:data_organization}
\end{table*}
In addition, a second ASCII text file, '\texttt{HOWTO\_extract\_data\_from\_file.txt}', is included. This file contains a one-line command that allows for extraction of the data of a single pole trajectory from the file, together with an example that shows the application explicitly. The command should work in any linux/UNIX environment that has the program '\texttt{sed}' installed. The data can then be used to find good parametrizations for the infrared Maris-Tandy modeled quark propagator.
Assuming that a certain bare mass value $m_0$ is desired, as well as a certain bound state mass $M_{q\bar{q}}$, the following steps can be performed to identify the relevant poles and residues.
\begin{itemize}
\item\textbf{STEP 1: Consult Figure \ref{fig:all_pole_trajectories}}\\
For given values of $m_0$ and $M_{q\bar{q}}$, choose the relevant parabola in Figure \ref{fig:all_pole_trajectories} and note down all trajectories that intersect the parabola. 
\item\textbf{STEP 2: Extract the trajectories from the data file}\\
Extract the data of the trajectories that have been identified in \textbf{STEP 1}.
The data contains the real and imaginary part of the pole location, as well as the residues in $\sigma_S$ and $\sigma_V$ for bare mass values with a separation of 5 MeV. However, for very small residues it can happen that the a certain pole has not been tracked for every mass value. In that case, one can interpolate the data to find the desired values.
\item\textbf{STEP 3: Solve and fit the quark propagator DSE on the real axis}\\
In this study, a plethora of data containing complex solutions of the quark propagator has been produced. However, the amount of data is too overwhelming to be made available even as supplemental material. In order to still take advantage of the pole properties provided here, one can follow Section III B of \cite{Dorkin:2013rsa} to produce fits for the quark propagator, together with the representation that takes the pole positions and residues into account.
\end{itemize}
\section{\label{sec:summary}Summary and outlook}
In this study I presented results for the rainbow truncated quark propagator Dyson-Schwinger equation in the Landau gauge. The interaction has been modeled using the infrared part of the Maris-Tandy model, which renders the self-energy integrand analytic on the cut-plane $-\pi<\arg(x)<\pi$ of the square of external momenta. The angular integral can be solved analytically, and the contour of the remaining (radial) integral can be maintained on the real axis, even for complex external momenta. This scenario is thus an ideal basis to develop new techniques based on non-perturbative Dyson-Schwinger equations that are capable of providing robust and accurate solutions in the complex domain, as required for bound state equation such as the Bethe-Salpeter equation. Using an automated algorithm for the extraction of poles and residues, all poles for bound state masses of up to $M_{q\bar{q}}=4.5$ GeV and for a large range of bare quark masses have been identified and provided in form of plots and in form of raw data published as supplemental material.   
Several possibilities for future calculations exist. The next natural step is to extend the framework further to allow for a proper inclusion of the ultraviolet term of the Maris-Tandy interaction model. A first step towards this goal has been performed in \cite{Dorkin:2013rsa}, however, a full treatment that accounts for the non-analyticities arising in the complex plane of the radial integration variable as the external momentum is driven to complex values remains elusive. The status of this ongoing work is, that the obstructive branch cuts have been successfully identified analytically. The analytic prediction agrees perfectly with the numerical analysis, and I am currently working on the implementation of the necessary contour deformations. These findings will be made available in a future publication.  Once the framework has been set up, a wider range of interactions can be studied, like the Qin-Chang interaction model \cite{Qin:2011dd}, or more complicated vertex constructions that go beyond the tree-level tensor structure. This, in turn, bears the potential to answer more profound questions, such as the positivity properties of the Landau gauge quark propagator.

\section{Acknowledgments}
I thank Craig Roberts for discussions. Furthermore, I would like to express my gratitude towards my advisor Mark G. Alford, and I also thank David Hall, Sai Iyer and Richard Schmaeng from the Physics Department at Washington University in St.~Louis for providing the computational resources in form of a GPU cluster node that made these calculations possible. Support through the U.S. Department of Energy, Office of Science, Office of Nuclear Physics under Award Number \#DE-FG-02-05ER41375, as well as through the Austrian Science Fund (FWF), Schr\"odinger Fellowship J3800-N27 is acknowledged.  

\appendix
\section{\label{app:A}Some details of the rainbow truncation}
\subsection{Convention}
All calculations are performed in Euclidean space.
The standard representation for Gamma matrices is
\begin{equation}
\gamma^{k}=\left(\begin{array}{cc}
0 & -i\sigma^{k}\\
i\sigma^{k} & 0
\end{array}\right),\ \gamma^{4}=\left(\begin{array}{cc}
\mathbbm{1} & 0\\
0 & -\mathbbm{1}
\end{array}\right),\ \gamma^{5}=\left(\begin{array}{cc}
0 & \mathbbm{1}\\
\mathbbm{1} & 0
\end{array}\right),\label{eq:Convention1}
\end{equation}
 with $\sigma^{k}$ the Pauli matrices, such that
\begin{eqnarray}
\gamma^{\mu} & = & \left(\gamma^{\mu}\right)^{\dagger},\label{eq:Convention2}
\end{eqnarray}
\begin{eqnarray}
\gamma^{5} & = & -\gamma^{1}\gamma^{2}\gamma^{3}\gamma^{4},\label{eq:Convention3}
\end{eqnarray}
and
\begin{eqnarray}
\left(\gamma^{1}\right)^{2}=\left(\gamma^{2}\right)^{2}=\left(\gamma^{3}\right)^{2}= & \left(\gamma^{4}\right)^{2} & =\mathbbm{1}_D.\label{eq:Convention4}
\end{eqnarray}
In Euclidean space, the Clifford algebra is defined through
\begin{equation}
\{\gamma^{\mu},\gamma^{\nu}\}=2\delta^{\mu\nu}.\label{eq:Convention5}
\end{equation}

\subsection{Color space}

The color space part of the quark DSE is very simple. On the left, the inverse dressed quark propagator is diagonal in color space, $\delta_{\alpha\beta}$. On the right hand side, there are two terms, the bare inverse propagator, which also is diagonal in color space and thus just yields a delta function, and the quark self-energy, which gives a non-trivial contribution via the bare and dressed quark-gluon vertices. The vertices are connected through a dressed quark and a dressed gluon, which are both diagonal in color space, such that the color structure of the quark self-energy can be written as follows,

\begin{eqnarray}
\Sigma_{color}&=&\delta_{\gamma\delta}\delta^{ab}\left(t^a\right)_{\alpha\gamma}\left(t^b\right)_{\delta\beta}\\
&=&\left(t^a\right)_{\alpha\gamma}\left(t^a\right)_{\gamma\beta}.\nonumber
\end{eqnarray}

Making use of the Fiertz identity (see e.g. equation (8.4) in \cite{Borodulin:1995xd})

\begin{equation}
\left(t^a\right)_{\alpha\beta}\left(t^a\right)_{\gamma\delta}=\frac{1}{2}\left(\delta_{\alpha\delta}\delta_{\beta\gamma}-\frac{1}{N_c}\delta_{\alpha\beta}\delta_{\gamma\delta}\right),
\end{equation}

the color structure becomes

\begin{eqnarray}
\Sigma_{color}&=&\left(t^a\right)_{\alpha\gamma}\left(t^a\right)_{\gamma\beta}\\
&=&\frac{1}{2}\left(\delta_{\alpha\beta}\underbrace{\delta_{\gamma\gamma}}_{=N_c}-\frac{1}{N_c}\delta_{\alpha\gamma}\delta_{\gamma\beta}\right)\nonumber\\
&=&\frac{1}{2}\left(N_c-\frac{1}{N_c}\right)\delta_{\alpha\beta}\nonumber\\
&=&\frac{N_c^2-1}{2N_c}\delta_{\alpha\beta}.\nonumber
\end{eqnarray}

Since both, the left and the right hand side of the quark propagator are thus proportional to $\delta_{\alpha\beta}$, the trace thereof cancels and the quark self-energy integral is modified by the factor of the quadratic Casimir, $(N_c^2-1)/(2N_c)$, which, for $N_c=3$, evaluates to $\frac{4}{3}$.

\subsection{Dirac space}

The trivial traces on the left hand side of the equation, as well as on the bare inverse propagator on the right are not explicitly discussed here. However, for the sake of completeness I also present the Dirac traces of the quark self energy. The first trace of interest appears in the projection on the vector part of the propagator,

\begin{eqnarray}
&&\Sigma_A\left(p^2\right)\\
&=&\frac{1}{4p^{2}}\mbox{Tr}_{D}\left\{-i\cancel{p}\Sigma \right\}\nonumber\\
&=&\frac{1}{4p^{2}}\frac{4}{3}\int\frac{d^4q}{(2\pi)^4}\frac{\mathcal{G}\left((p-q)^2\right)}{q^2A^2\left(q^2\right)+B^2\left(q^2\right)}\mbox{Tr}_{D}\left\{-i\cancel{p}\gamma^\mu\right.\nonumber\\
&&\times\left.\left(-i\cancel{q}A\left(q^2\right)\right)\gamma^\nu\left(\delta^{\mu\nu}-\frac{(p-q)^\mu(p-q)^\nu}{(p-q)^2}\right) \right\}\nonumber\\
&=&-\frac{1}{4p^{2}}\frac{4}{3}\int\frac{d^4q}{(2\pi)^4}\frac{A\left(q^2\right)\mathcal{G}\left((p-q)^2\right)}{q^2A^2\left(q^2\right)+B^2\left(q^2\right)}\nonumber\\
&&\times\mbox{Tr}_D\left\{\cancel{p}\gamma^\mu\cancel{q}\gamma^\nu\left(\delta^{\mu\nu}-\frac{(p-q)^\mu(p-q)^\nu}{(p-q)^2}\right) \right\}\nonumber.
\end{eqnarray} 

Considering the trace only yields

\begin{eqnarray}
&&\mbox{Tr}_D\left\{\cancel{p}\gamma^\mu\cancel{q}\gamma^\nu\left(\delta^{\mu\nu}-\frac{(p-q)^\mu(p-q)^\nu}{(p-q)^2}\right) \right\}\\
&=&\mbox{Tr}_D\left\{\cancel{p}\gamma^\mu\cancel{q}\gamma^\mu\right\}-\frac{1}{(p-q)^2}\mbox{Tr}_D\left\{\cancel{p}(\cancel{p}-\cancel{q})\cancel{q}(\cancel{p}-\cancel{q})\right\}\nonumber\\
&=&p^\rho q^\sigma\mbox{Tr}_D\left\{\gamma^\rho\gamma^\mu\gamma^\sigma\gamma^\mu\right\}-\frac{1}{(p-q)^2}\bigg[\mbox{Tr}_D\left\{\cancel{p}\cancel{p}\cancel{q}\cancel{p}\right\}\nonumber\\
&&-\mbox{Tr}_D\left\{\cancel{p}\cancel{p}\cancel{q}\cancel{q}\right\}-\mbox{Tr}_D\left\{\cancel{p}\cancel{q}\cancel{q}\cancel{p}\right\}+\mbox{Tr}_D\left\{\cancel{p}\cancel{q}\cancel{q}\cancel{q}\right\}\bigg]\nonumber\\
&=&p^\rho q^\sigma\mbox{Tr}_D\left\{\gamma^\rho\left(2\delta^{\mu\sigma}-\gamma^\sigma\gamma^\mu\right)\gamma^\mu\right\}\nonumber\\
&&-\frac{1}{(p-q)^2}\bigg[\mbox{Tr}_D\left\{p^2\cancel{q}\cancel{p}\right\}-\mbox{Tr}_D\left\{p^2q^2\mathbbm{1}_D\right\}\nonumber\\
&&-\mbox{Tr}_D\left\{p^2q^2\mathbbm{1}_D\right\}+\mbox{Tr}_D\left\{q^2\cancel{p}\cancel{q}\right\}\bigg]\nonumber\\
&=&2\mbox{Tr}_D\left\{\cancel{p}\cancel{q}\right\}-\mbox{Tr}_D\{\cancel{p}\cancel{q}\underbrace{\gamma^\mu\gamma^\mu}_{=4\ \mathbbm{1}_D}\}\nonumber\\
&&-\frac{1}{(p-q)^2}\bigg[p^2(q.p)-2p^2q^2+q^2(q.p)\bigg]\mbox{Tr}_D\left\{\mathbbm{1}_D\right\}\nonumber\\
&=&-8(q.p)-4\frac{(p^2+q^2)(q.p)-2p^2q^2}{(p-q)^2}\nonumber.
\end{eqnarray}

The projection of the quark self-energy is then given by

\begin{eqnarray}
\label{sigma_A_1}
&&\Sigma_A\left(p^2\right)\\
&=&\frac{4}{3p^2}\int\frac{d^4q}{(2\pi)^4}\frac{A\left(q^2\right)\mathcal{G}\left((p-q)^2\right)}{q^2A^2\left(q^2\right)+B^2\left(q^2\right)}\nonumber\\
&&\times\left(2(q.p)+\frac{(p^2+q^2)(q.p)-2p^2q^2}{(p-q)^2}\right)\nonumber
\end{eqnarray}

On the other hand, the projection on the scalar part gives rise to the self-energy contribution

\begin{eqnarray}
\label{sigma_B_1}
&&\Sigma_B\left(p^2\right)\\
&=&\frac{1}{4}\mbox{Tr}_{D}\left\{\Sigma \right\}\nonumber\\
&=&\frac{1}{4}\frac{4}{3}\int\frac{d^4q}{(2\pi)^4}\frac{\mathcal{G}\left((p-q)^2\right)}{q^2A^2\left(q^2\right)+B^2\left(q^2\right)}\mbox{Tr}_{D}\left\{\gamma^\mu\right.\nonumber\\
&&\times\left.\left(B\left(q^2\right)\right)\gamma^\nu\left(\delta^{\mu\nu}-\frac{(p-q)^\mu(p-q)^\nu}{(p-q)^2}\right) \right\}\nonumber\\
&=&\frac{1}{4}\frac{4}{3}\int\frac{d^4q}{(2\pi)^4}\frac{B\left(q^2\right)\mathcal{G}\left((p-q)^2\right)}{q^2A^2\left(q^2\right)+B^2\left(q^2\right)}\nonumber\\
&&\times\bigg[\mbox{Tr}_{D}\left\{\gamma^\mu\gamma^\mu\right\}-\mbox{Tr}_D\left\{\frac{(\cancel{p}-\cancel{q})(\cancel{p}-\cancel{q})}{(p-q)^2}\right\}\bigg]\nonumber\\
&=&\frac{1}{4}\frac{4}{3}\int\frac{d^4q}{(2\pi)^4}\frac{B\left(q^2\right)\mathcal{G}\left((p-q)^2\right)}{q^2A^2\left(q^2\right)+B^2\left(q^2\right)}\nonumber\\
&&\times(16-4)\nonumber\\
&=&\frac{4}{3}\int\frac{d^4q}{(2\pi)^4}\frac{3B\left(q^2\right)\mathcal{G}\left((p-q)^2\right)}{q^2A^2\left(q^2\right)+B^2\left(q^2\right)}\nonumber.
\end{eqnarray}

\subsection{Hyperspherical coordinates}

The integration over the 4-momentum $q$ can be expressed through hyperspherical coordinates as follows,

\begin{eqnarray}
&&\int_{\mathbb{R}^{4}}d^{4}q \rightarrow\\ 
&&\int_{0}^{2\pi}d\phi\int_{0}^{\infty}dq\ q^{3}\int_{0}^{\pi}d\theta_{1}\sin^{2}\theta_{1}\int_{0}^{\pi}d\theta_{2}\sin\theta_{2}\label{eq:Convention7}\nonumber\\
 & = & \Bigg|\begin{array}{c}
y\equiv q^{2}\rightarrow dy=2qdq\nonumber\\
\theta_{1}\equiv\arccos z\rightarrow d\theta_{1}=-\frac{dz}{\sqrt{1-z^{2}}}\\
\theta_{2}\equiv\arccos w\rightarrow d\theta_{2}=-\frac{dw}{\sqrt{1-w^{2}}}
\end{array}\Bigg|\\
 & = & \frac{1}{2}\int_{0}^{2\pi}d\phi\int_{0}^{\infty}dy\ y\int_{-1}^{1}dz\sqrt{1-z^{2}}\int_{1}^{1}dw.\nonumber 
\end{eqnarray}

Since there are only two different momenta in the quark self-energy integral, the 'external' momentum $p$ and the 'internal' (loop) momentum $q$, there is only the radial and one angular integral that is non-trivial. Performing the two trivial integrations, and introducing an IR cutoff $\varepsilon$, as well as an UV cutoff $\Lambda$, the self-energy integral becomes

\begin{eqnarray}
&&\int_{\mathbb{R}^{4}}\frac{d^{4}q}{(2\pi)^4} \rightarrow\\ 
&&\frac{1}{(2\pi)^3}\int_{\varepsilon}^{\Lambda}dy\ y\int_{-1}^{1}dz\sqrt{1-z^{2}}\nonumber.
\end{eqnarray}

Introducing the variable $x$ for the square of the external momentum $p^2$,

\begin{eqnarray}
x := p^2,
\end{eqnarray}

the scalar products appearing in the integrand of the self-energy can be rewritten,

\begin{eqnarray}
p.q = \sqrt{x}\sqrt{y}z.
\end{eqnarray}

Switching to hyperspherical coordinates, as well as using the variable $x$, the quark-self energy contributions $\Sigma_A$ and $\Sigma_B$ become

\begin{eqnarray}
\label{sigma_A_2}
&&\Sigma_A\left(x\right) = \frac{1}{6\pi^3}\int_\varepsilon^\Lambda dy y\frac{A\left(y\right)}{yA^2\left(y\right)+B^2\left(y\right)}\\
&&\times\int_{-1}^{+1}dz\sqrt{1-z^2}\mathcal{G}\left(x+y-2\sqrt{x}\sqrt{y}z\right)\nonumber\\
&&\times\left(2\frac{\sqrt{y}}{\sqrt{x}}z+\frac{(1+\frac{y}{x})\sqrt{x}\sqrt{y}z-2y}{x+y-2\sqrt{x}\sqrt{y}z}\right)\nonumber,
\end{eqnarray}

\begin{eqnarray}
\label{sigma_B_2}
&&\Sigma_B\left(x\right) = \frac{1}{6\pi^3}\int_\varepsilon^\Lambda dy y\frac{3B\left(y\right)}{yA^2\left(y\right)+B^2\left(y\right)}\\
&&\times\int_{-1}^{+1}dz\sqrt{1-z^2}\mathcal{G}\left(x+y-2\sqrt{x}\sqrt{y}z\right)\nonumber.
\end{eqnarray}


\section{\label{app:ang_int}The angular integral of the IR part of the interaction model}
In this section I present the analytic solution of the angular integral of the IR part of the Maris-Tandy interaction. The first integral in  question is the angular integral in the quark self-energy contribution $\Sigma_A$. The integral reads
\begin{eqnarray}
&&\int_{-1}^{+1}dz\sqrt{1-z^2}\bigg[\underbrace{-\frac{2}{3}y}_{=:c1}+\underbrace{\left(1+\frac{y}{x}\right)\sqrt{x}\sqrt{y}}_{=:c_2}z\underbrace{-\frac{4}{3}y}_{=:c_3}z^2\bigg]\nonumber\\
&&\times\exp\bigg\{\underbrace{-\frac{x+y}{\omega^2}}_{=:c_4}+\underbrace{\frac{2\sqrt{x}\sqrt{y}}{\omega^2}}_{=:c_5}z \bigg\}\\
&=&\bigg|\begin{array}{c}
z=\cos\theta\\
\int_{-1}^{+1}dz\sqrt{1-z^2}\rightarrow\int_0^\pi d\theta\sin^2\theta
\end{array}\bigg|\nonumber\\
&=&\int_0^\pi d\theta\sin^2\theta\bigg[c_1+c_2\cos\theta+c_3\cos^2\theta\bigg]\underbrace{\exp\{c_4\}}_{=:c_6}\nonumber\\
&&\times\exp\{c_5\cos\theta\}.\nonumber
\end{eqnarray}
These integrals can be rewritten in such a way that they correspond to integral representations of the modified Bessel functions of the first kind. Those functions are holomorphic on the cut-plane $\pi>\arg(x)>-\pi$, for $x\in\mathbb{C}$.

The integral representations of the modified Bessel functions of the first kind can be found in the book of Abramowitz and Stegun, page 376 \cite{Abramowitz:1972mf}. Here I just present the two relevant relations \textbf{9.6.18} and \textbf{9.6.19},
\begin{equation}
\label{Bessel_int_1}
I_\nu(z) =\frac{\left(\frac{1}{2}z\right)^\nu}{\pi^{\frac{1}{2}}\Gamma\left(\nu+\frac{1}{2}\right)}\int_0^\pi d\theta\exp\{\pm z\cos\theta\}\sin^{2\nu}\theta,
\end{equation}
\begin{equation}
\label{Bessel_int_2}
I_n(z) =\frac{1}{\pi}\int_0^\pi d\theta\exp\{z\cos\theta\}\cos{n\theta},
\end{equation}
where $z\in\mathbb{C}$, $\Re\nu>-\frac{1}{2}$ and $n\in\mathbb{Z}_+\cup\{0\}$. $\Gamma$ is the Gamma function, which reduces to the shifted factorial $(n-1)!$ for $n\in\mathbb{Z}_+$.
Now the integrals have to be identified,
\begin{eqnarray}
&&\int_0^\pi d\theta\sin^2\theta\bigg[c_1+c_2\cos\theta+c_3\cos^2\theta\bigg]c_6\exp\{c_5\cos\theta\}\nonumber\\
&=&c_1c_6\int_0^\pi d\theta\sin^2\theta\exp\{c_5\cos\theta\}\\
&&+c_2c_6\int_0^\pi d\theta\sin^2\theta\cos\theta\exp\{c_5\cos\theta\}\nonumber\\
&&+c_3c_6\int_0^\pi d\theta\sin^2\theta\cos^2\theta\exp\{c_5\cos\theta\}\nonumber\\
&=:&T_1 + T_2 + T_3.\nonumber
\end{eqnarray}
$T_1$ is already in the form of (\ref{Bessel_int_1}). Using $\nu=1$ yields
\begin{eqnarray} 
\label{T1_int}
T_1&=&c_1c_6\int_0^\pi d\theta\sin^2\theta\exp\{c_5\cos\theta\}\\
&\stackrel{(\ref{Bessel_int_1})}{=}&c_1c_6\frac{\pi^\frac{1}{2}\overbrace{\Gamma\left(\frac{3}{2}\right)}^{=\sqrt{\pi}/2}}{\frac{1}{2}c_5}I_1\left(\frac{2\sqrt{x}\sqrt{y}}{\omega^2}\right)\nonumber\\
&=&-\frac{\pi\omega^2}{3}\frac{\sqrt{y}}{\sqrt{x}}\exp\left\{-\frac{x+y}{\omega^2}\right\}I_1\left(\frac{2\sqrt{x}\sqrt{y}}{\omega^2}\right).\nonumber
\end{eqnarray} 
For $T_2$, the following relation can be exploited,
\begin{eqnarray}
\cos3\theta = -3\sin^2\theta\cos\theta+\cos^3\theta,
\end{eqnarray}
such that
\begin{eqnarray}
\sin^2\theta\cos\theta &=& -\frac{1}{3}\left(\cos3\theta-\cos^3\theta\right)\\
&=& -\frac{1}{3}\left(\cos3\theta-\cos\theta+\cos\theta\sin^2\theta\right)\nonumber\\
&=& -\frac{1}{4}\left(\cos3\theta-\cos\theta\right)\nonumber,
\end{eqnarray}
which takes integral $T_2$ into the form of (\ref{Bessel_int_2}),
\begin{eqnarray} 
T_2&=&c_2c_6\int_0^\pi d\theta\sin^2\theta\cos\theta\exp\{c_5\cos\theta\}\\
&=&c_2c_6\frac{1}{4}\int_0^\pi d\theta\bigg[\cos\theta-\cos3\theta\bigg]\exp\{c_5\cos\theta\}\nonumber\\
&\stackrel{(\ref{Bessel_int_2})}{=}&c_2c_6\frac{\pi}{4}\left[I_1\left(\frac{2\sqrt{x}\sqrt{y}}{\omega^2}\right)-I_3\left(\frac{2\sqrt{x}\sqrt{y}}{\omega^2}\right)\right]\nonumber\\
&=&\frac{\pi}{4}\left(1+\frac{y}{x}\right)\sqrt{x}\sqrt{y}\exp\left\{-\frac{x+y}{\omega^2}\right\}\nonumber\\
&&\times\left[I_1\left(\frac{2\sqrt{x}\sqrt{y}}{\omega^2}\right)-I_3\left(\frac{2\sqrt{x}\sqrt{y}}{\omega^2}\right)\right]\nonumber.
\end{eqnarray} 
The difference of the two Bessel functions can be rewritten as follows. Using the recurrence relation (see \textbf{9.6.26} in the book of Abramowitz and Stegun \cite{Abramowitz:1972mf})
\begin{eqnarray}
I_{\nu-1}(z)-I_{\nu+1}(z)=\frac{2\nu}{z}I_\nu(z),
\end{eqnarray}
with $\nu=2$, $T_2$ can be simplified to
\begin{eqnarray} 
T_2&=&c_2c_6\int_0^\pi d\theta\sin^2\theta\cos\theta\exp\{c_5\cos\theta\}\\
&=&\frac{\omega^2\pi}{2}\left(1+\frac{y}{x}\right)\exp\left\{-\frac{x+y}{\omega^2}\right\}I_2\left(\frac{2\sqrt{x}\sqrt{y}}{\omega^2}\right)\nonumber.
\end{eqnarray} 
Finally, $T_3$ can be rewritten to yield
\begin{eqnarray} 
T_3&=&c_3c_6\int_0^\pi d\theta\sin^2\theta\cos^2\theta\exp\{c_5\cos\theta\}\\
&=&c_3c_6\int_0^\pi d\theta\bigg[\cos^2\theta-\cos^4\theta\bigg]\exp\{c_5\cos\theta\}\nonumber\\
&=&c_3c_6\int_0^\pi d\theta\bigg[1-\sin^2\theta-\left(1-\sin^2\theta\right)^2\bigg]\nonumber\\
&&\times\exp\{c_5\cos\theta\}\nonumber\\
&=&c_3c_6\int_0^\pi d\theta\bigg[\sin^2\theta-\sin^4\theta\bigg]\exp\{c_5\cos\theta\}\nonumber,
\end{eqnarray} 
which is of the form of \ref{Bessel_int_1}. The first integral is the same as \ref{T1_int}, and $T_3$ becomes
\begin{eqnarray} 
T_3&=&c_3c_6\int_0^\pi d\theta\bigg[\sin^2\theta-\sin^4\theta\bigg]\exp\{c_5\cos\theta\}\nonumber\\
&=&c_3c_6\frac{\pi}{c_5}I_1\left(\frac{2\sqrt{x}\sqrt{y}}{\omega^2}\right)-c_3c_6\frac{3\pi}{c_5^2}I_2\left(\frac{2\sqrt{x}\sqrt{y}}{\omega^2}\right)\nonumber\\
&=&\frac{c_3c_6\pi}{c_5}\bigg[I_1\left(\frac{2\sqrt{x}\sqrt{y}}{\omega^2}\right)-\frac{3}{c_5}I_2\left(\frac{2\sqrt{x}\sqrt{y}}{\omega^2}\right)\bigg]\nonumber\\
&=& -\frac{4}{3}y\frac{\pi\omega^2}{2\sqrt{x}\sqrt{y}}\exp\left\{-\frac{x+y}{\omega^2}\right\}I_1\left(\frac{2\sqrt{x}\sqrt{y}}{\omega^2}\right)\\
&& +2\omega^4\frac{\sqrt{y}}{\sqrt{x}}\frac{\pi}{2\sqrt{x}\sqrt{y}}\exp\left\{-\frac{x+y}{\omega^2}\right\}I_2\left(\frac{2\sqrt{x}\sqrt{y}}{\omega^2}\right)\nonumber\\
&=& -\frac{2\pi}{3}\frac{\sqrt{y}}{\sqrt{x}}\omega^2\exp\left\{-\frac{x+y}{\omega^2}\right\}I_1\left(\frac{2\sqrt{x}\sqrt{y}}{\omega^2}\right)\nonumber\\
&& +\frac{\pi}{x}\omega^4\exp\left\{-\frac{x+y}{\omega^2}\right\}I_2\left(\frac{2\sqrt{x}\sqrt{y}}{\omega^2}\right).\nonumber
\end{eqnarray} 
Combining the three individual integration results finally yields
\begin{eqnarray}
&&\int_{-1}^{+1}dz\sqrt{1-z^2}\bigg[-\frac{2}{3}y+\left(1+\frac{y}{x}\right)\sqrt{x}\sqrt{y}z-\frac{4}{3}yz^2\bigg]\nonumber\\
&&\times\exp\bigg\{-\frac{x+y}{\omega^2}+\frac{2\sqrt{x}\sqrt{y}}{\omega^2}z \bigg\}\\
&&=T_1 + T_2 + T_3\nonumber\\
&&= -\frac{\pi\omega^2}{3}\frac{\sqrt{y}}{\sqrt{x}}\exp\left\{-\frac{x+y}{\omega^2}\right\}I_1\left(\frac{2\sqrt{x}\sqrt{y}}{\omega^2}\right)\nonumber\\
&&+\frac{\omega^2\pi}{2}\left(1+\frac{y}{x}\right)\exp\left\{-\frac{x+y}{\omega^2}\right\}I_2\left(\frac{2\sqrt{x}\sqrt{y}}{\omega^2}\right)\nonumber\\
&&-\frac{2\pi}{3}\frac{\sqrt{y}}{\sqrt{x}}\omega^2\exp\left\{-\frac{x+y}{\omega^2}\right\}I_1\left(\frac{2\sqrt{x}\sqrt{y}}{\omega^2}\right)\nonumber\\
&&+\frac{\pi}{x}\omega^4\exp\left\{-\frac{x+y}{\omega^2}\right\}I_2\left(\frac{2\sqrt{x}\sqrt{y}}{\omega^2}\right),\nonumber
\end{eqnarray}
which can be simplified to give
\begin{eqnarray}
&&\int_{-1}^{+1}dz\sqrt{1-z^2}\bigg[-\frac{2}{3}y+\left(1+\frac{y}{x}\right)\sqrt{x}\sqrt{y}z-\frac{4}{3}yz^2\bigg]\nonumber\\
&&\times\exp\bigg\{-\frac{x+y}{\omega^2}+\frac{2\sqrt{x}\sqrt{y}}{\omega^2}z \bigg\}\\
&=&\frac{\pi\omega^2}{2}\exp\left\{-\frac{x+y}{\omega^2}\right\}\bigg[-\frac{\sqrt{y}}{\sqrt{x}}I_1\left(\frac{2\sqrt{x}\sqrt{y}}{\omega^2}\right)\nonumber\\
&&+\left(1+\frac{\sqrt{y}}{\sqrt{x}}+2\frac{\omega^2}{x}\right)I_2\left(\frac{2\sqrt{x}\sqrt{y}}{\omega^2}\right)\bigg].\nonumber
\end{eqnarray}
The angular integral of $\Sigma_B$ can be treated in a similar fashion,
\begin{eqnarray}
&&\int_{-1}^{+1}dz\sqrt{1-z^2}\bigg[\underbrace{x+y}_{=:c_7}\underbrace{-2\sqrt{x}\sqrt{y}}_{=:c_8}z\bigg]\nonumber\\
&&\times\exp\bigg\{\underbrace{-\frac{x+y}{\omega^2}}_{=:c_4}\underbrace{+\frac{2\sqrt{x}\sqrt{y}}{\omega^2}}_{=:c_5}z \bigg\}\\
&=&\int_{0}^{\pi}d\theta\sin^2\theta\bigg[ c_7+c_8\cos\theta\bigg]\underbrace{\exp\{c_4\}}_{=:c_6}\exp\{c_5\cos\theta\}\nonumber\\
&=&c_6c_7\int_{0}^{\pi}d\theta\sin^2\theta \exp\{c_5\cos\theta\}\nonumber\\
&=&c_6c_8\int_{0}^{\pi}d\theta\underbrace{\sin^2\theta\cos\theta}_{=-\frac{1}{4}(\cos3\theta-\cos\theta)} \exp\{c_5\cos\theta\}\nonumber\\
&=&c_6c_7\int_{0}^{\pi}d\theta\sin^2\theta \exp\{c_5\cos\theta\}\nonumber\\
&&+c_6c_8\frac{1}{4}\int_{0}^{\pi}d\theta\cos\theta \exp\{c_5\cos\theta\}\nonumber\\
&&-c_6c_8\frac{1}{4}\int_{0}^{\pi}d\theta\cos3\theta \exp\{c_5\cos\theta\}\nonumber\\
&=:&T_4+T_5+T_6\nonumber.
\end{eqnarray}
These integrals are the same as the ones above, and it follows that 
\begin{eqnarray}
T_4&=&\exp\left\{-\frac{x+y}{\omega^2}\right\}(x+y)\frac{\omega^2\pi}{2\sqrt{x}\sqrt{y}}I_1\left(\frac{2\sqrt{x}\sqrt{y}}{\omega^2}\right)\nonumber\\
&=&\exp\left\{-\frac{x+y}{\omega^2}\right\}\frac{\omega^2\pi}{2}\left(\frac{\sqrt{x}}{\sqrt{y}}+\frac{\sqrt{y}}{\sqrt{x}}\right)\\
&&\times I_1\left(\frac{2\sqrt{x}\sqrt{y}}{\omega^2}\right)\nonumber.
\end{eqnarray}
\begin{eqnarray}
T_5&=&-\exp\left\{-\frac{x+y}{\omega^2}\right\}\frac{\sqrt{x}\sqrt{y}\pi}{2}I_1\left(\frac{2\sqrt{x}\sqrt{y}}{\omega^2}\right),\nonumber\\
\end{eqnarray}
\begin{eqnarray}
T_6&=&\exp\left\{-\frac{x+y}{\omega^2}\right\}\frac{\sqrt{x}\sqrt{y}\pi}{2}I_3\left(\frac{2\sqrt{x}\sqrt{y}}{\omega^2}\right),\nonumber\\
\end{eqnarray}
such that the last two Bessel functions can be combined by using the recurrence relation as before. The final result is then given by
\begin{eqnarray}
&&\int_{-1}^{+1}dz\sqrt{1-z^2}\bigg[x+y-2\sqrt{x}\sqrt{y}z\bigg]\nonumber\\
&&\times\exp\bigg\{-\frac{x+y}{\omega^2}+\frac{2\sqrt{x}\sqrt{y}}{\omega^2}z \bigg\}\\
&=&\frac{\omega^2\pi}{2}\exp\left\{-\frac{x+y}{\omega^2}\right\}\bigg[\left(\frac{\sqrt{x}}{\sqrt{y}}+\frac{\sqrt{y}}{\sqrt{x}}\right)\nonumber\\
&&\times I_1\left(\frac{2\sqrt{x}\sqrt{y}}{\omega^2}\right)-2I_2\left(\frac{2\sqrt{x}\sqrt{y}}{\omega^2}\right)\bigg]\nonumber.
\end{eqnarray}

\bibliographystyle{utphys_mod}
\bibliography{citations}

\end{document}